%% file: mipsgal_bubbles_low_resolution.tex
\documentclass[apj]{emulateapj}

\usepackage[squaren]{SIunits}
\usepackage{multirow}
\usepackage{float}
\usepackage{epsfig}
\usepackage{graphicx}
\usepackage[caption=false]{subfig}
\usepackage{natbib}
\usepackage[T1]{fontenc}
\usepackage[figuresright]{rotating}

\newcommand{\mic}{~\micro\meter}

\begin{document}

\title{Spitzer/InfraRed Spectrograph Investigation of MIPSGAL 24\mic\ Compact Bubbles : low resolution observations}

\author{M. Nowak\altaffilmark{1,2}}
\author{N. Flagey\altaffilmark{3,4}}
\author{A. Noriega-Crespo\altaffilmark{2,5}}
\author{N. Billot\altaffilmark{6}}
\author{S. J. Carey\altaffilmark{2}}
\author{R. Paladini\altaffilmark{7}}
\author{S. D. Van Dyk\altaffilmark{2}}

\email{mathias.nowak@ens-cachan.fr}

\altaffiltext{1}{D\'epartement de Physique, \'Ecole Normale Sup\'erieure de Cachan, 61 avenue du Pr\'esident Wilson, 94235 Cachan, France}
\altaffiltext{2}{Spitzer Science Center, California Institute of Technology, 1200 East California Boulevard, MC 314-6, Pasadena, CA 91125, USA}
\altaffiltext{3}{Jet Propulsion Laboratory, California Institute of Technology, 4800 Oak Grove Drive, Pasadena, CA 91109, USA}
\altaffiltext{4}{Institute for Astronomy, 640 North A'ohoku Place, Hilo, HI 96720-2700, USA}
\altaffiltext{5}{Space Telescope Science Institute, 3700 San Martin Dr, Baltimore, MD 21218, USA}
\altaffiltext{6}{Instituto de Radio Astronom\'ia Milim\'etrica, Avenida Divina Pastora, 7, Local 20, 18012 Granada, Spain}
\altaffiltext{7}{NASA Herschel Science Center, California Institute of Technology, Pasadena, CA 91125, USA}
%\altaffiltext{8}{SOFIA Science Center, NASA Ames Research Center, MS 232-12, Moffett Field, CA 94035, USA}

\begin{abstract}

We present Spitzer/IRS low resolution observations of 11 compact circumstellar bubbles from the MIPSGAL 24\mic\ Galactic Plane Survey. We find that this set of MIPSGAL bubbles (MBs) is divided into two categories, and that this distinction correlates with the morphologies of the MBs in the mid-IR. The four MBs with central sources in the mid-IR exhibit dust-rich, low excitation spectra, and their 24\mic\ emission is accounted for by the dust continuum. The seven MBs without central sources in the mid-IR have spectra dominated by high excitation gas lines (e.g., [O~\textsc{iv}]~26.0\mic, [Ne~\textsc{v}]~14.3 and 24.3\mic, [Ne~\textsc{iii}]~15.5\mic), and the [O~\textsc{iv}] line accounts for 50 to almost 100\% of the 24\mic\ emission in five of them. In the dust-poor MBs, the [Ne~\textsc{v}] and [Ne~\textsc{iii}] line ratios correspond to high excitation conditions. Based on comparisons with published IRS spectra, we suggest that the dust-poor MBs are highly excited planetary nebulae with peculiar white dwarfs (e.g., [WR], novae) at their centers. The central stars of the four dust-rich MBs are all massive star candidates. Dust temperatures range from 40 to 100~K in the outer shells. We constrain the extinction along the lines of sight from the IRS spectra. We then derive distance, dust masses, and dust production rate estimates for these objects. 
%estimates, determine dust masses between a few $10^{-4}$ and a few $10^{-3}M_\odot$, and dust production rates of a few $10^{-6}M_\odot/$yr to $10^{-4}M_\odot/$yr. 
These estimates are all consistent with the nature of the central stars. We summarize the identifications of MBs made to date and discuss the correlation between their mid-IR morphologies and natures. Candidate Be/B[e]/LBV and WR stars are mainly ``rings'' with mid-IR central sources, whereas PNe are mostly ``disks'' without mid-IR central sources. Therefore we expect that most of the 300 remaining unidentified MBs will be classified as PNe.

\end{abstract}

\section{Introduction}

Massive stars only represent a small fraction of all the stars in our Galaxy. However, they have an important impact on the interstellar medium (ISM). Stars from the asymptotic giant branch are among the main producers of dust in the Galaxy and beyond. The strong winds and radiations generated by supernovae, Wolf-Rayet stars (WR), and Luminous Blue Variables (LBV)  convey considerable amounts of kinetic energy, ionizing photons, and materials to the ISM. As such, these stars are the sources of significant chemical and energetic feedback mechanisms.

The \citet{Mizuno2010} catalog revealed more than 400 {objects that look like shells, rings, and disks}, and which have been discovered from visual inspection of the MIPSGAL~24\mic~Galactic plane survey \citep{Carey2009}. These MIPSGAL {``bubbles''} (MBs) are scattered all over the Galactic plane, exhibit many different morphologies, and span a large range of sizes and fluxes. The very large majority of the MBs are detected at 24\mic\ only. Around 85\% of them were unidentified at the time of their discovery, and still more than 70\% remain unidentified at the time of this paper. All the {identified} MBs are {associated with} massive and/or evolved stars: OB stars, red giants and supergiants, planetary nebulae (PNe), WR stars, LBV candidates, or other type of emission line stars (e.g., Be, B[e]). Their mid-infrared (IR) emission may trace either warm dust or ionized gas in the circumstellar environment of the stars. About 300 MBs remain to be identified and there is thus a great potential to discover many more massive and/or evolved stars that have been hidden until now in the Galactic plane.

Nearly 3000 PNe are known in the Galaxy \citep{Frew2010} but only a few hundred central stars with spectral types are reported in the Strasbourg-ESO Catalogue of Galactic PNe \citep{Acker1992}. A few novae were found every year (149 novae brighter than V=10 between 1900 and 2000), while at least a factor of a few more are expected \citep[$\sim30$~yr$^{-1}$, e.g][]{Shafter2002}. Several thousand WRs are expected to be located within the Galaxy, but only a few hundreds of them have been found so far \citep{vanderHucht2001, vanderHucht2006, Mauerhan2011, Kanarek2014, Faherty2014, Shara2012}. Only a few tens of LBVs are known or suggested as candidates \citep{Clark2005}. The MBs are thus good candidates to account for part of these ``missing'' stars in the Milky-Way.

\begin{figure*}
  \begin{center}
    \subfloat[4384]{\includegraphics[height=4.5cm]{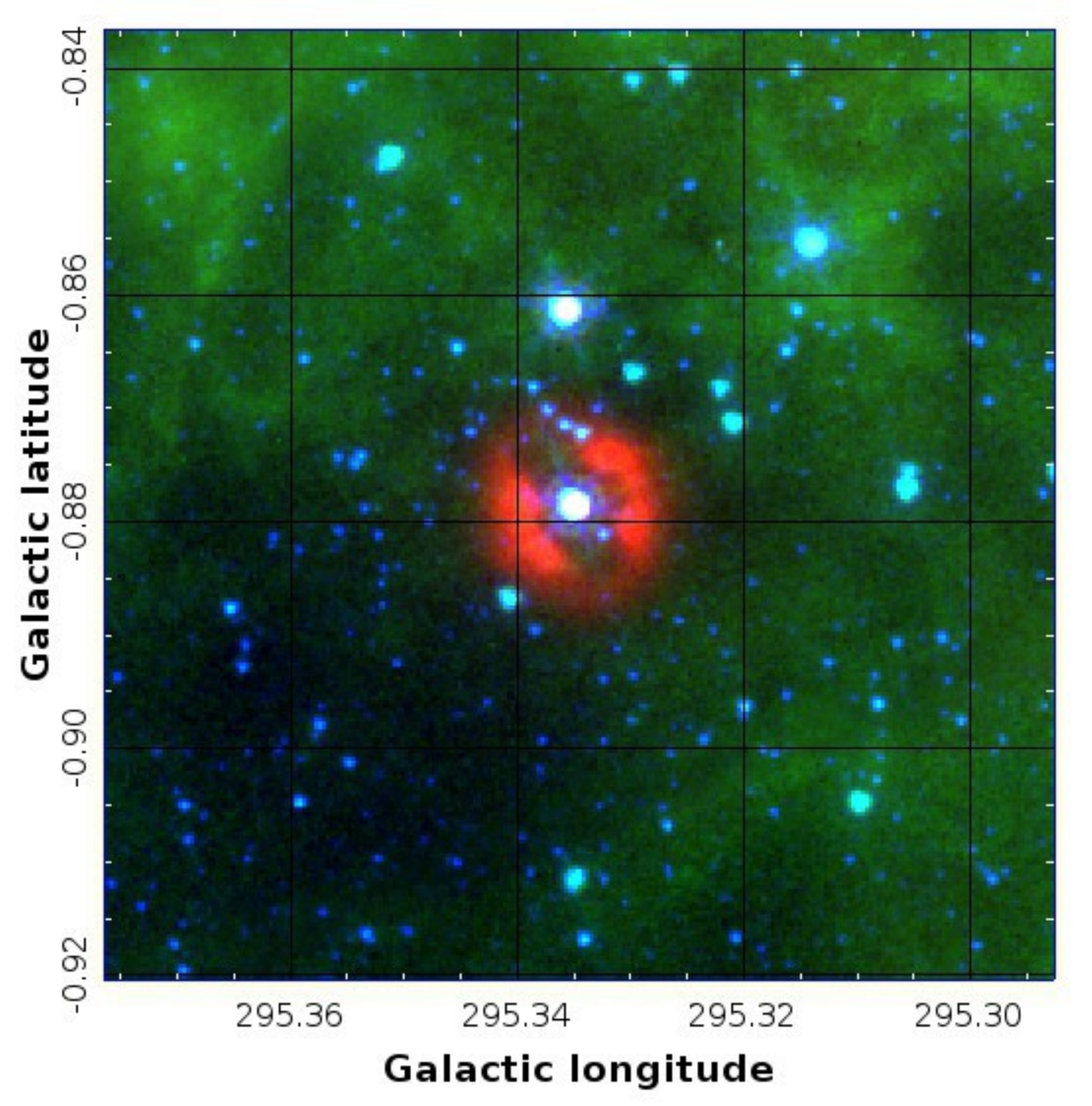}}
    \subfloat[4376]{\includegraphics[height=4.5cm]{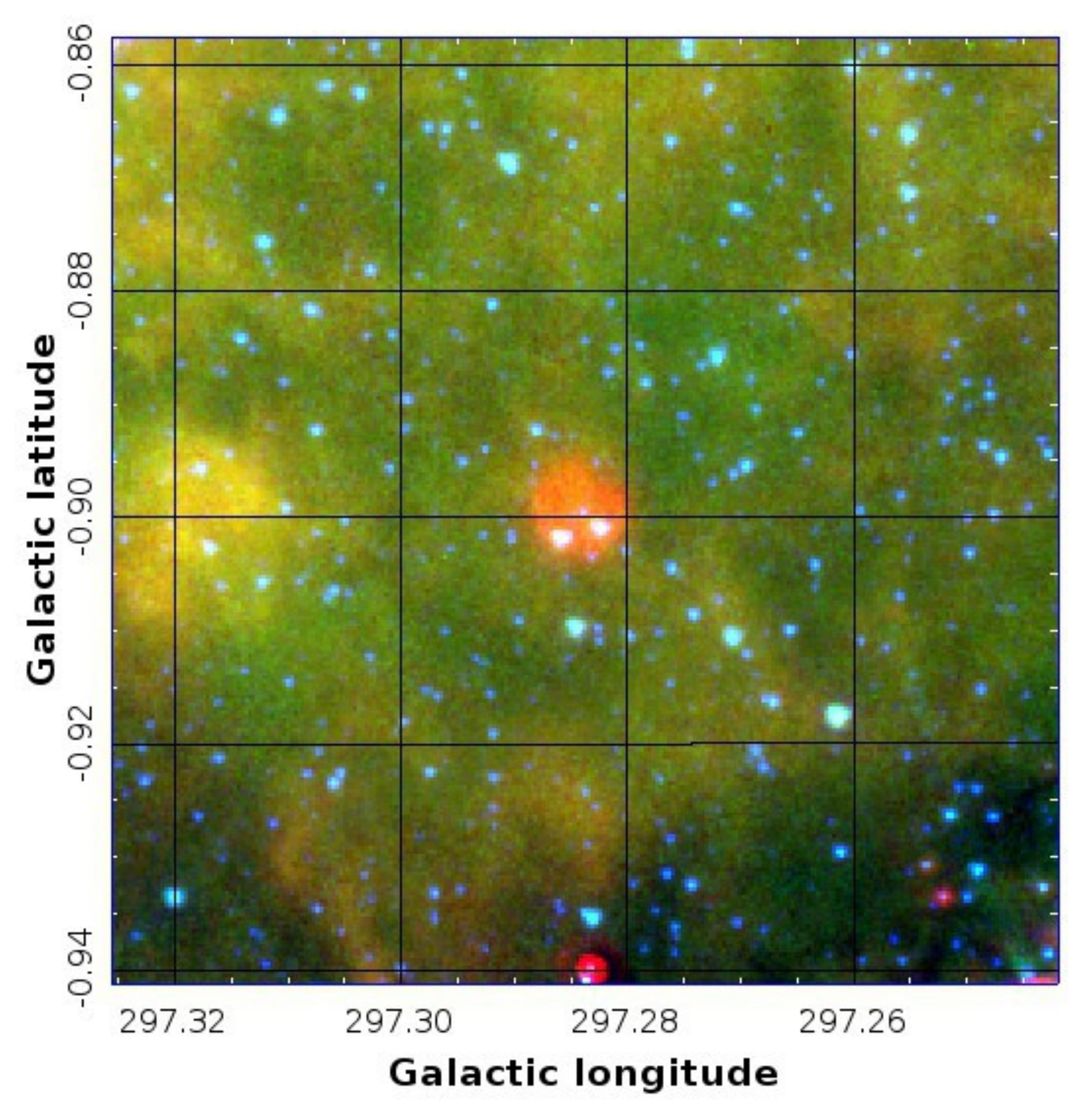}}
    \subfloat[3944]{\includegraphics[height=4.5cm]{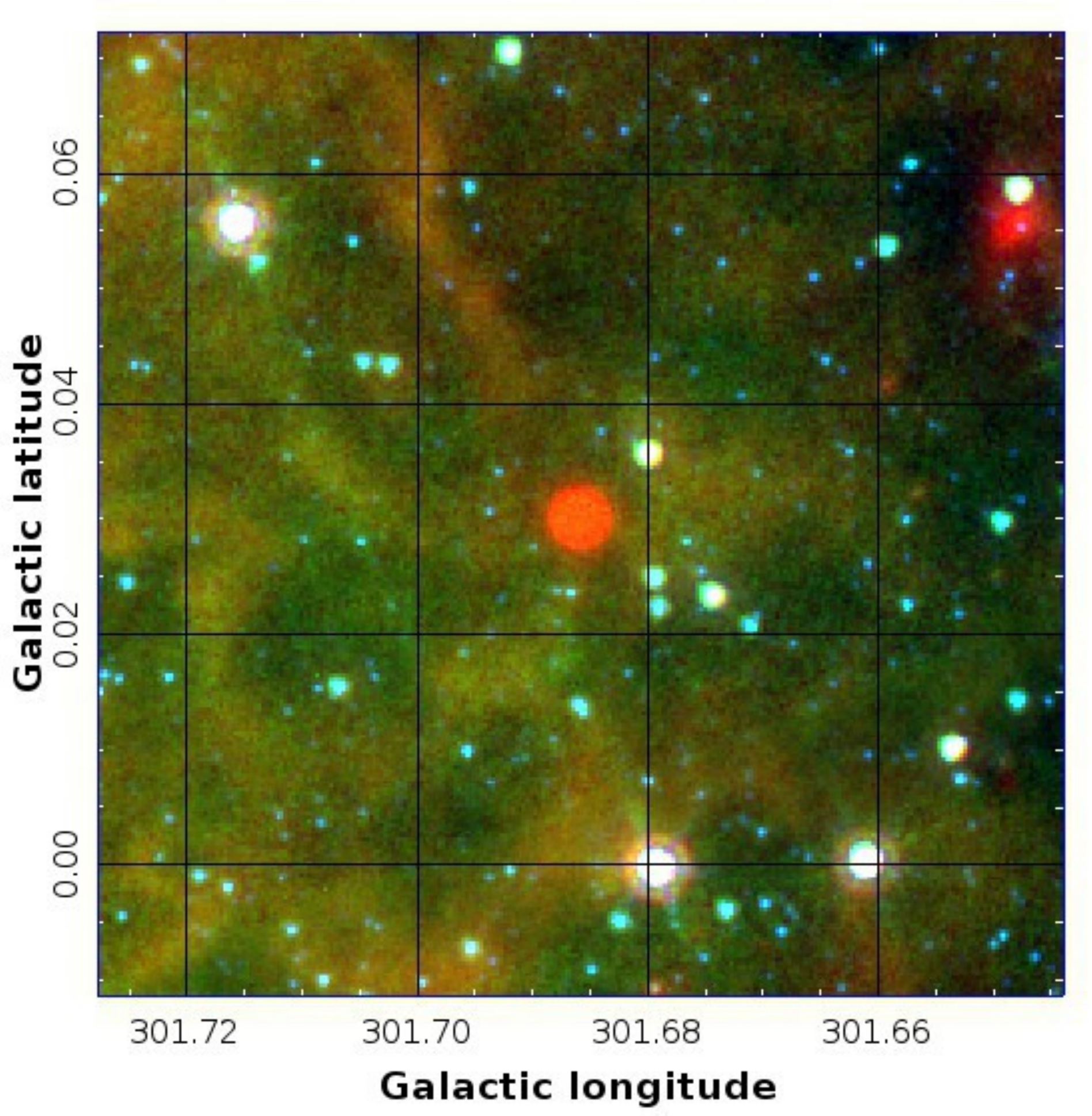}}
    \subfloat[3955]{\includegraphics[height=4.5cm]{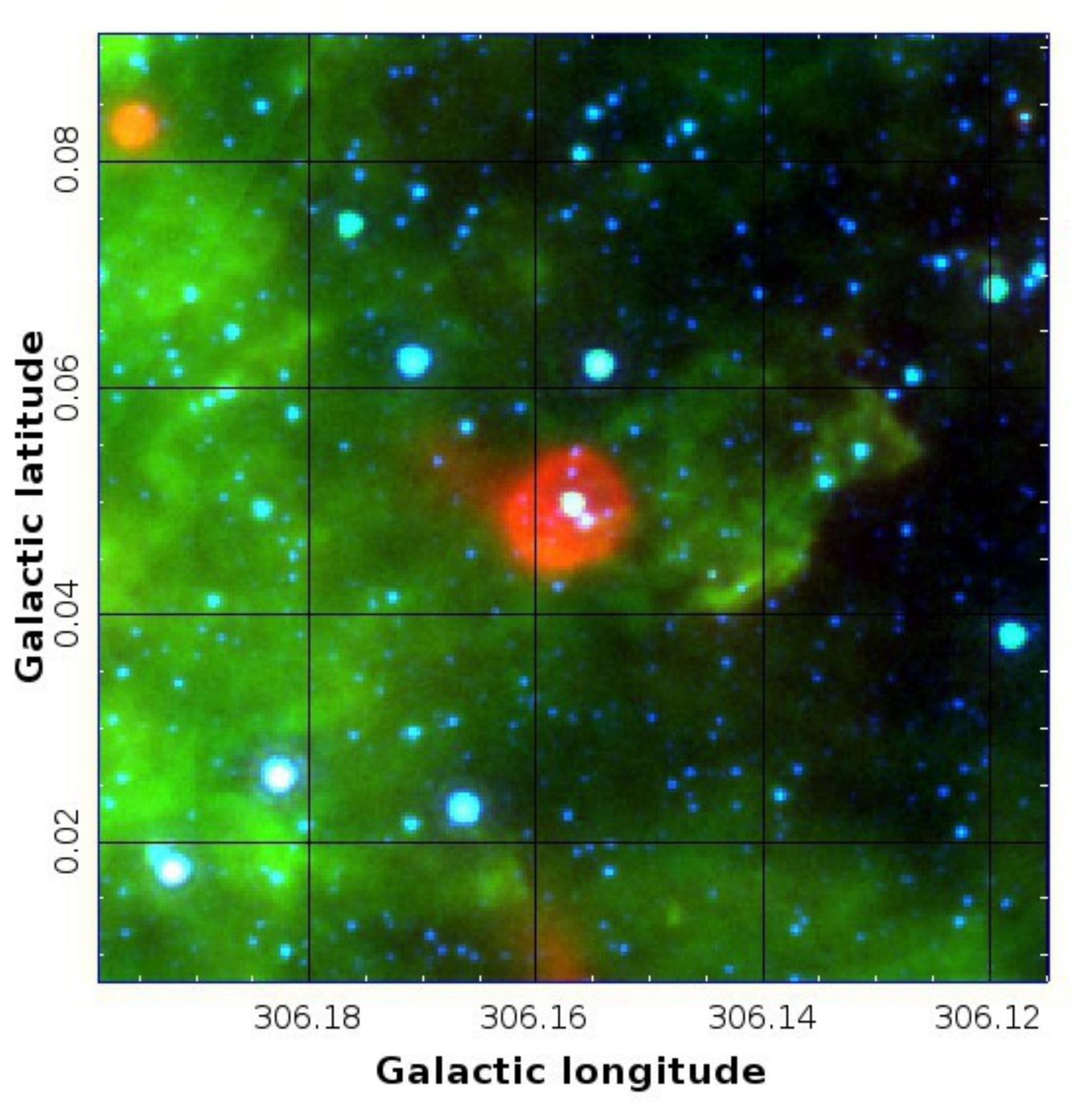}}\\
    \subfloat[4021]{\includegraphics[height=4.5cm]{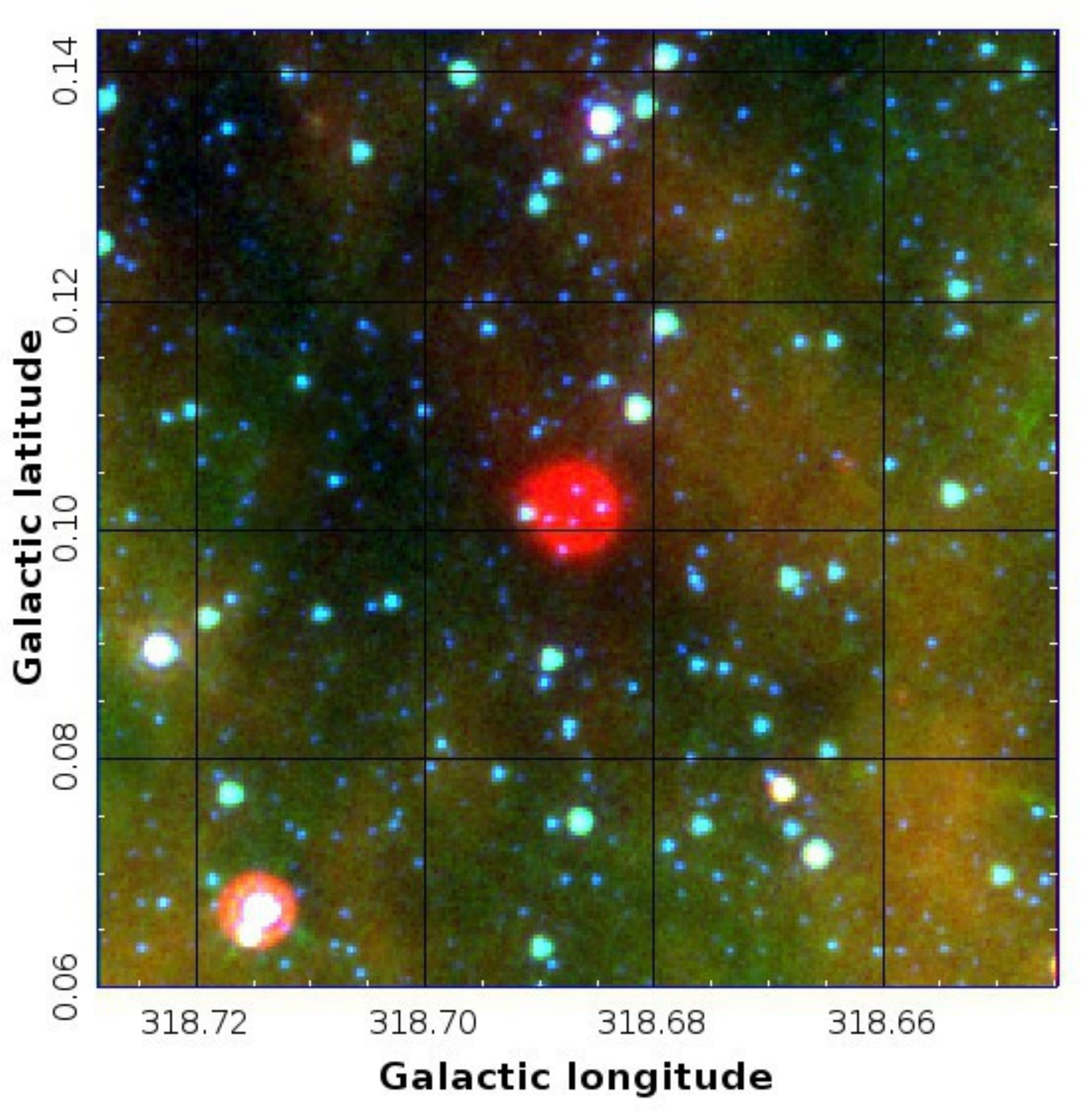}}
    \subfloat[4017]{\includegraphics[height=4.5cm]{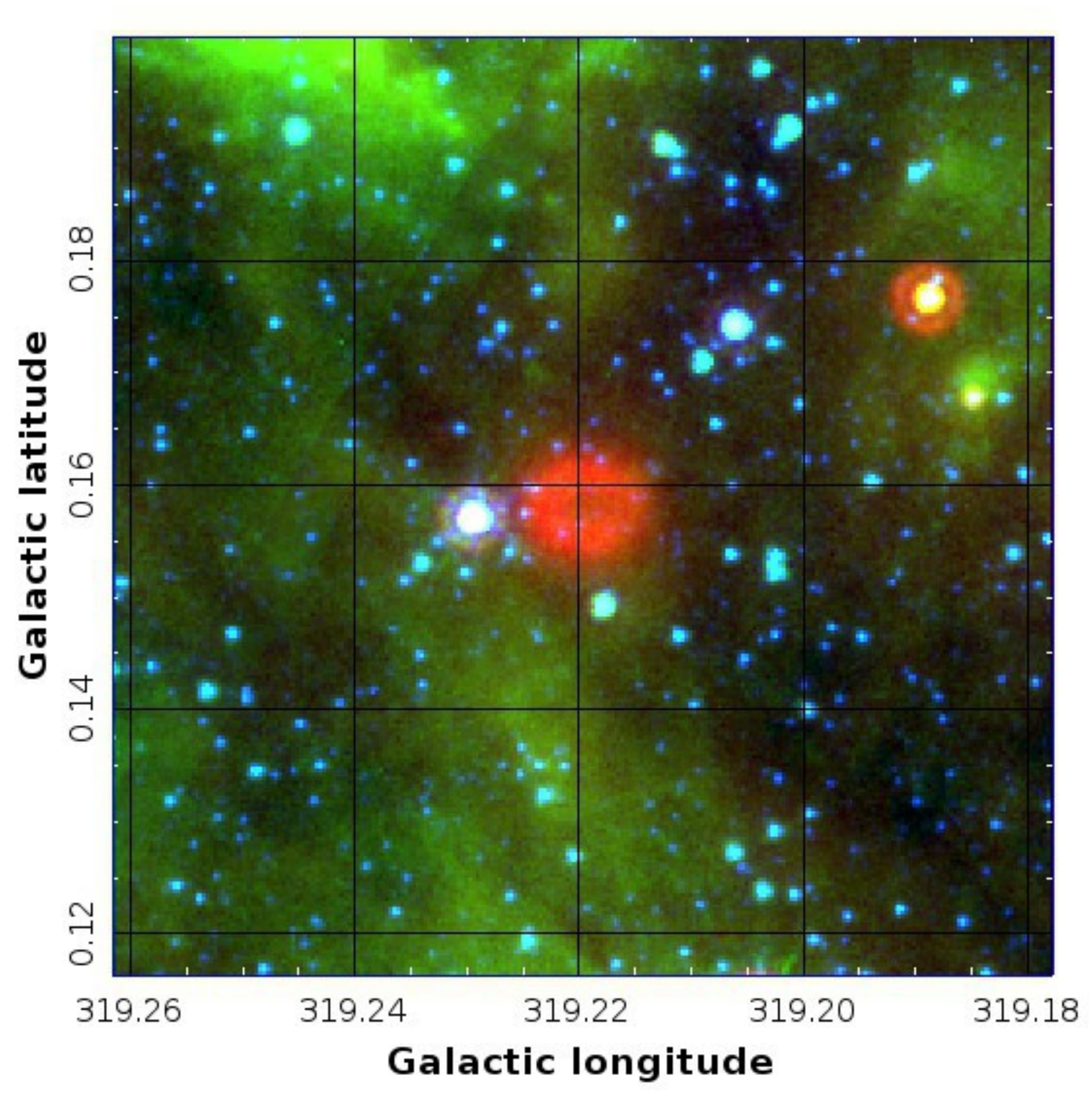}}
    \subfloat[4066]{\includegraphics[height=4.5cm]{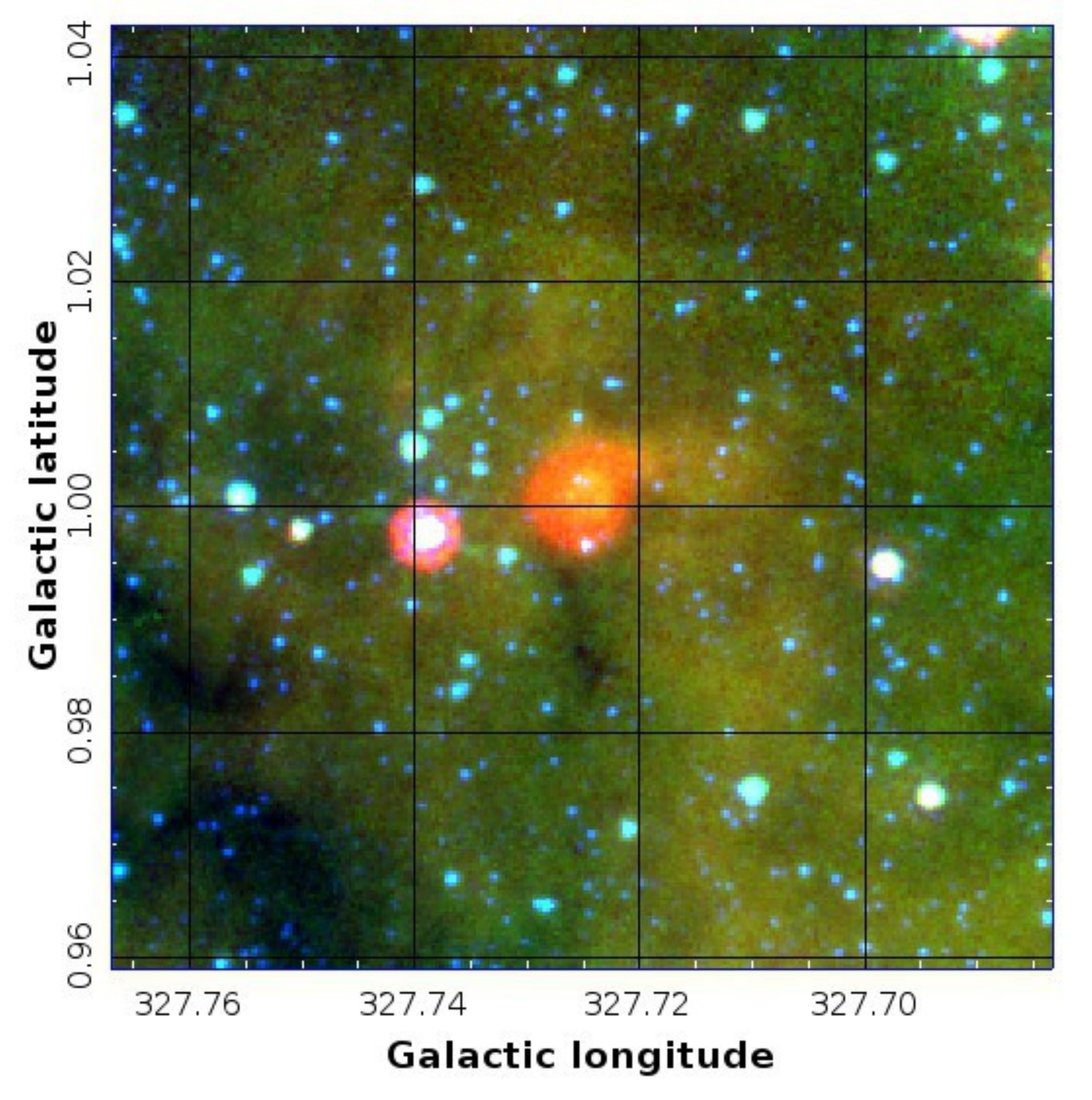}}
    \subfloat[4076]{\includegraphics[height=4.5cm]{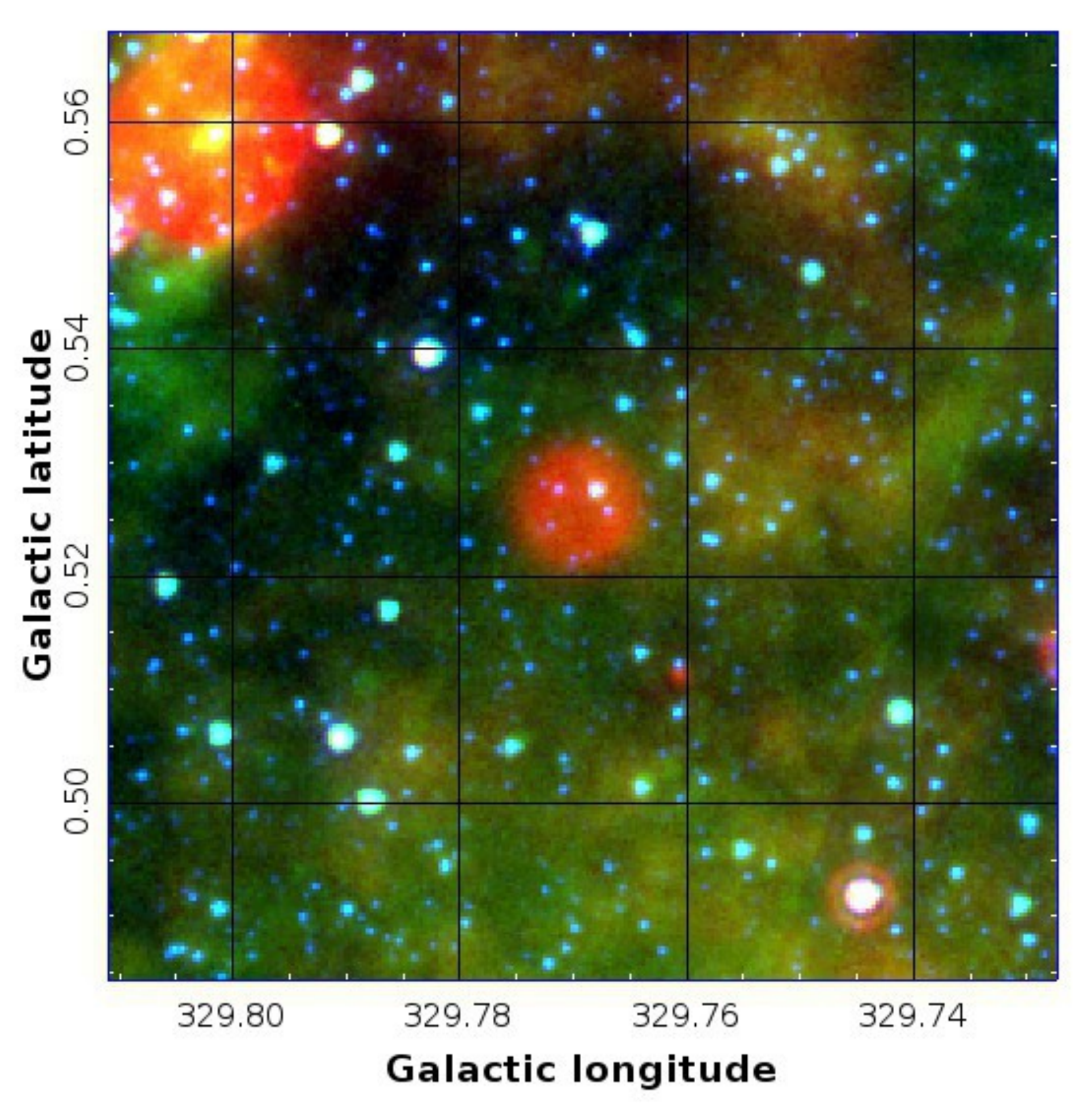}}\\
    \subfloat[4121]{\includegraphics[height=4.5cm]{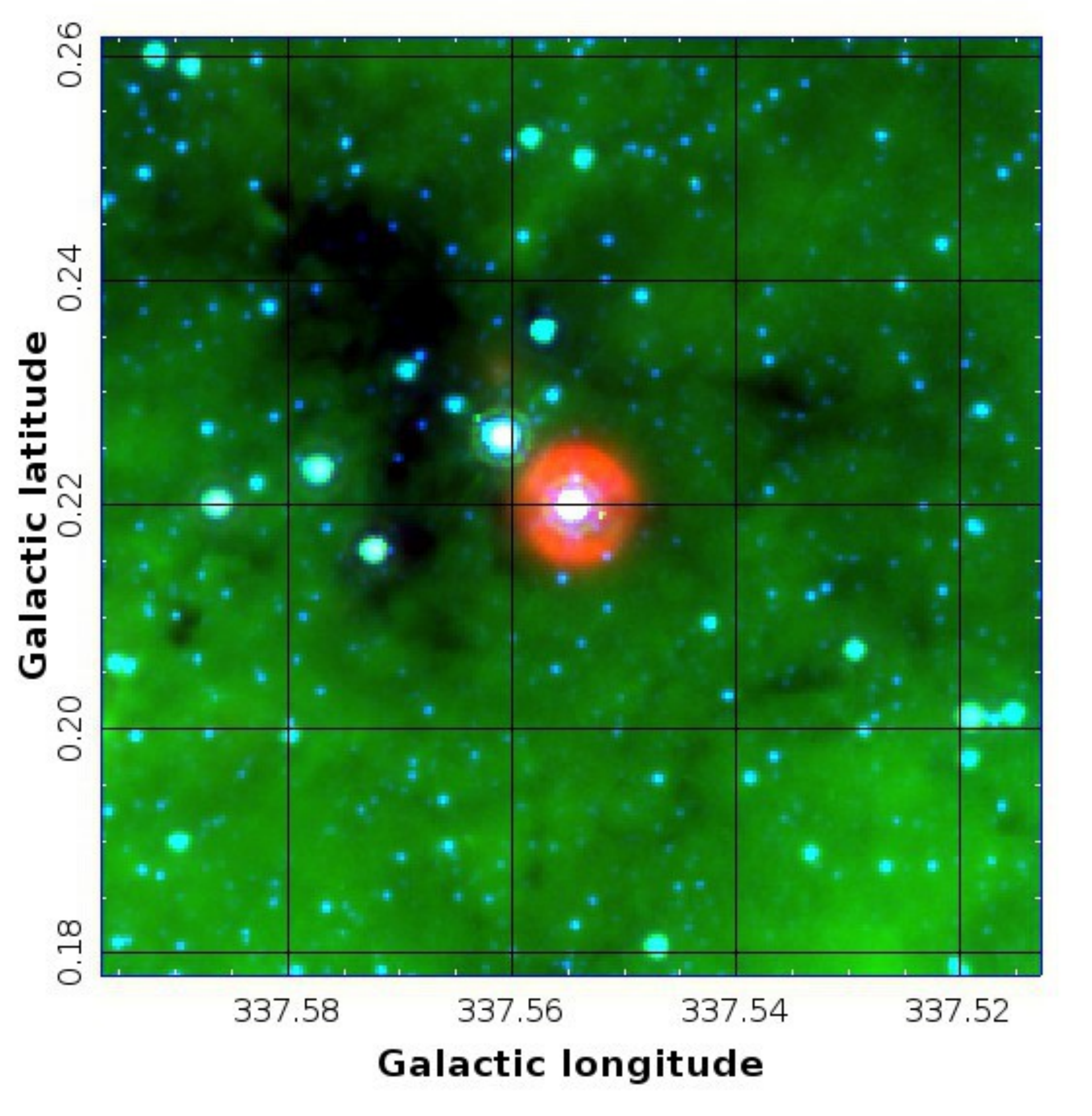}}
    \subfloat[4124]{\includegraphics[height=4.5cm]{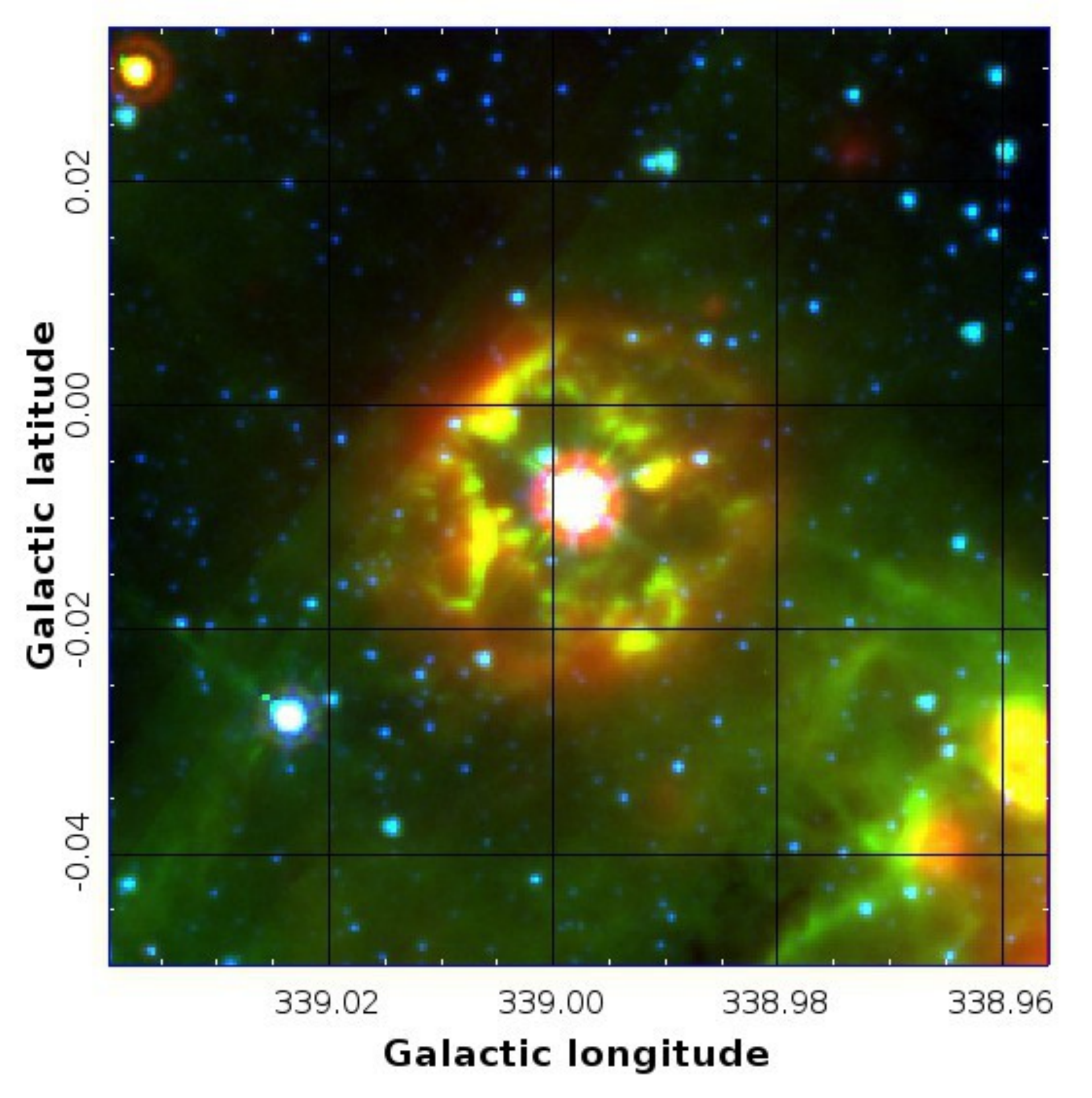}}
    \subfloat[3575]{\includegraphics[height=4.5cm]{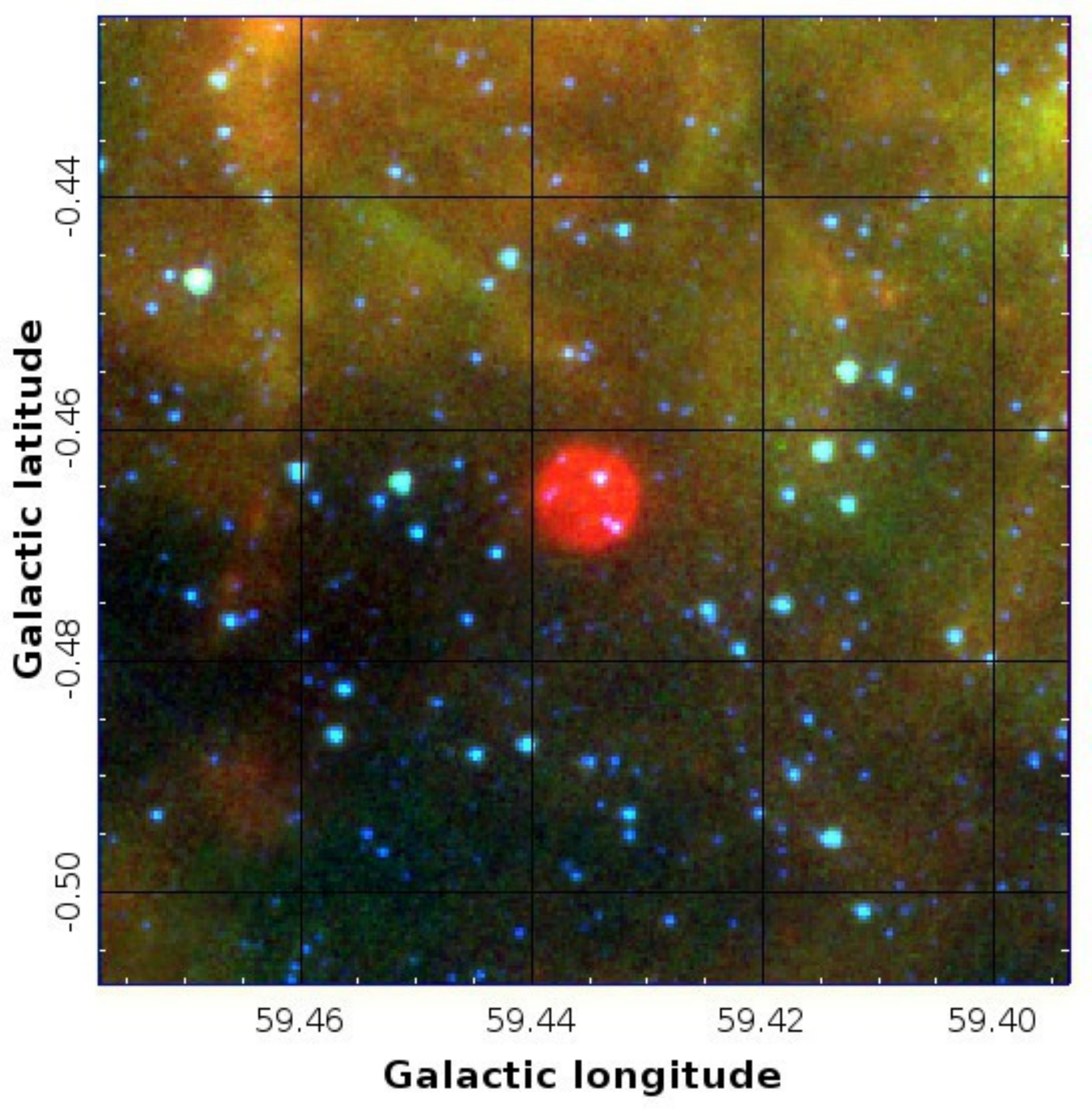}}
    \caption{3-color images of the MBs analyzed in this paper (red is MIPS~24\mic, green is IRAC~8\mic, and blue is IRAC~4.5\mic). Each image is 5' by 5'.}
    \label{bubbles_3colors}
  \end{center}
\end{figure*}

\citet{Wachter2010, Wachter2011}, \citet{Gvaramadze2010}, and \citet{Flagey2014a} have recently confirmed this potential by identifying the central sources in tens of MBs, thanks to optical and near-IR spectroscopic observations, and have found about 50\% of them to be new WR and LBV candidates. Such observations have targeted MBs of a specific morphology: those with central sources in the Multiband Imaging Photometer \citep[MIPS,][]{Rieke2004} 24\mic\ images and that represent only about 15\% of the MBs. They cannot give much information about the remaining 85\% of the catalog. Moreover, these observations, made in optical and near-IR, do not answer the question of the origin of the MIPS 24\mic\ emission. 

\citet{Flagey2011} followed a different approach, based on mid-IR spectroscopic observations made with the high-resolution module ($R\sim600$) of the InfraRed Spectrograph \citep[IRS,][]{Houck2004}, on-board the {\it Spitzer Space Telescope} \citep{Werner2004}. They analyzed the mid-IR spectra of four MBs, and modeled the gas lines and dust emission to conclude that: (1) two MBs are dust-poor, highly excited, PNe, where the 24\mic\ emission mainly comes from the [O~\textsc{iv}]~25.9\mic\ line ; (2) two MBs are dust-rich, where the 24\mic\ emission mainly comes from a warm dust continuum. However, with only four MBs in their sample, they could not draw any statistical conclusions on the entire catalog.

\begin{table*}
  \begin{center}
    \caption{List of the observed MBs.\label{observation_table}}
    \begin{tabular}{c c c c c c c c c c}
      \hline
      \hline
      MB & Name & \multicolumn{2}{c}{J2000 coordinates} & {Diameter} & \multicolumn{3}{c}{Detection flags} & \multicolumn{2}{c}{Identification} \\
      & & $\alpha{}$ & $\delta{}$ & (``) & 24\mic & 12\mic & 8\mic & Nature & Ref \\
      \hline
      MB4384 & MGE295.3343-00.8786 & 11h44m18s & -62d45m21s &  63 & SC &  C &  C & Oe/WN & \citet{Wachter2010} \\
      MB4376 & MGE297.2836-00.8995 & 12h00m59s & -63d12m59s &  35 &  S &  - & C? & & \\
      MB3944 & MGE301.6857+00.0304 & 12h40m31s & -62d48m54s &  16 &  S &  - &  - & & \\
      MB3955 & MGE306.1565+00.0494 & 13h19m35s & -62d38m55s &  41 &  S & SC &  C & OB & \citet{Wachter2011} \\
      MB4021 & MGE318.6864+00.1018 & 14h58m02s & -58d50m31s &  26 &  S &  S &  - & & \\
      MB4017 & MGE319.2193+00.1581 & 15h01m27s & -58d32m24s &  42 &  S &  S &  - & & \\
      MB4066 & MGE327.7248+01.0008 & 15h48m41s & -53d07m02s &  25 &  S &  S &  S & & \\
      MB4076 & MGE329.7690+00.5262 & 16h01m03s & -52d10m38s &  50 &  S &  - &  - & & \\
      MB4121 & MGE337.5544+00.2198 & 16h36m42s & -46d56m20s &  50 & SC &  C &  C & LBVc & \citet{Wachter2010} \\
      MB4124 & MGE338.9975-00.0082 & 16h43m16s & -46d00m41s & 149 & SC & SC & SC & LBVc & \citet{Gvaramadze2010} \\
      MB3575 & MGE059.4884-00.5093 & 19h44m26s & +23d10m07s &  34 &  S &  - &  - & & \\
      \tableline
    \end{tabular}
    \tablecomments{Names follow the {\it Spitzer} nomenclature for the MBs, as given in \citet{Mizuno2010}. Radius is given at 24\mic. Detection flags indicate if the shell (S) and/or the central source (C) is detected. A question mark suggests a possible detection.}
  \end{center}
\end{table*}
 
In this paper, we present the observations of 11 MBs acquired with the low resolution module ($R\sim60-120$) of IRS. These observations complement the work of \citet{Flagey2011} by increasing the sample by a factor 4, which will help us draw conclusions relevant for the entire catalog, and pave the way for observations with future mid-IR facilities. Because observations from the ground in these wavelengths ($\sim$8 to 35\mic) are limited by the atmospheric transmission, and because the next space telescope with mid-IR capabilities will not be launched before 2018, this set of observations is unique, and critical to characterize the mid-IR emission of the MBs. 

This paper is divided as follows. In section~\ref{sec:observations} and \ref{sec:data_reduction}, we present the observations and the data reduction process. In section~\ref{results}, we present the spectra, the measured fluxes, and the contributions of the gas and dust to the mid-IR emission in the MBs. In sections~\ref{sec:disc_dust_poor}, \ref{sec:disc_dust_rich} and \ref{sec:disc_extrap}, we discuss separately the dust-poor MBs, the dust-rich MBs, and the extrapolation of our findings and others to the whole catalog. Our conclusions are given in section~\ref{sec:ccl}.

\section{Observations}
\label{sec:observations}

The observations have been obtained in 2008 and 2009, during the {\it Spitzer} cryogenic mission. They are registered under the program ID\#50808. S.J.~Carey is the PI of this 3.7 hours program.

The targets have been selected among the numerous MBs in the \citet{Mizuno2010} catalog to sample diversity among the MBs, both in terms of fluxes and morphologies. In Figure \ref{bubbles_3colors}, we show mid-IR 3-color images for the MBs in this sample. Table \ref{observation_table} lists some basic information about the observed MBs: names, coordinates, apparent radii at 24\mic. We also indicate whether or not the shells or central sources are detected in the MIPS~24\mic, WISE~12\mic\ \citep{Wright2010}, and IRAC~8\mic\ images. The last two columns give the natures of the central sources and their references for four MBs. The stars at the center of MB4121 and MB4124 are LBV candidates, an Oe/WN star is the central source of MB4384, and MB3955 has a central star identified as an OB type star \citep{Gvaramadze2010, Wachter2010, Wachter2011}.

\begin{table*}
  \begin{center}
    \caption{List of the observed positions.\label{aor_table}}
    \begin{tabular}{c c c c c c | c c c c c c}
      \hline
      \hline
      MB & Module & Angle & Pos. & RA & DEC & MB & Module & Angle & Pos. & RA & DEC \\
      \hline
      MB4384 & SL & 143 & 1 & 11h44m18.0s & -62d45m21.0s & MB4121 & SL & 179 & 1 & 16h36m42.8s & -46d56m20.6s \\
             &    &     & 2 & 11h44m16.0s & -62d45m10.8s &        &    &     & 2 & 16h36m41.3s & -46d56m20.3s \\
             &    &     & 3 & 11h44m20.0s & -62d45m31.2s &        &    &     & 3 & 16h36m44.2s & -46d56m21.0s \\

             & LL & -133& 1 & 11h44m18.0s & -62d45m20.9s &         & LL & -98 & 1 & 16h36m42.8s & -46d56m20.6s \\
             &    &     & 2 & 11h44m16.3s & -62d45m33.3s &         &    &     & 2 & 16h36m42.6s & -46d56m35.5s \\
             &    &     & 3 & 11h44m19.7s & -62d45m08.5s &         &    &     & 3 & 16h36m43.0s & -46d56m05.7s \\
      \hline
      MB4376 & SL & 115 & 1 & 12h00m59.1s & -62d12m59.9s & MB4021 & SL & 162 & 1 & 14h58m02.8s & -58d50m31.3s \\
             & LL & -161& 1 & 12h00m59.1s & -62d12m59.9s &         & LL & -114& 1 & 14h58m02.8s & -58d50m31.2s \\
      \hline
      MB4124 & SL & 176 & 1 & 16h43m16.3s & -46d00m41.7s & MB4076 & SL & 174 & 1 & 16h01m03.9s & -52d10m38.0s \\
             &    &     & 2 & 16h43m15.5s & -45d59m58.2s &        & LL & -102& 1 & 16h01m03.9s & -52d10m37.9s \\ \cline{7-12}
             &    &     & 3 & 16h43m18.5s & -46d01m25.8s & MB4066 & SL & 172 & 1 & 15h48m41.8s & -53d07m02.4s \\
             &    &     & off  & 16h42m59.9s & -46d02m27.4s &     & LL & -105& 1 & 15h48m41.8s & -53d07m02.3s \\ \cline{7-12}
             & LL & -100& 1 & 16h43m16.3s & -46d00m41.7s & MB4017 & SL & 162 & 1 & 15h01m27.7s & -58d32m24.2s \\
             &    &     & 2 & 16h43m15.5s & -45d59m58.1s &        & LL & -115& 1 & 15h01m27.7s & -58d32m24.1s \\ \cline{7-12}
             &    &     & 3 & 16h43m18.5s & -46d01m25.7s & MB3944 & SL & 125 & 1 & 12h40m31.4s & -62d48m54.4s \\
             &    &     & off  & 16h42m59.9s & -46d02m27.4s &     & LL & -151& 1 & 12h40m31.5s & -62d48m54.4s \\
      \hline
      MB3955 & SL & 138 & 1 & 13h19m33.9s & -62d38m44.3s & MB3575 & SL & 162 & 1 & 19h44m26.2s & +23d10m07.1s \\
             &    &     & 2 & 13h19m35.6s & -62d38m55.0s &        & LL & -115& 1 & 19h44m26.2s & +23d10m07.2s \\
             & LL & -138& 1 & 13h19m33.9s & -62d38m44.2s \\
      \tableline
    \end{tabular}
    \tablecomments{Angles are given in degrees and measured East from North.}
  \end{center}
\end{table*}

The spectra on which this paper is based have been acquired using the low resolution modules of IRS. We used the long wavelengths module {(LL slit, 168\arcsec\ by 10.5\arcsec)}, featuring a resolution of 57 to 126, in its first order (LL1: from 19.9 to 38.0\mic), and second order (LL2: from 13.9 to 21.3\mic), and the short wavelengths module {(SL slit, 57\arcsec\ by 3.6\arcsec)}, with a resolution of 60 to 127, in its first order only (SL1: 7.4 to 14.5\mic). The observations have been performed in staring mode, meaning that for each order, an exposure has been obtained with the MBs centered at {one third of the way along the slit} (nod 1), and another with the MBs centered at {two thirds of the way along the slit} (nod 2). Each exposure has been obtained with a ramp duration of 14 seconds, with 3 repetitions for the SL module, and 4 for the LL module.

\begin{figure*}
\begin{center}
\subfloat[4384]{\includegraphics[width=0.225\linewidth]{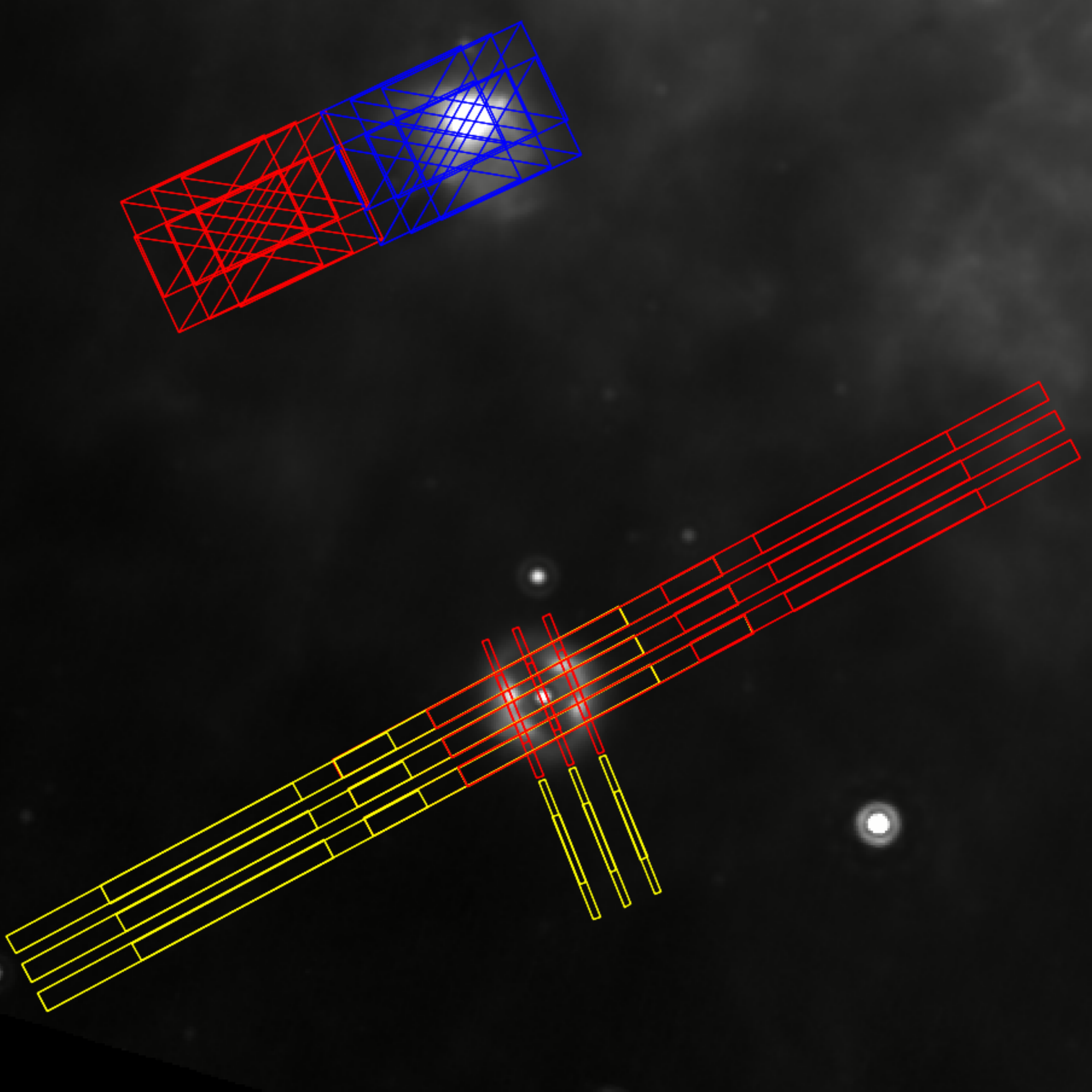}}
\hfill
\subfloat[4376]{\includegraphics[width=0.225\linewidth]{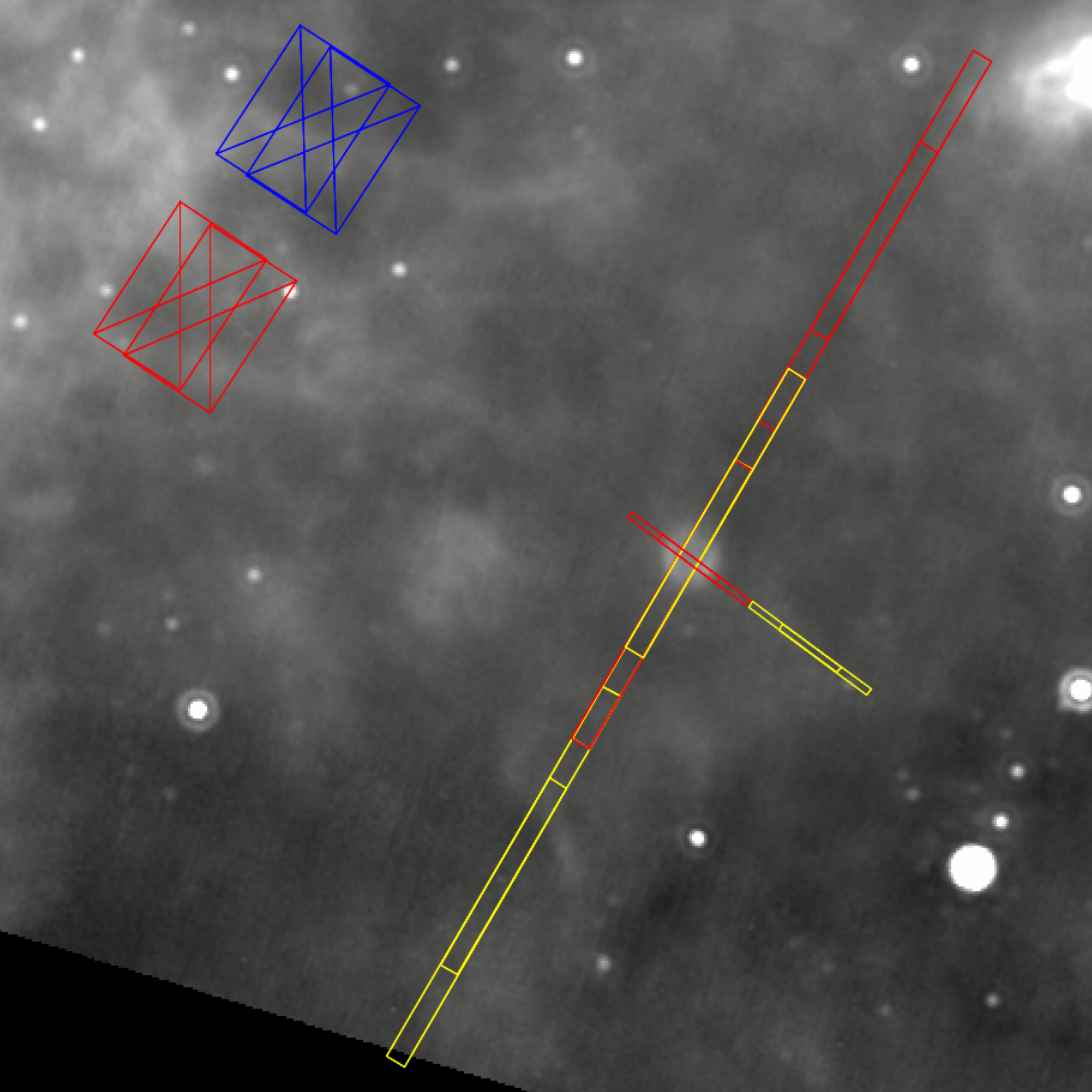}}
\hfill
\subfloat[3944]{\includegraphics[width=0.225\linewidth]{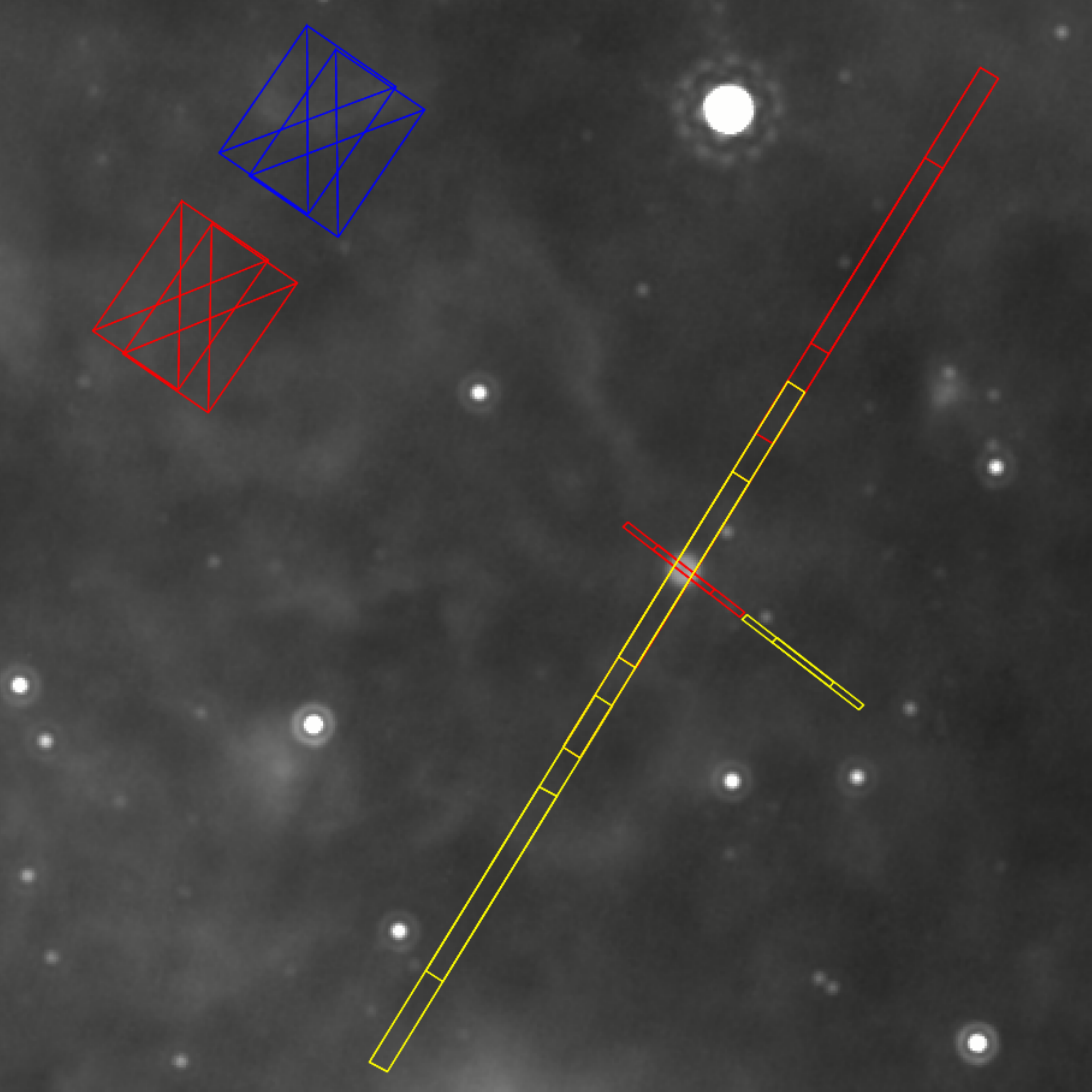}}
\hfill
\subfloat[3955]{\includegraphics[width=0.225\linewidth]{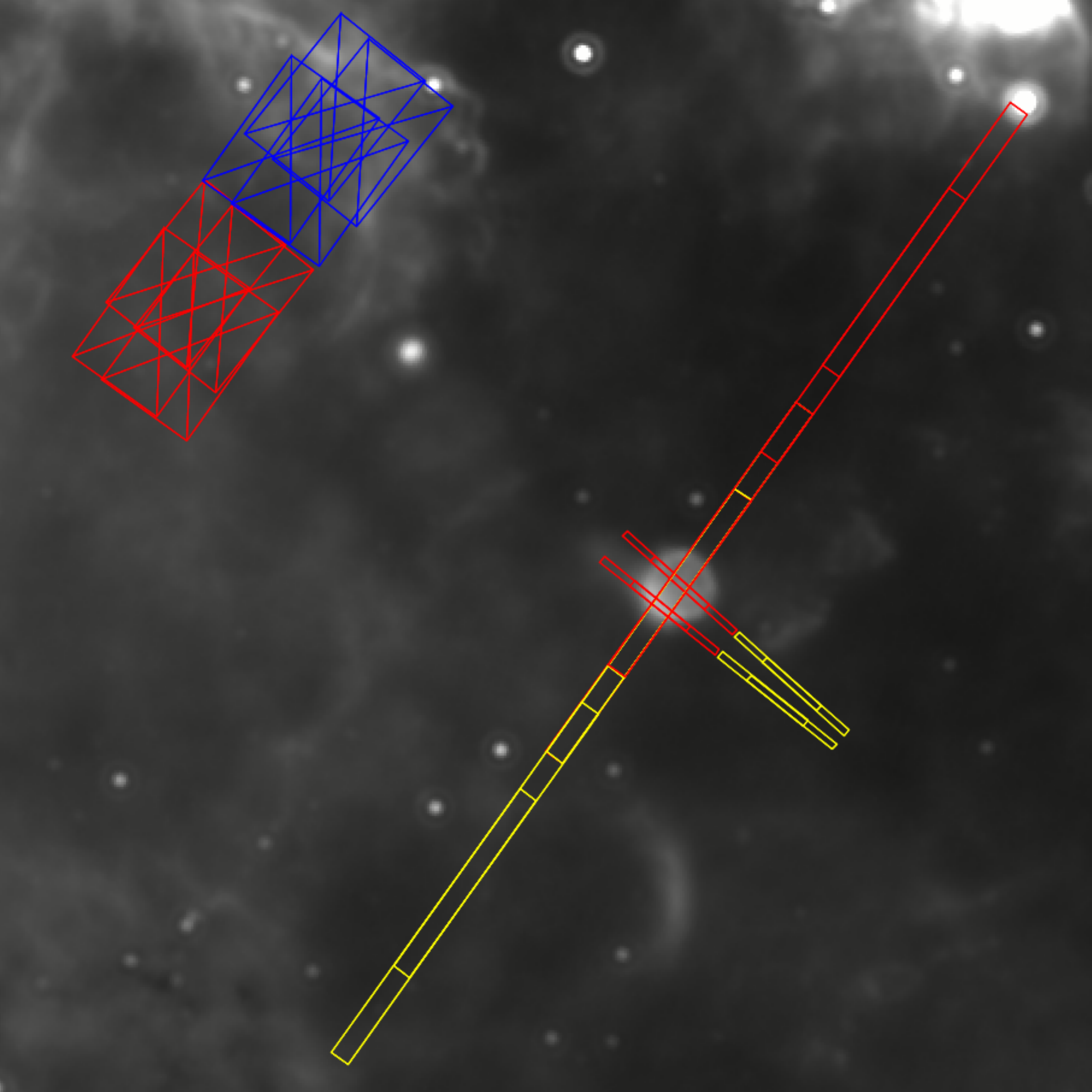}}\\
\subfloat[4021]{\includegraphics[width=0.225\linewidth]{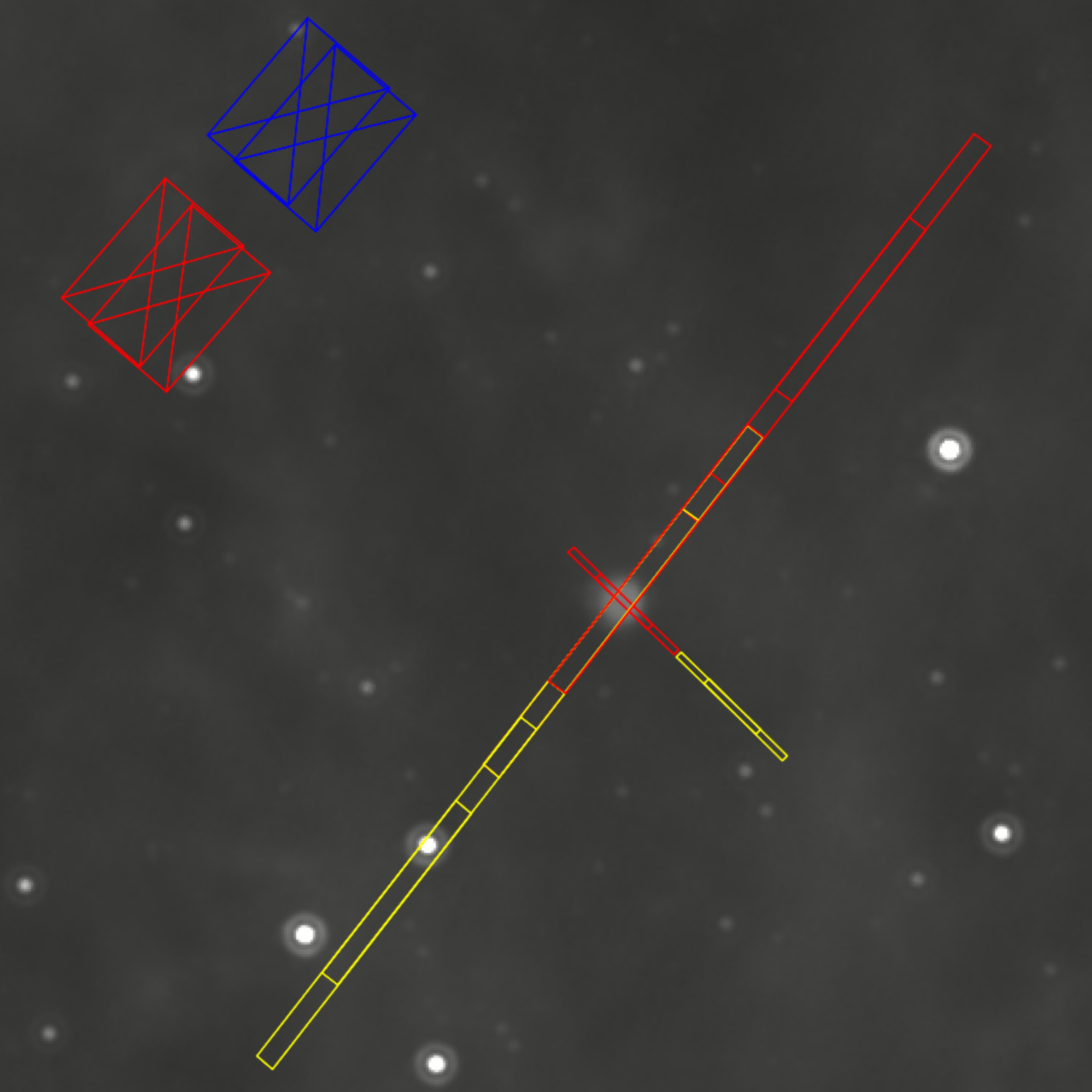}}
\hfill
\subfloat[4017]{\includegraphics[width=0.225\linewidth]{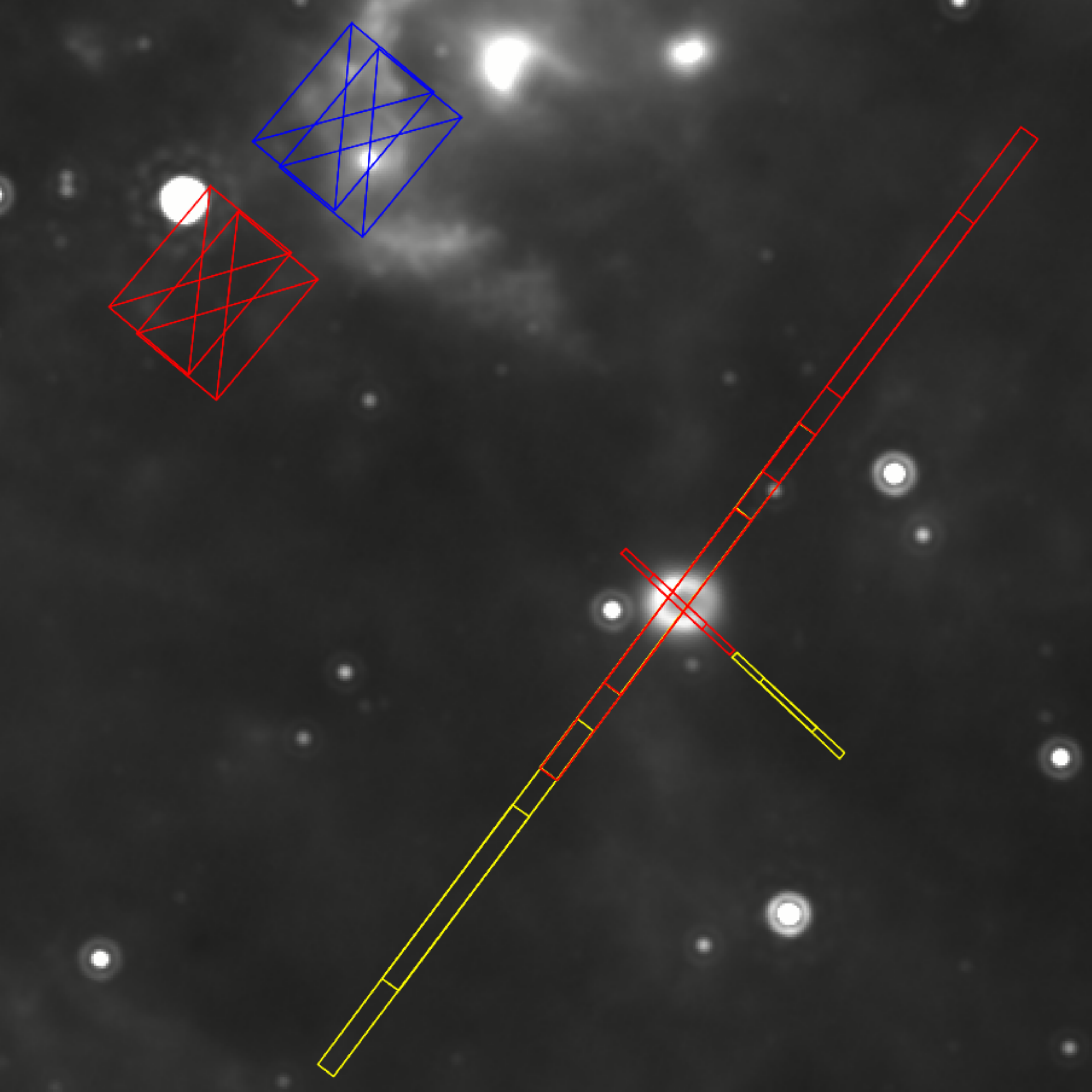}}
\hfill
\subfloat[4066]{\includegraphics[width=0.225\linewidth]{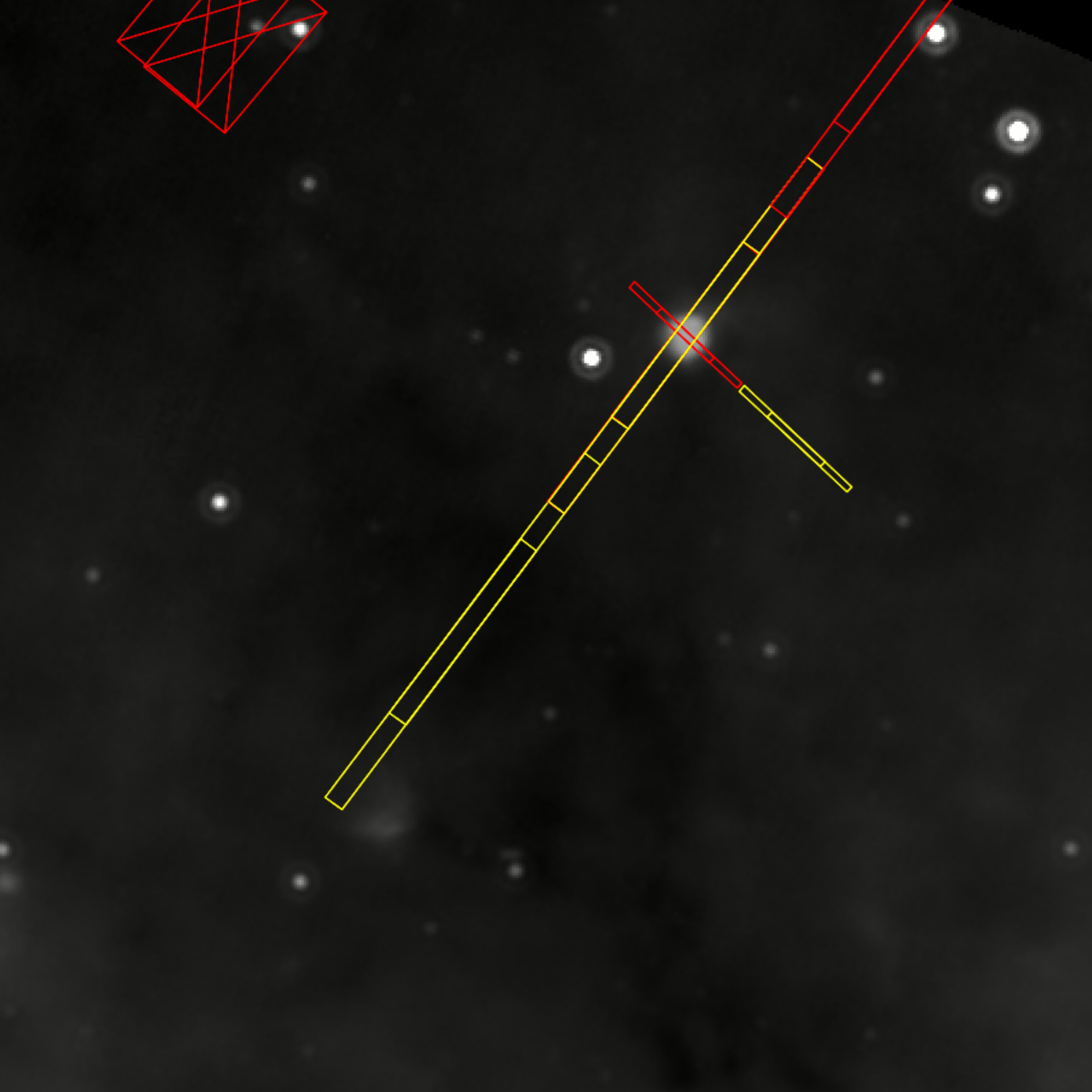}}
\hfill
\subfloat[4076]{\includegraphics[width=0.225\linewidth]{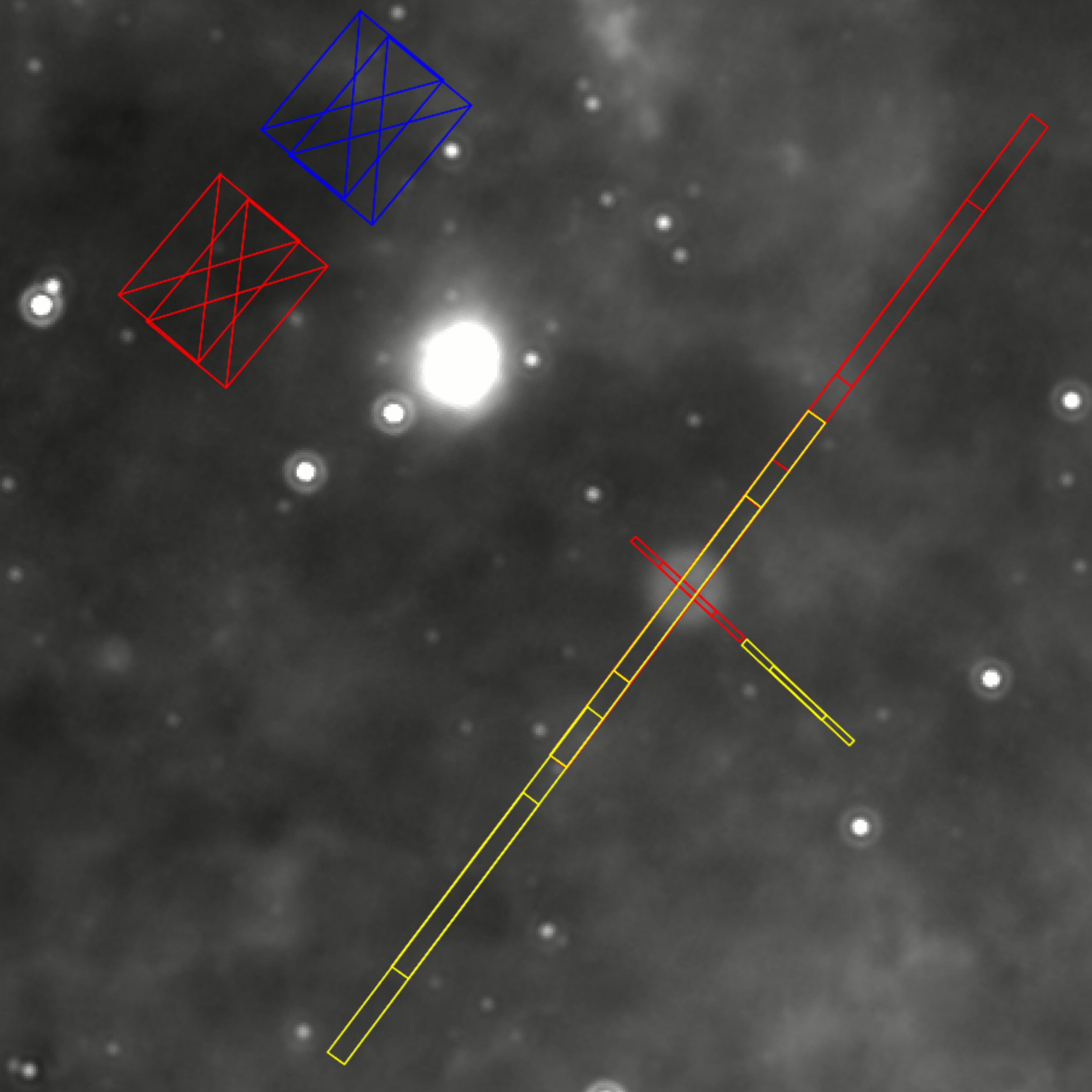}}\\
\subfloat[4121]{\includegraphics[width=0.225\linewidth]{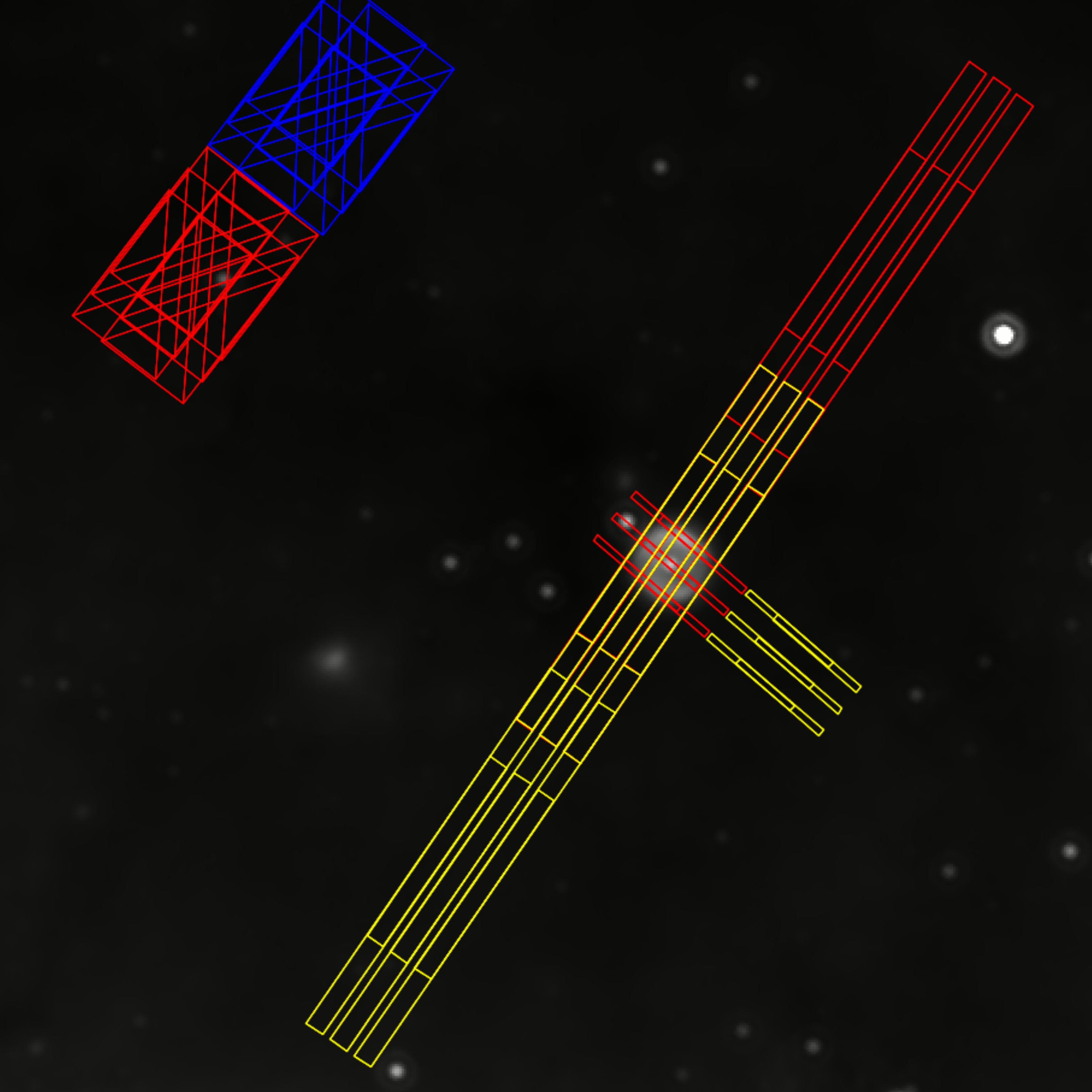}}
\hspace{.25in}
\subfloat[4124]{\includegraphics[width=0.225\linewidth]{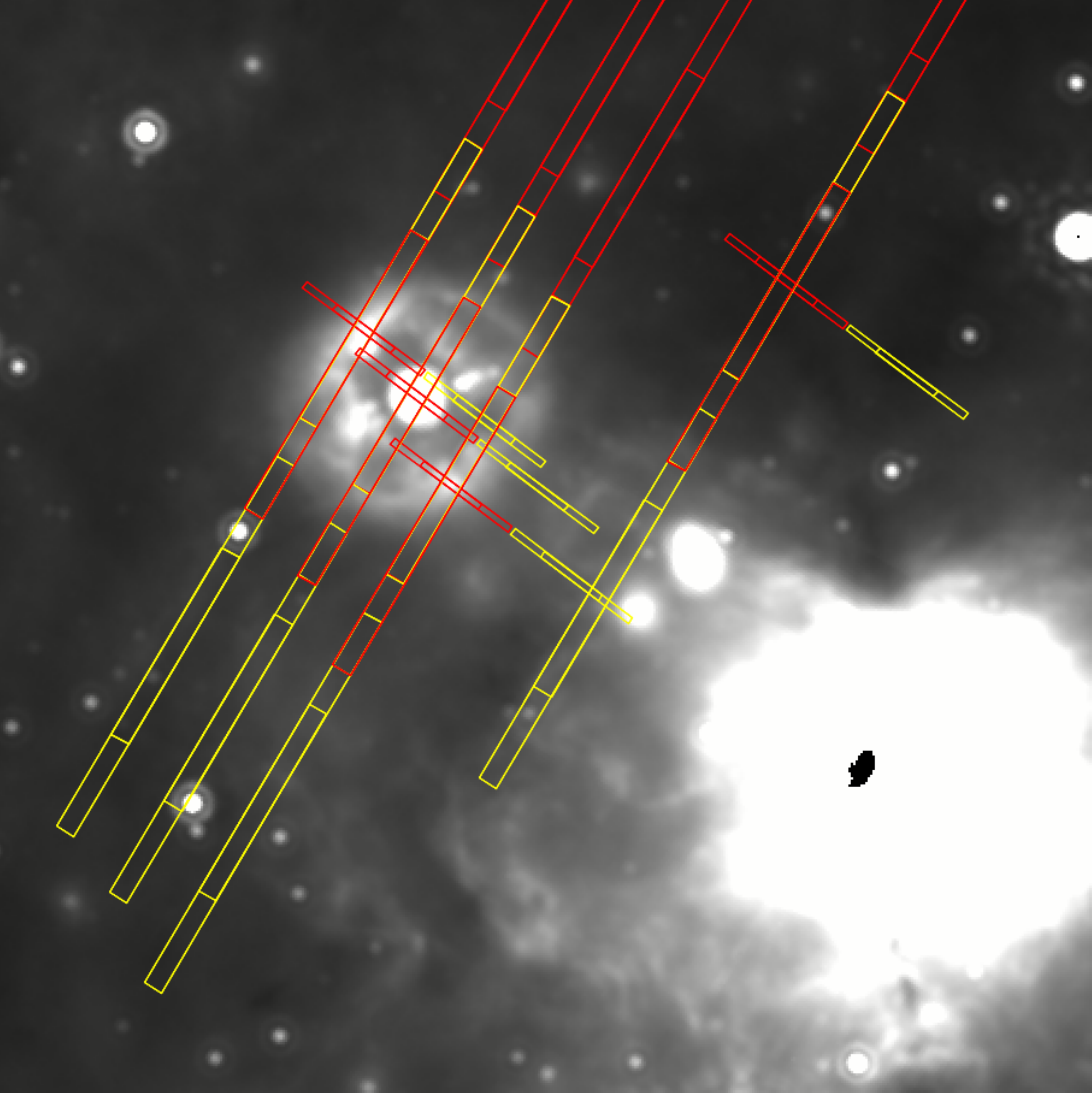}}
\hspace{.25in}
\subfloat[3575]{\includegraphics[width=0.225\linewidth]{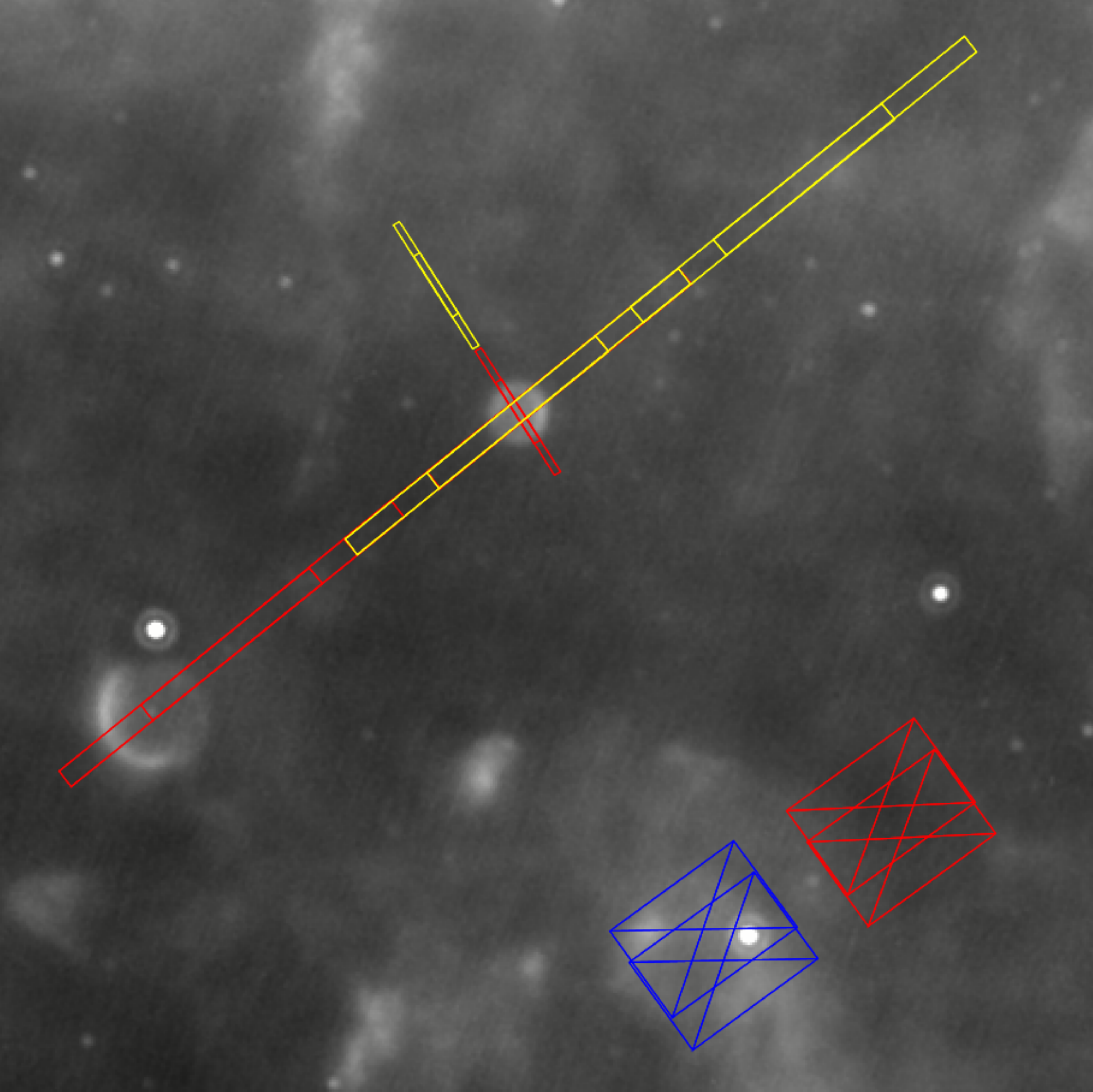}}
\caption{Positions of the slits of the IRS instrument plotted over MIPS 24\mic\ images. First orders are yellow, second orders are red ; LL slits are long (\unit{168}{\arcsecond}), and SL slits are short (\unit{57}{\arcsecond}). Red and blue rectangles are the peak-up arrays observations, that we did not use. Each panel has a different angular size.}
\label{bubbles_slits}
\end{center}
\end{figure*}

For the four MBs which are relatively extended and show central sources in the mid-IR, multiple positions were observed: (1) towards a bright rim and (2) towards the center, to derive spectra of both the shells and the central sources. Figure~\ref{bubbles_slits} shows the position of the slits over the MIPS 24\mic\ images of the MBs. {The positions and orientations of the slits are given in Table \ref{aor_table}.}

The design of both IRS modules (SL and LL) features two slits side by side (one for each order) that acquire data simultaneously. Thanks to this particular design, when acquiring on-source data with one order, we always get a close-by observation for the other order, at about \unit{3}{\arcminute} from the MBs in the LL module, and \unit{1}{\arcminute} in the SL module. Since we requested both LL1 and LL2, we therefore automatically obtain a dedicated background for these orders. The dedicated background for LL1 has been obtained when observing the MBs with LL2, and vice-versa. For the SL module, we only requested SL1 observations, since most of the MBs are not detected in the IRAC~8\mic\ images. Therefore we do not have a close-by SL1 dedicated background for each target. Instead, we defined a dedicated background set of observations using all four orders (SL1, SL2, LL1, LL2) in a region a few arcminutes away from one target (MB4124). We use these data as the SL1 background of all the targets (see section \ref{background} for details).

\section{Data reduction}
\label{sec:data_reduction}

Our data reduction follows the usual, recommended steps for {\it Spitzer}/IRS observations, {but} because of the nature of the targets, and the observation strategy, some refinements are required. We thus detail hereafter the whole data reduction process.

\subsection{Cleaning the images}

The first step of the data reduction process is to use the IRSCLEAN\footnote{http://irsa.ipac.caltech.edu/data/SPITZER/docs/dataanaly\-sistools/tools/irsclean/} algorithm on the post-basic calibrated data (post-BCD) images to remove bad pixels. Post-BCD are the images obtained by coadding the repeated exposures at each observed positions, and are directly provided by the {\it Spitzer}/IRS data reduction pipeline. We use the rogue-mask that indicates the pixels known to have been faulty during the campaign of the observations. The algorithm of IRSCLEAN then cleans those pixels by interpolating over them. For the SL1 module, this cleaning is not sufficient, and we have to use the second IRSCLEAN algorithm that automatically detects and interpolates all the pixels that are judged faulty.

\subsection{Extracting the spectra}

The second step of the data reduction is to use SPICE\footnote{http://irsa.ipac.caltech.edu/data/SPITZER/docs/dataanaly\-sistools/tools/spice/} to extract the spectra from the post-BCD images. We use SPICE version 2.5. For each MB, we extract the spectra for the on-source and the background positions with exactly the same parameters (i.e. width and relative position of extraction within the slit). We remind the reader that the background for the MB4124 SL1 observations was used for the SL1 observations as well. We discuss the subtraction of the background in section~\ref{background}.

%We extract the bubble and background spectra separately to measure the fluxes of the lines in both positions and estimate the influence of the background in the measurements. This is particularly important since the background always comes from a position at least \unit{3}{\arcminute} away from the bubble and we expect the background to vary over such spatial scale (see fig~\ref{bubbles_slits}). 

SPICE extracts the spectra in arbitrary units, and then applies a calibration tuning to scale them to astrophysical units. Two different tunings are possible. The \textit{point source} tuning is used for sources significantly smaller than the size of the slit, whereas the \textit{extended source} tuning is used for sources significantly larger than the slit. The observed MBs are compact-extended structures that represent intermediary cases between \textit{point sources} and \textit{extended sources}. The ratio between both tunings is stable at the 5\% level over most of the wavelength range but reaches 20\% at wavelengths of about 10\mic. The uncertainty in the spectra are therefore not significantly affected by the choice of a tuning. We decide to use the {\it extended source} tuning for all the MBs, except for those four with central sources in the mid-IR. In this case, we use the {\it point source} tuning for the central sources and both tunings for the shells. The {\it extended source} tuning is used to extract the spectra of the shells in surface brightness, and the {\it point source} tuning is used to ensure a correct subtraction of the outer shell's contribution to the observations towards the central sources.

\subsection{Matching orders and scaling the spectra}
\label{sec:match}

The LL1 and LL2 spectra usually match very well without any particular adjustment. Since it is not always the case, we use the mean value of the three last spectral elements of LL2 and that of the three first spectral elements of LL1 to automatically scale {LL2 onto LL1}. The factors introduced here range from 0.9 to 1.1. We then scale the SL1 spectrum to the LL spectrum using the same method. We do this for the on-source and the background spectra independently. The SL1 background always comes from a region nearby MB4124 (see section~\ref{sec:observations}). Therefore, we assume that for each MB, this part of the background's spectrum has the correct \emph{spectral shape}, and that it differs from the SL background's spectrum near the MB only by a scaling factor. This is an approximation as spectral features (e.g., polycyclic aromatic hydrocarbons or PAH features) are known to vary over such distances \citep[e.g., for PAHs,][and references therein]{Peeters2011b}.

After scaling the different modules with each other, we calibrate the whole spectrum, from 8 to 35\mic, with the MIPSGAL~24\mic\ images. We compute the MIPS 24\mic\ surface brightness of the spectrum using the filter's spectral response and taking into account the color correction inherent to the spectrum. We then scale the whole spectrum so that this value matches the mean MIPSGAL~24\mic\ surface brightness within the area of the slit where we extracted the spectrum. We do this for the on-source and the background spectra independently.

Since the background was obtained in a region along the slit that is at least 1' away from the targets, and since the MIPS~24\mic\ surface brightness of the ISM can vary over such angular scales, the scaled background might be under or over-estimated with respect to the true background towards the target. This is discussed in section~\ref{background}.
 
We finally derive the mean on-source and background spectra by averaging {the spectra from the two nod positions} for the on-source and the background positions, separately. We do this for display purposes only, and use both nodes independently when measuring the flux of gas lines (see section \ref{line_fluxes}).

\subsection{Subtracting the background}
\label{background}

\begin{figure}
\begin{center}
\includegraphics[trim=75 50 75 50, clip, width=\linewidth, angle=180]{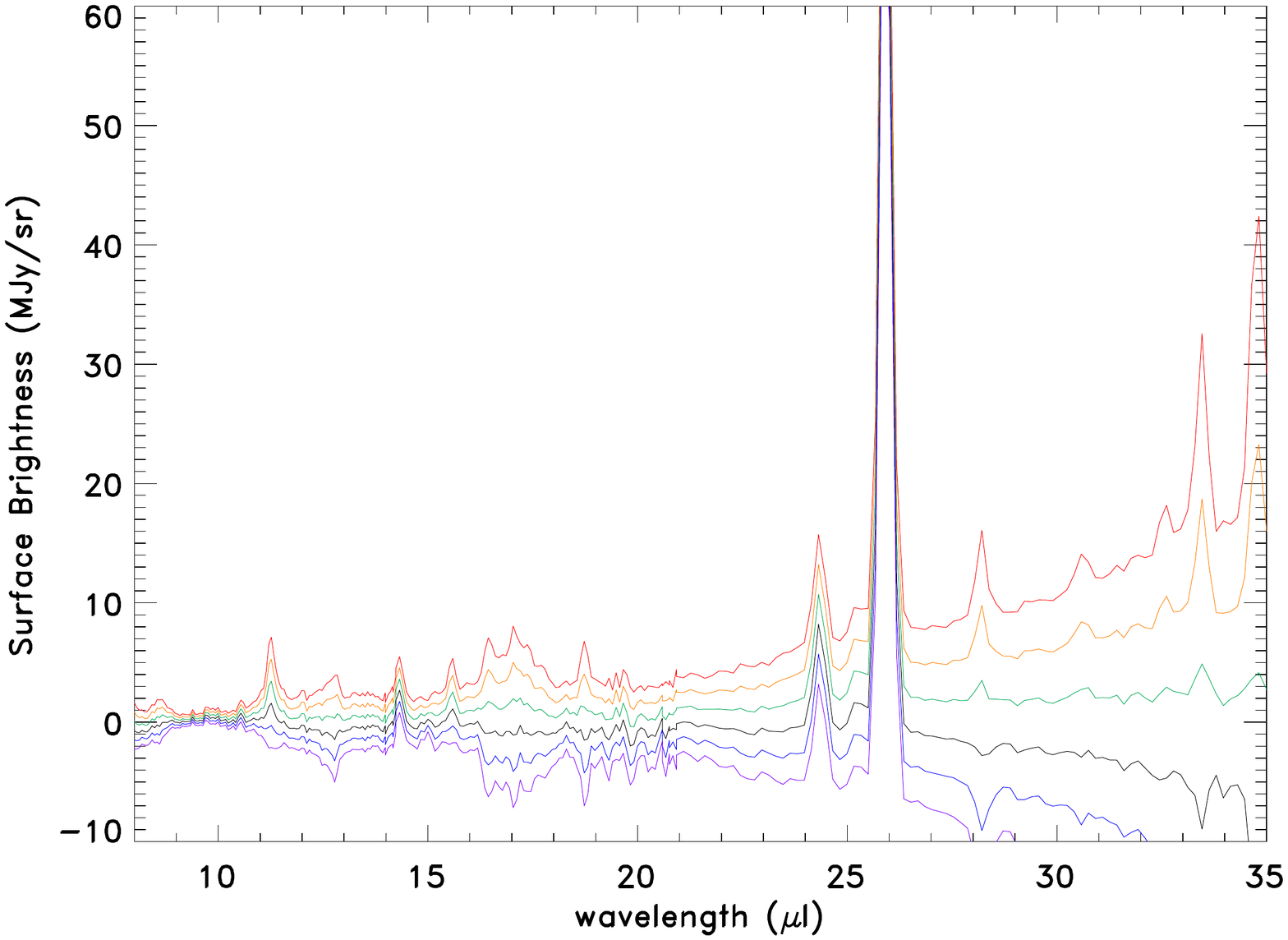}
\caption{Background subtracted spectra for MB3575, with different scale factors applied on the background's spectrum (purple is 1.2, blue is 1.1, black is 1.0, green is 0.9, orange is 0.8, and red is 0.7).}
\label{plusieurs_coeff}
\end{center}
\end{figure}

The background is obtained in a region of the sky at least 1' away from the sources (see section~\ref{sec:observations}). Variations in the spectral features of the gas and the dust are expected over such angular scales, especially since we are looking at highly structured fields of the low latitude Galactic plane. We apply a scaling factor to the background spectrum to achieve the best background subtraction. Figure~\ref{plusieurs_coeff} shows the effect of such variations in the scaling of the background spectrum on the background subtracted spectrum of MB3575.

Several spectral features are directly affected by a change in the scaling factor (e.g., PAH features between 11 and 12.5\mic\ and between 16 and 18\mic, the H$_2$ line at 28.2\mic, the [S~\textsc{iii}]~33.5\mic\ line), while others remain unchanged (e.g. the [Ne~\textsc{iii}]~15.5\mic, [Ne~\textsc{v}]~14.3 and 24.3\mic\ lines). This allows us to disentangle the lines that are emitted by the circumstellar envelopes from those that, at least in part, originate in the ISM. To determine the best factor and its uncertainty, we assume that, in most cases, the PAH features and the H$_2$ line are only due to the ISM, and we adjust the coefficient to minimize these features. In the case of MB3575, we thus use a factor of 0.9$\pm$0.1. For all the MBs, we find that the best scaling factors for the background subtraction are determined with a precision of about $\pm$0.1 and range between 0.75 and 1.15. We checked that this factor is in agreement with the variations of the MIPS~24\mic\ surface brightness along the slit, from the location where the background spectrum was actually observed to the immediate vicinity of the MBs. A 10\% uncertainty is significantly larger than the 5\% quoted for the MIPS~24\mic~photometric uncertainty on point sources. We will thus take into account a 10\% uncertainty in the background when we measure the line fluxes (see section \ref{line_fluxes}). {The longer wavelength continuum of the background subtracted spectra may drop below zero (e.g. see Figure \ref{fig:4376}). However, we made sure that the best coefficients and the resulting uncertainties on the spectra were consistent with a positive continuum.}

\section{Results}
\label{results}

We show in Figures~\ref{spectra_shell} and \ref{spectra_other} the IRS spectra of the MBs with and without central sources, respectively. In Figure~\ref{spectra_shell} we show both the spectra towards the shells and towards the central sources of the MBs. In both Figures, we indicate the impact of the background subtraction uncertainty with dashed lines. {In this section, we first present the general characteristics of the two groups of spectra. We then discuss the gas line identification and their flux measurements, before quantitatively answering the question about the nature of the mid-IR emission in the MBs.}

\subsection{Two groups of MBs}
\label{spectra}

\begin{figure*}
\begin{center}
\subfloat[4121]{\includegraphics[trim=75 50 75 50, clip, width=0.5\linewidth, angle=180]{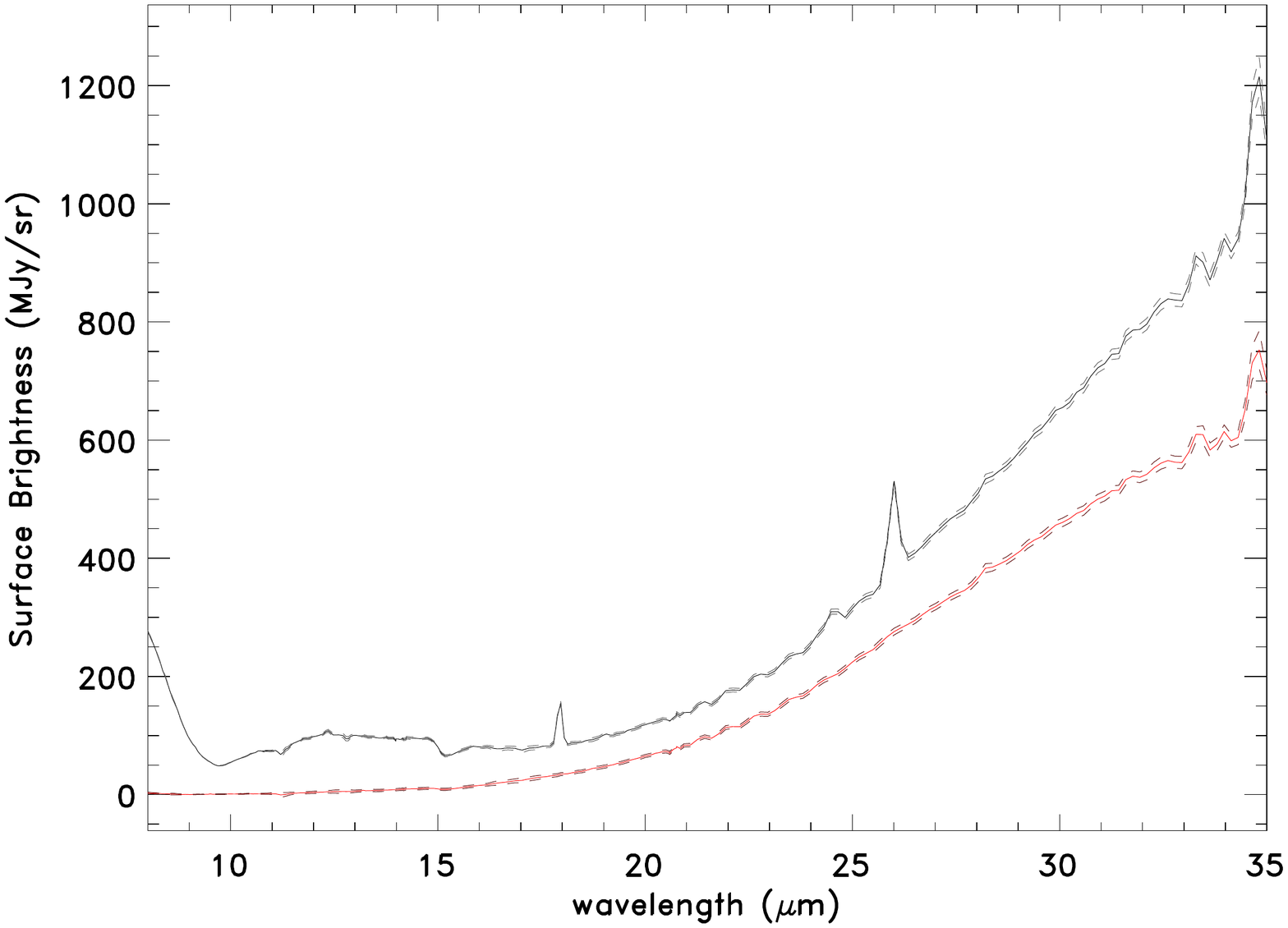}}
\subfloat[4124]{\includegraphics[trim=75 50 75 50, clip, width=0.5\linewidth, angle=180]{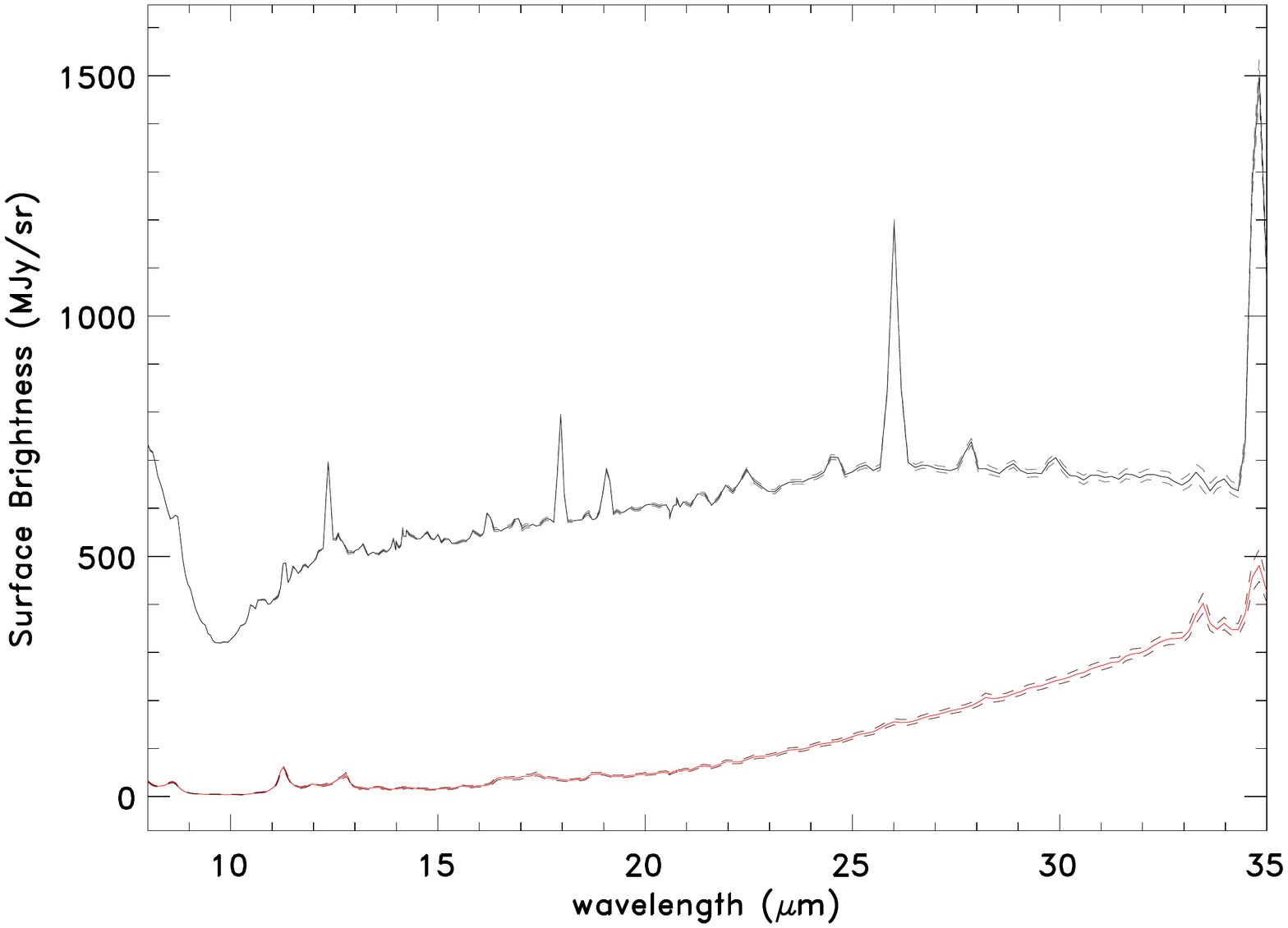}}\\
\subfloat[4384]{\includegraphics[trim=75 50 75 50, clip, width=0.5\linewidth, angle=180]{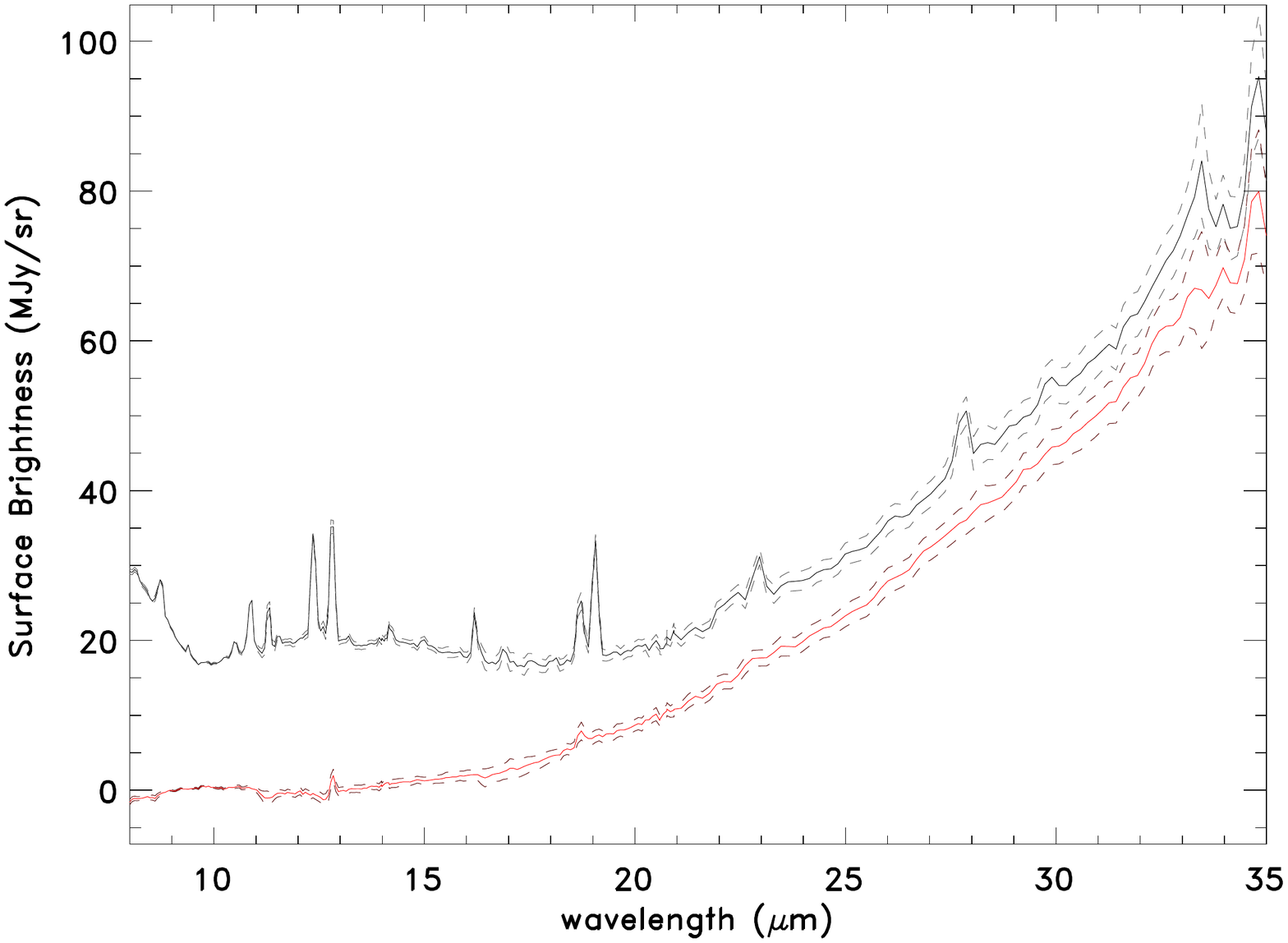}}
\subfloat[3955]{\includegraphics[trim=75 50 75 50, clip, width=0.5\linewidth, angle=180]{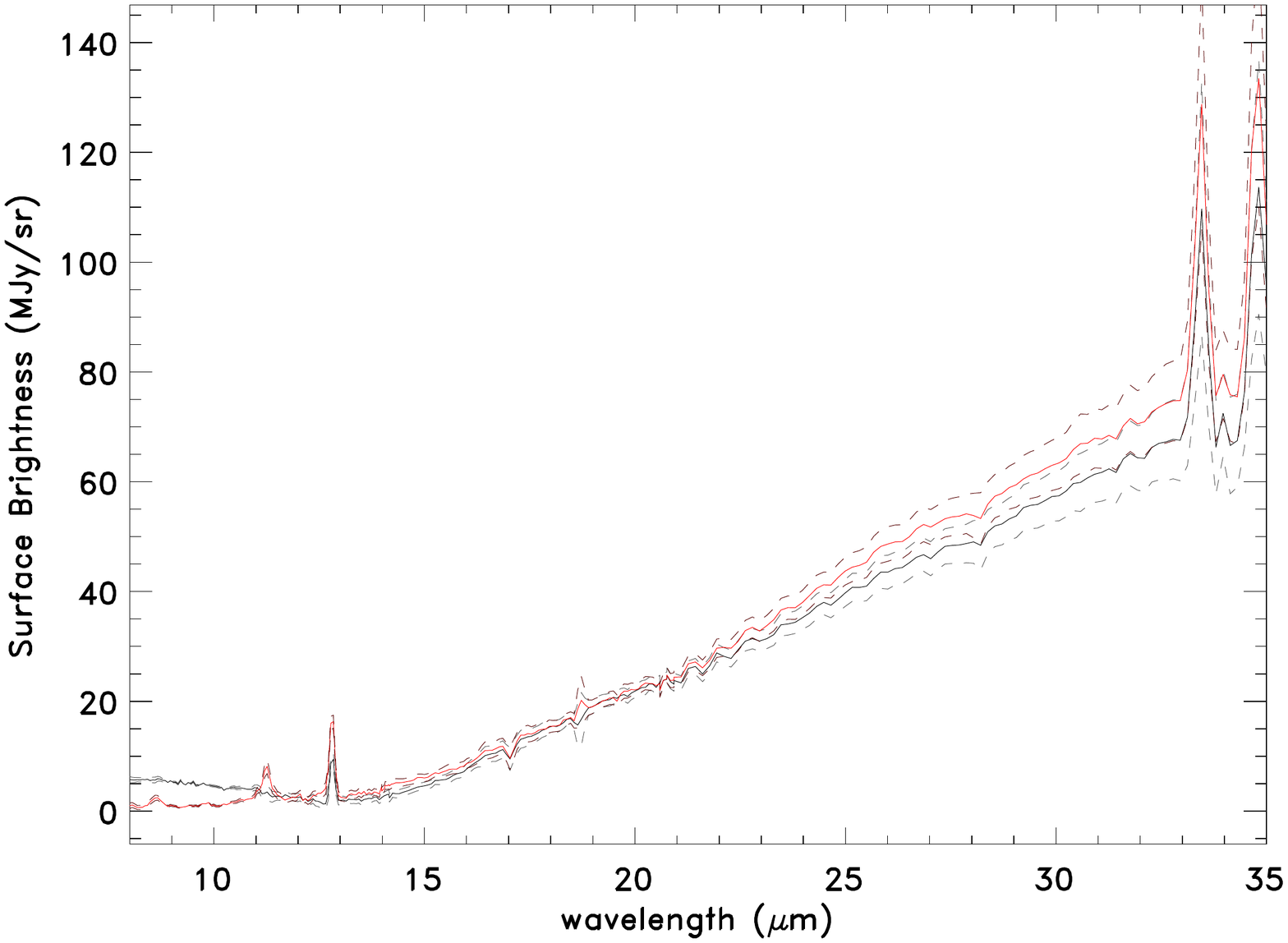}}
\caption{IRS spectra of the MBs with central sources in the mid-IR (red is the spectra of the shells, and black is the spectra towards the central sources ; dashed lines show what happens if we increase or decrease the background by 10\%)}
\label{spectra_shell}
\end{center}
\end{figure*}

A first look at the spectra suggests a division of the sample in two different subsets. The spectra of the four MBs with central sources detected in the mid-IR (MB3955, MB4121, MB4124, and MB4384), shown in Figure~\ref{spectra_shell}, are dominated by a dust continuum, with almost no gas lines. These MBs form the ``dust-rich'' subset. {On the other hand}, the spectra of the seven MBs without central sources detected in the mid-IR (MB3575, MB3944, MB4017, MB4021, MB4066, MB4076, and MB4376) show little or no continuum, and are dominated by strong lines of highly ionized gas (e.g., [O~\textsc{iv}]~26\mic, [Ne~\textsc{V}]~14.3 and 24.3\mic, [Ne~\textsc{iii}]~15.5\mic, [S~\textsc{iv}]~10.5\mic). These MBs constitute the ``highly excited'' subset. MB4066, and to a lesser extent MB4021, are not straightforward to classify. They show the presence of a continuum and some emission lines. Since those gas lines are mostly the same as those found in the spectra of the other ``highly excited'' MBs, we decided to add them to this group. 

This separation based on the IRS spectra is thus highly correlated with the morphologies of the MBs (see Figure~\ref{bubbles_3colors}), {as it has initially been suggested by \citet{Mizuno2010}, and later by \citet{Flagey2011} on a sample of four MBs}. We also note that the shells in the ``dust-rich'' group are significantly larger (radius at 24\mic\ between 40 and 149\arcsec) than those in the ``highly excited'' set (radius at 24\mic\ between 16 and 50\arcsec). {This also confirms what was initially suggested by \citet{Mizuno2010}}.

\begin{table}
\caption{Ionization potentials of Neon, Sulfur, Oxygen and Iron for their first few ionization states}
\label{ionization}
\begin{center}
\begin{tabular}{l c c c c}
\hline
\hline
Element & \textsc{ii} & \textsc{iii} & \textsc{iv} & \textsc{v}\\
\hline
Neon & 21.56 & 40.96 & 63.45 & 97.12\\
Sulfur & 10.36 & 23.34 & 34.76 & 47.22\\
Oxygen & 13.62 & 35.12 & 54.93 & 77.41\\
Iron & 7.9 & 16.19 & 30.65 &\\
\hline
\end{tabular}
\end{center}
\end{table}

\begin{figure*}
\begin{center}
\subfloat[3944]{\includegraphics[trim=75 50 75 50, clip, width=0.5\linewidth, angle=180]{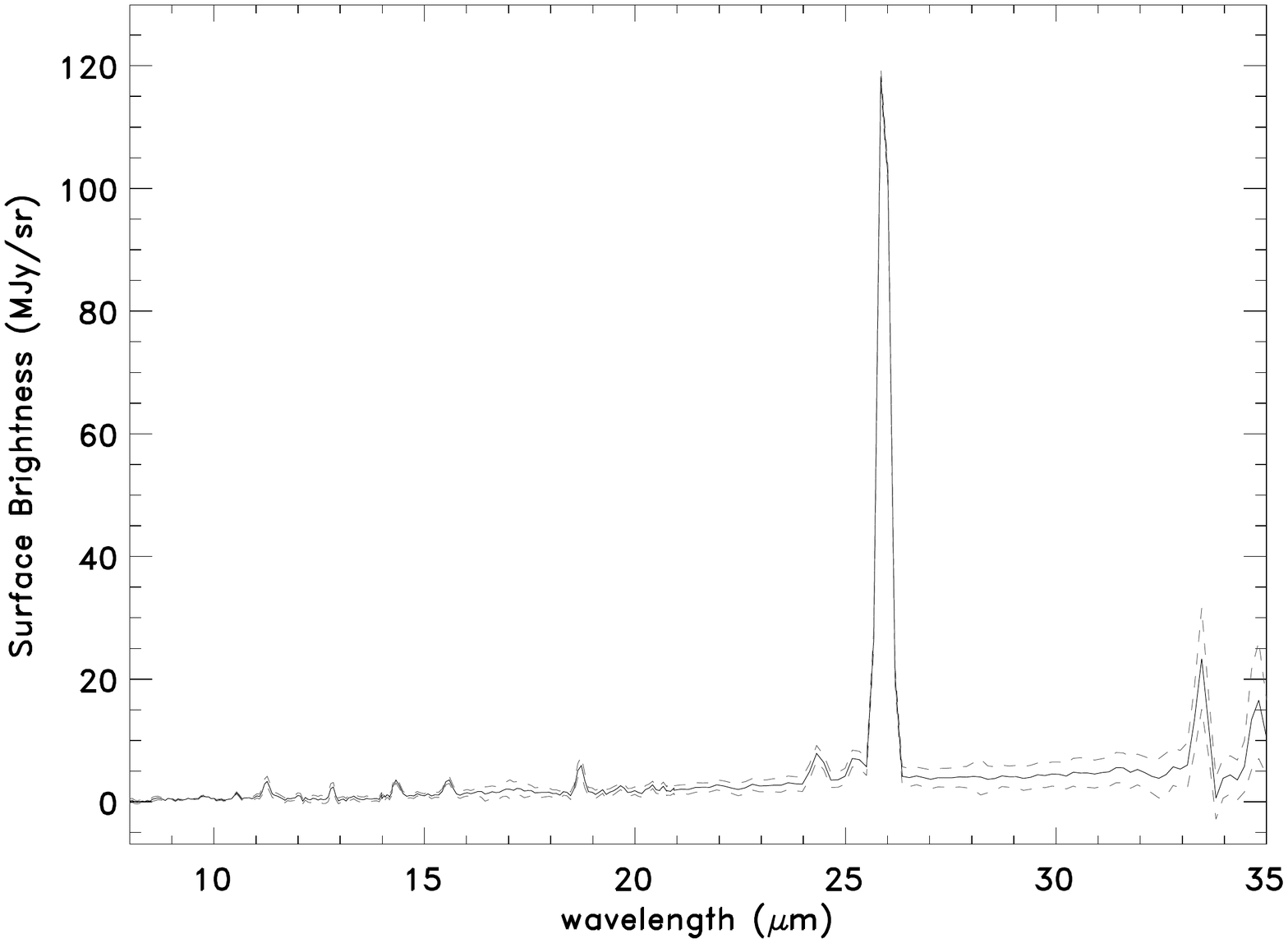}}
\subfloat[3575]{\includegraphics[trim=75 50 75 50, clip, width=0.5\linewidth, angle=180]{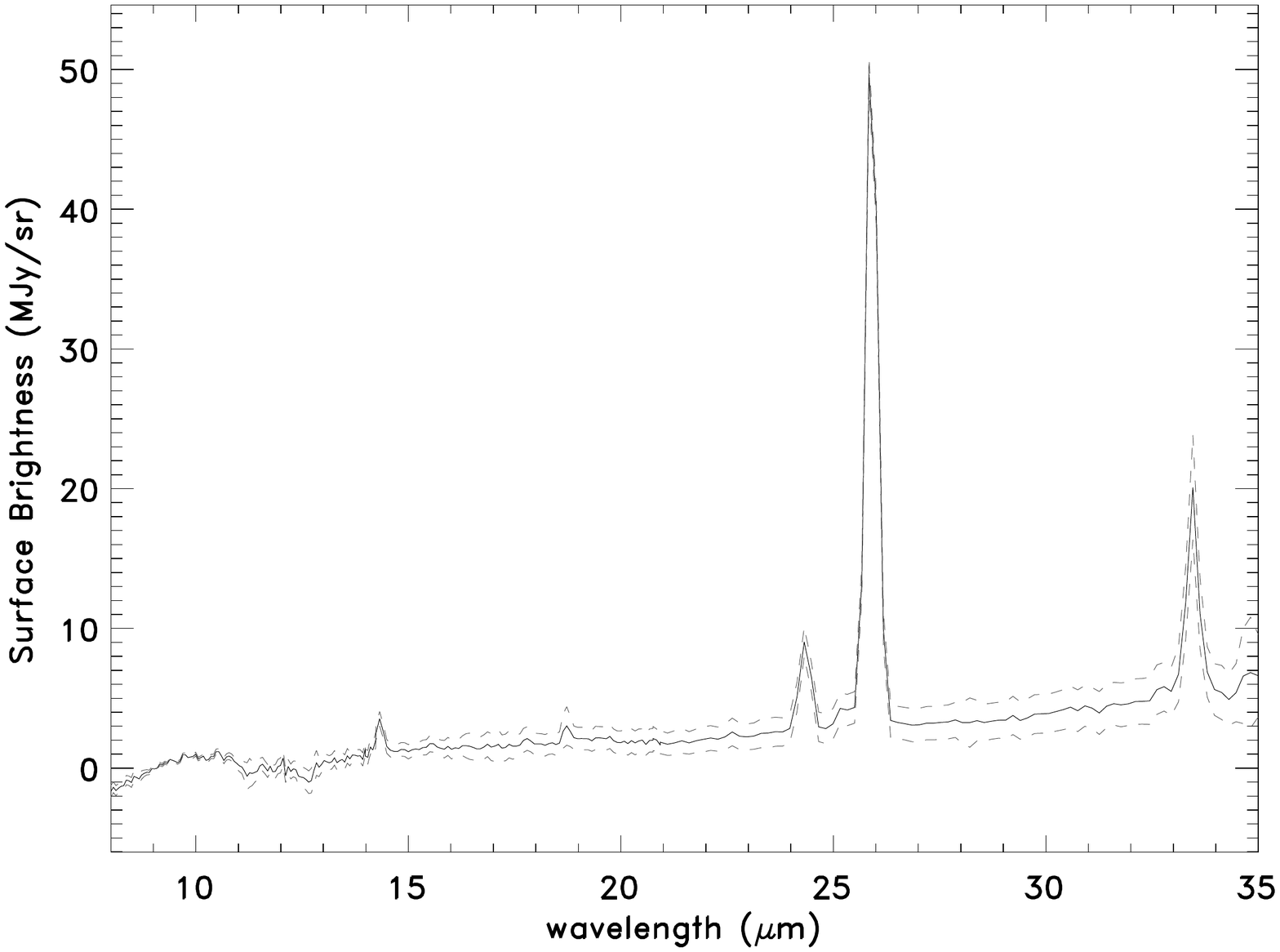}}\\
\subfloat[4021]{\includegraphics[trim=75 50 75 50, clip, width=0.5\linewidth, angle=180]{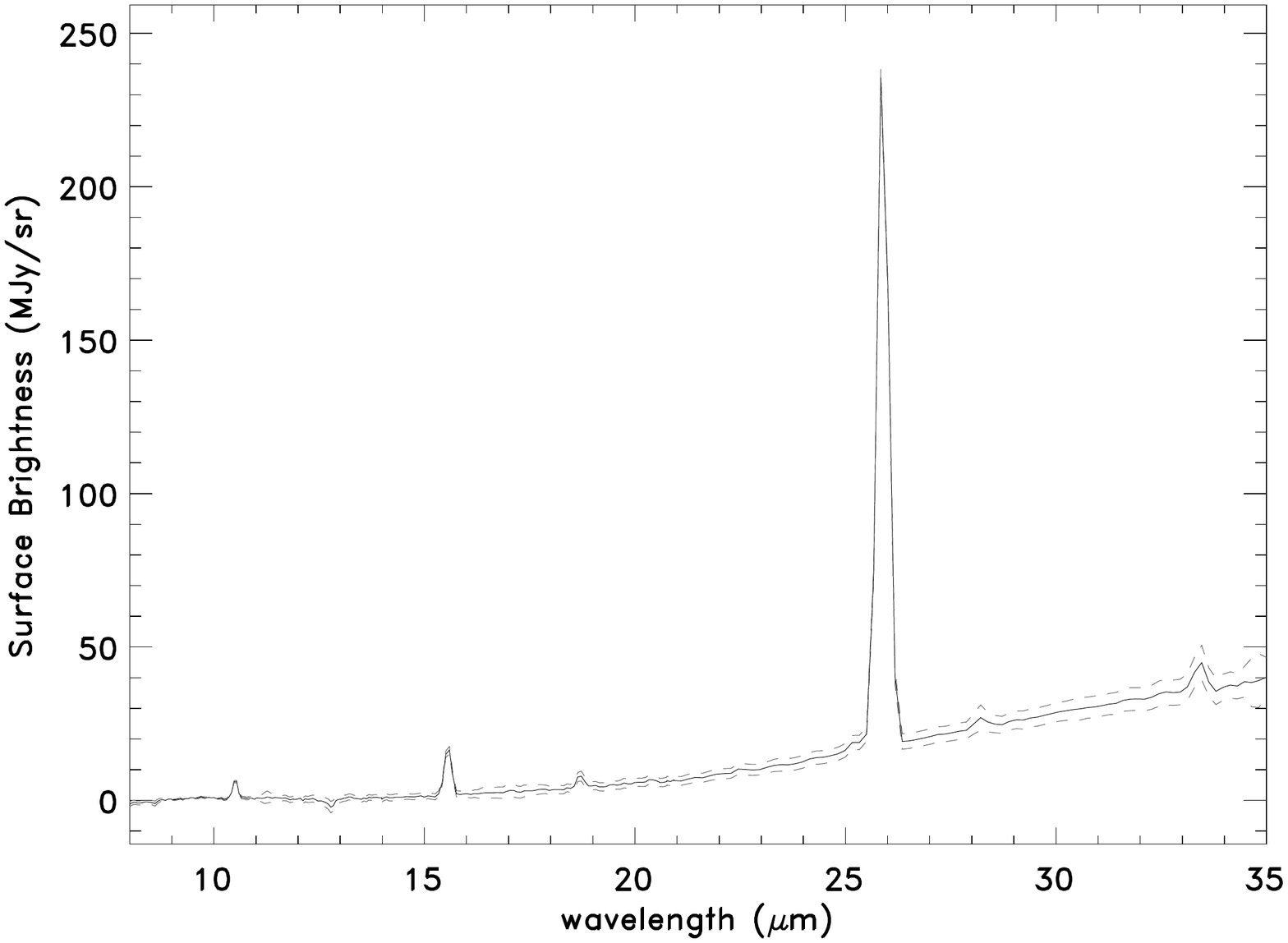}}
\subfloat[4017]{\includegraphics[trim=75 50 75 50, clip, width=0.5\linewidth, angle=180]{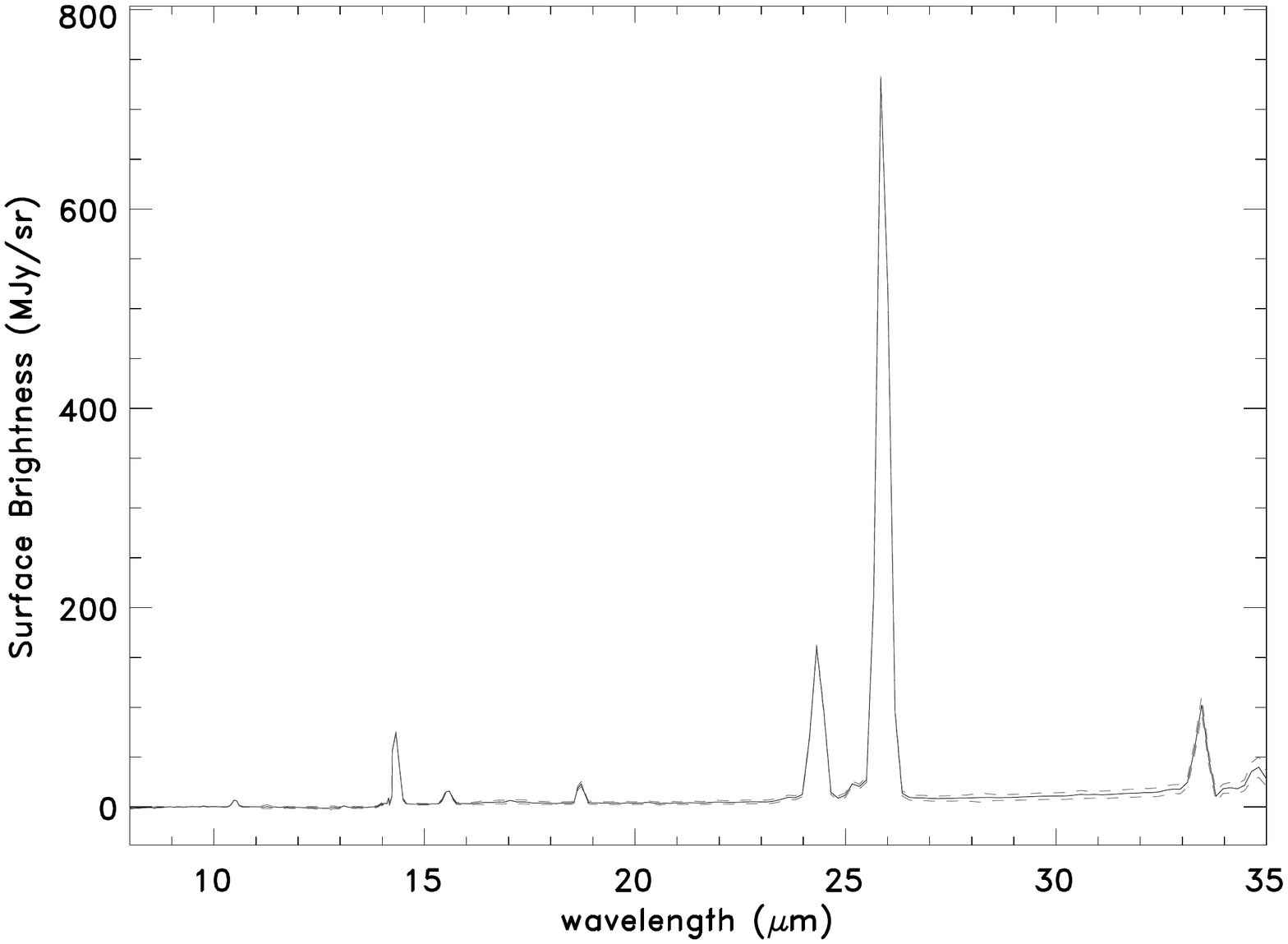}}\\
\caption{Same as Figure \ref{spectra_shell}, except for the MBs without central sources in the mid-IR.}
\label{spectra_other}
\end{center}
\end{figure*}

\begin{figure*}
\ContinuedFloat
\begin{center}
\subfloat[4076]{\includegraphics[trim=75 50 75 50, clip, width=0.5\linewidth, angle=180]{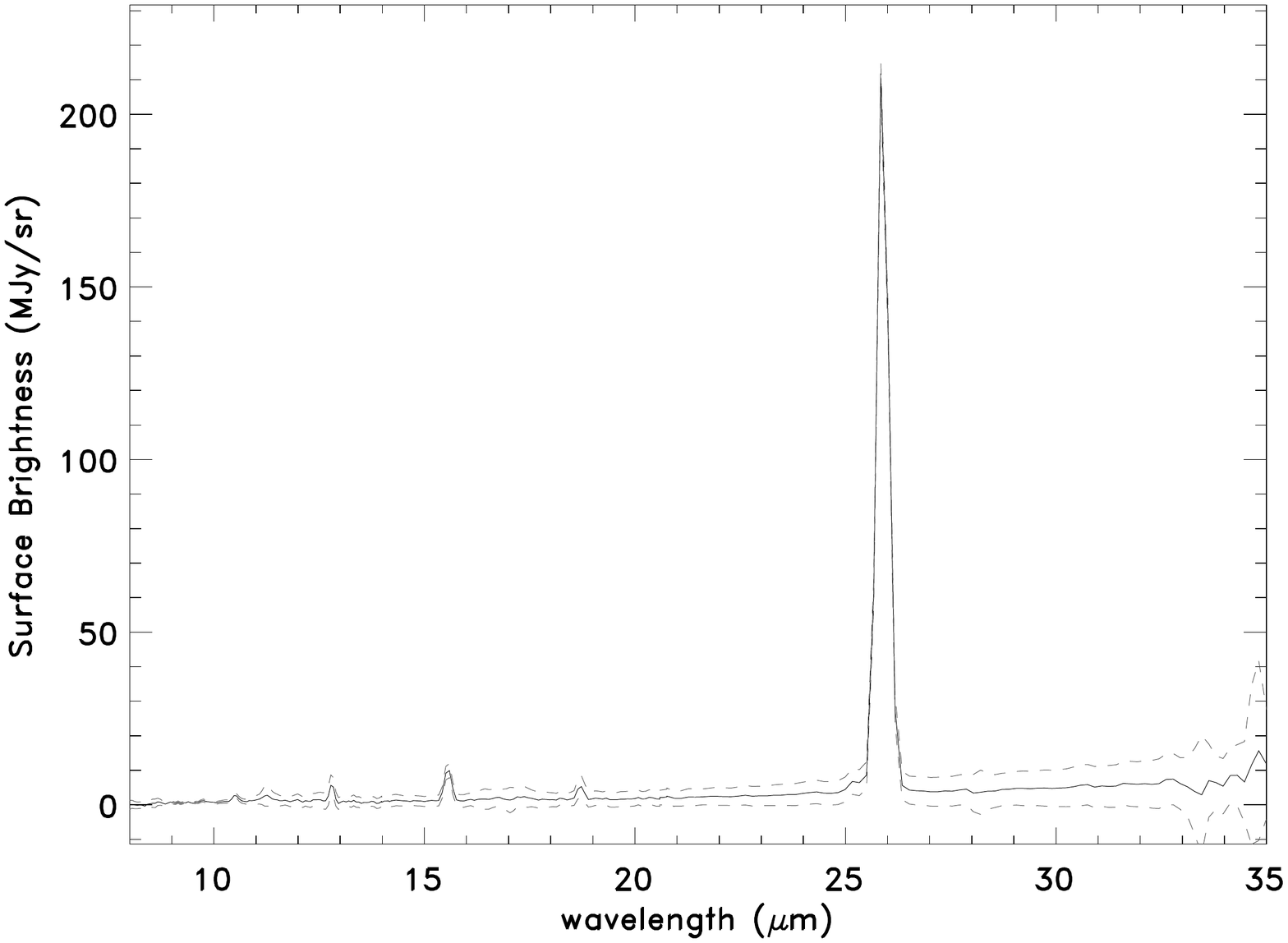}}
\subfloat[4066]{\includegraphics[trim=75 50 75 50, clip, width=0.5\linewidth, angle=180]{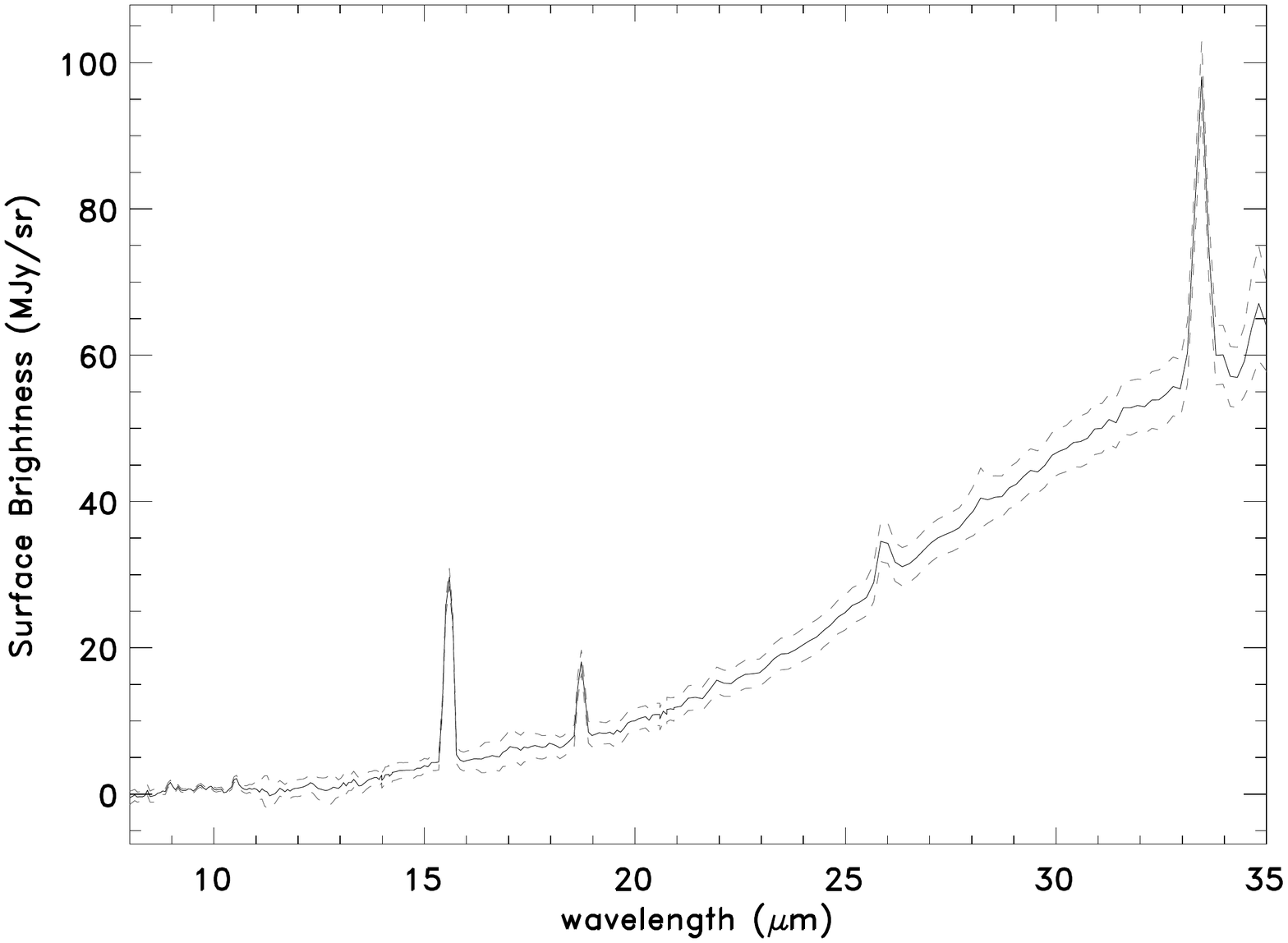}}\\
\subfloat[4376]{\includegraphics[trim=75 50 75 50, clip, width=0.5\linewidth, angle=180]{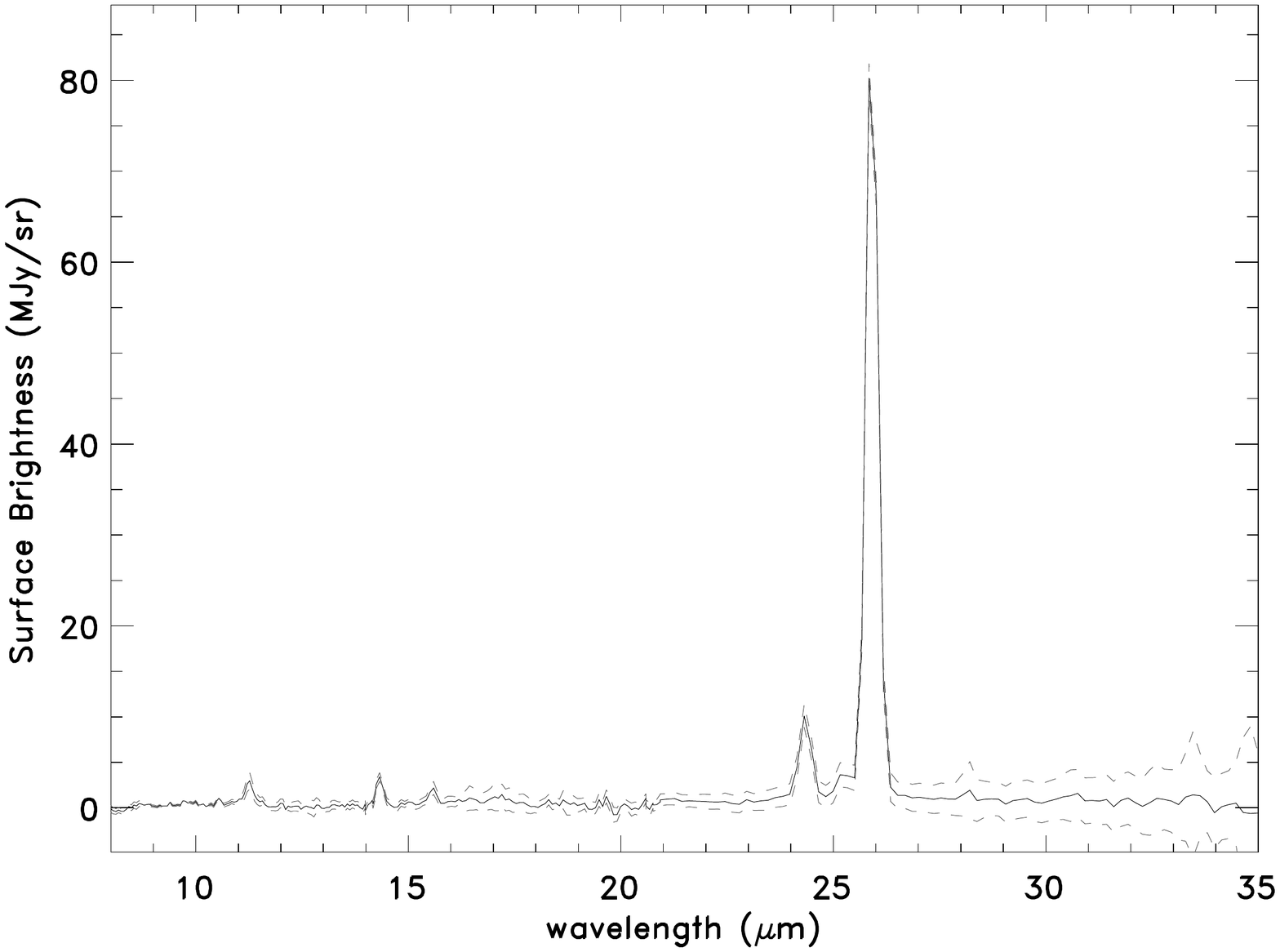}\label{fig:4376}}
\caption{continued}
\label{spectra_other}
\end{center}
\end{figure*}

\subsection{Gas line identification and flux measurement}
\label{line_fluxes}

To measure the fluxes of the gas lines detected in emission in the MBs, we use all four spectra we have for each position: on-source and background, nod1 and nod2. We use PAHFIT\footnote{tir.astro.utoledo.edu/jdsmith/research/pahfit.php} \citep[see][for details]{Smith2007} on each spectrum to fit all the detected gas lines. Some of the lines we observe in the IRS spectra are not listed in PAHFIT (e.g., the [Ne~\textsc{v}]~14.3 and 24.3\mic\ lines), so we add them to the code. The inferred fluxes and uncertainties of the gas lines are the means and the quadratic means of the values found for each nod. We do this separately for the sources and background positions. We then subtract from the on-source fluxes those measured on the background, taking into account the best coefficient for the background subtraction. The uncertainties are the quadratic sum of those inferred from the variations between the two nods and those from the background subtraction (see section~\ref{background}). Table~\ref{table_flux_shells_1} and \ref{table_flux_shells_2} list the fluxes for all the gas lines detected in the spectra of the MBs. Some lines in the tables are subject to large uncertainties, which makes their detection in the MBs uncertain. For example, the lines of [S~\textsc{iii}] and [Si~\textsc{ii}] beyond 33\mic\ are almost always associated with large uncertainties. The [Si~\textsc{ii}] line, being far too unreliable, is not presented in the table. There are also a few lines in MB4066, MB4076, and MB3944 that are uncertain (mainly [Ne \textsc{ii}] and the sulfur lines). We report the fluxes of these lines but suggest that some of them are, at least in part, of interstellar origin. For the four central sources in the sample, we derive the gas line fluxes after subtraction of those from their surrounding shells.

With a resolution of around 100, the IRS low resolution modules cannot distinguish the [O~\textsc{iv}]~25.9\mic\ from the [Fe \textsc{ii}]~26.0\mic\ lines, the [Ne~\textsc{v}]~24.3\mic\ from the [Fe~\textsc{ii}]~24.5\mic\ lines, or the [Ne~\textsc{v}]~14.3\mic\ from the [Cl~\textsc{ii}]~14.4\mic\ lines. Hence, when using PAHFIT to fit the whole spectra, it happens that the 26\mic, the 24.5\mic, and the 14.3\mic\ emission features are fitted using either or both lines. To determine which line dominates the emission in each feature, we use several methods. First, we look at the other gas lines detected in the spectrum. Considering the ionization potentials of the main species (see Table \ref{ionization}), and the presence or absence of, for instance, the [S~\textsc{iv}]~10.5\mic, [Ne~\textsc{ii}]~12.8\mic, or [Ne~\textsc{iii}]~15.5\mic\ line, we can usually disentangle a low ionization transition (e.g., [Fe \textsc{ii}]~26.0\mic) from a high ionization transition (e.g., [O~\textsc{iv}]~25.9\mic). Additionally, we can rule out in most cases the [Fe~\textsc{ii}]~26 and 24.5\mic\ lines, since in that case, we should also detect the [Fe~\textsc{ii}]~17.9\mic\ line. We finally can use the IRS high resolution observations of the four MBs in \citet{Flagey2011}, which correspond to templates of low- to high-excitation line spectra and that are similar to some of the observations in this paper. Having determined which of the two lines is responsible for the emission feature, we simply add the fluxes of the two lines used by PAHFIT to get the flux of the whole feature.

Except for the four dust-rich MBs (see section~\ref{sec:disc_dust_rich}), we do not have any way to constrain the extinction. Therefore we do not correct the line fluxes for this effect. However, we discuss the impact of the extinction on the measurements. The two main ratios that we use for the identification of the MBs are the [Ne~\textsc{iii}]~15.5\mic\ to [Ne~\textsc{v}]~24.3\mic\ ratio, and the [Ne~\textsc{v}]~14.3 to [Ne~\textsc{v}]~24.3\mic\ ratio (see section~\ref{sec:disc_dust_poor}). {We use the extinction curve for the diffuse ISM from \citet{Compiegne2011}}. With 10 magnitudes of visual extinction the first ratio increases {by less than 15\%}, while the second {increases by 6\%. To underestimate the [Ne~\textsc{iii}]~15.5 to [Ne~\textsc{v}]~24.3\mic\ and the [Ne~\textsc{v}]~14.3 to [Ne~\textsc{v}]~24.3\mic\ ratios by a significant amount (50\%) would require much larger amounts of extinction: at least 30 and almost 70~mag of visual extinction, respectively.}

% If the visual extinction reaches 40 magnitudes, the [Ne~\textsc{iii}]~15.5\mic~/~[Ne~\textsc{v}]~24.3\mic\ ratio would be underestimated by a factor 2, and the [Ne~\textsc{v}]~14.3\mic~/~[Ne~\textsc{v}]~24.3\mic\ ratio by about 30\%.

\input{tables_flux}

\input{table_flux_corrected}

\subsection{Origin of the 24\mic\ emission}
\label{sec:24emission}

{Before the launch of the next space mission with mid-IR spectroscopic capabilities ({\it the James Webb Space Telescope})}, the {\it Spitzer}/IRS spectra are the only observations capable of {determining the origin of the 24\mic\ emission}. We already mentioned that four MBs are ``dust-rich'' while the seven others have mid-IR spectra dominated by gas lines. Here we derive quantitative measurements of the gas and dust contributions to the 24\mic\ emission detected in the MIPSGAL images.

We use the MIPS~24\mic\ filter spectral response (22.8 to 26.2\mic{} at half-maximum) and color correction definition to derive, from the IRS spectra, the surface brightness that MIPS would have measured at 24\mic. For each spectrum, this value corresponds to the MIPSGAL 24\mic\ surface brightness within the area of the slit where we extracted the spectrum (see section \ref{sec:match}). We then remove the gas lines from the spectra using a median filter, and recompute the MIPS~24\mic\ surface brightness for each spectrum. The ratio of these two values corresponds to the continuum contribution to the MIPSGAL flux (see Table \ref{dust_vs_gas}). The uncertainties come from those on the background subtraction (see section~\ref{background}). The background spectra are usually continuum dominated and 10\% uncertainties on their mean surface brightnesses are responsible for large uncertainties in the dust and gas contributions to the MIPS~24\mic\ emission. For instance, towards MB4376, the uncertainties in the background subtraction may lead to unreaslistic negatives values in the spectrum at longer wavelengths (see Figure \ref{fig:4376} and section \ref{background}).

The contributions of the gas lines in the spectra of the ``dust-rich'' MBs to the 24\mic\ emission is not larger than those due to the uncertainties (1-2\% in most cases). On the contrary, the spectra of the ``highly excited'' MBs exhibit a wide range of dust and gas contributions. In MB4376 and MB4017, the dust continuum accounts for less than 26\% of the MIPS~24\mic\ emission, while in MB4066 the gas lines contribution to the mid-IR emission is less than 2\%. MB4376 and MB4017 are thus very similar to MB4001 and MB4006, where the dust contribution was 15 and 30\% respectively \citep{Flagey2011}, while MB4066 seems, from that vantage point, very similar to the ``dust-rich'' MBs. Even though the spectra of the central sources are significantly richer than that of its surrounding nebulae, in terms of gas lines emission, the contribution of the dust to the MIPS~24\mic\ emission remains greater than 98\%.\\

\begin{table}
\caption{Dust and gas contribution to the MIPS~24\mic\ flux}
\label{dust_vs_gas}
\begin{center}
\begin{tabular}{l c c}
\hline
\hline
MB & Dust contribution & Gas contribution \\
\hline
MB4376 & 0 - 26\% & 74 - 100\% \\
MB3944 & 42 - 57\% & 43 - 58\% \\
MB3575 & 16 - 59\% & 41 - 84\% \\
MB4021 & 63 - 70\% & 30 - 37\% \\
MB4017 & 14 - 24\% & 76 - 86\% \\
MB4066 & >98\% & <2\% \\
MB4076 & 26 - 46\% & 74 - 54\% \\
MB4121's shell & >98\% & <2\% \\
MB4121's central source & >98\% & <2\% \\
MB4124's shell & >98\% & <2\% \\
MB4124's central source & >98\% & <2\% \\
MB4384's shell & >98\% & <2\% \\
MB4384's central source & >98\% & <2\% \\
MB3955's shell & >98\% & <2\% \\
MB3955's central source & >98\% & <2\% \\
\hline
\end{tabular}
\end{center}
\end{table}

Hereafter we discuss in detail the two subsets in the sample. The ``highly excited'' MBs are all unknown and we thus use the IRS data to constrain their nature. The spectral types of the central sources in the ``dust-rich'' MBs are all known, and we thus focus on characterizing their mid-IR dust-dominated emission.

\section{Highly excited MBs}
\label{sec:disc_dust_poor}

\subsection{Different level of excitation}

The subset of ``highly excited'' MBs is composed of seven MBs that all share some common characteristics: ``disk''-like morphology in the MIPS~24\mic~images \citep[see][and section \ref{sec:disc_extrap} for details about morphologies]{Mizuno2010}, strong [O~\textsc{iv}]~26\mic, [Ne~\textsc{iii}]~15.5\mic\ and/or [Ne~\textsc{v}]~14.3 and 24.3\mic\ lines in the IRS spectra. However, we also notice variations within the subset: MB4021 and MB4066 exhibit a significantly brighter continuum than the five other MBs. This continuum could indicate that among the ``highly excited'' MBs, those two have a less harsh environment such that the dust grains are not completely destroyed. In Table~\ref{Ne_ratio} we give the ratios between the [Ne~\textsc{ii}]~12.8\mic, the [Ne~\textsc{iii}]~15.5\mic, and the [Ne~\textsc{v}]~14.3, and 24.3\mic\ line fluxes. These ratios are indicators of the ionization level in the MBs and do not depend on the element abundance. {However, despite our efforts to subtract the background emission from the IRS spectra (see section \ref{background}), a fraction of the [Ne~\textsc{ii}]~12.8\mic\ and the [Ne~\textsc{iii}]~15.5\mic\ lines fluxes could originate in the ISM.} In the following, we use the dust contributions to the mid-IR emission and the neon line ratios to guide the discussion from the less excited to the more excited MBs in the subset. 

MB4066, MB4021, and MB4076 are the three MBs that do not have any [Ne~\textsc{v}] lines in their spectra, and the first two have large dust contributions to their MIPS~24\mic\ emission: about 100\% for MB4066, and about 67\% for MB4021 (see Table \ref{dust_vs_gas}). The dust contribution to the 24\mic\ emission in MB4076 is significantly lower, at about 36\%. Another difference in the IRS spectra of these three MBs is the much clearer presence of the [S~\textsc{iv}] line at 10.5\mic, relative to the [S~\textsc{iii}] line at 18.7\mic, in MB4021 and MB4076, as compared to MB4066 (see Table \ref{table_flux_shells_1}). The only MBs for which we detect the [Ne~\textsc{ii}]~12.8\mic\ line (MB4076 and MB3944) are also among those with weak or no [Ne~\textsc{v}] lines at all. However, the detection of the [Ne~\textsc{ii}] line remains questionable. The dust contribution to the 24\mic\ emission in MB3944 is still significant, at about 50\%. Among the MBs with detected [Ne~\textsc{v}] lines, MB3944 is the one with the lower [Ne~\textsc{v}] to [Ne~\textsc{iii}] ratio.

The three other MBs in this paper with [Ne~\textsc{v}] lines in their IRS spectrum (MB4376, MB4017, and MB3575) all have [Ne~\textsc{v}]~14.3 to [Ne~\textsc{iii}]~15.5\mic\ line ratios greater than 1, and [Ne~\textsc{v}]~24.3\mic\ to [Ne~\textsc{iii}]~15.5\mic\ line ratios greater than a few. {In this respect,} they are similar to MB4001 and MB4006, for which \citet{Flagey2011} measured [Ne~\textsc{v}]~14.3\mic\ to [Ne~\textsc{iii}]~15.5\mic\ line ratios of 2.8 and 4.0, respectively. The gas contributions to the 24\mic{}~emission in these three MBs are significantly larger than 50\%, though the uncertainty on the background leads to a value of 62$\pm$22\% for MB3575. To this extent, MB4001 and MB4006 are similar MBs with dust contributions of about 30 and 15\% respectively.

\begin{figure}
\includegraphics[trim=75 50 75 50, clip, width=\linewidth, angle=180]{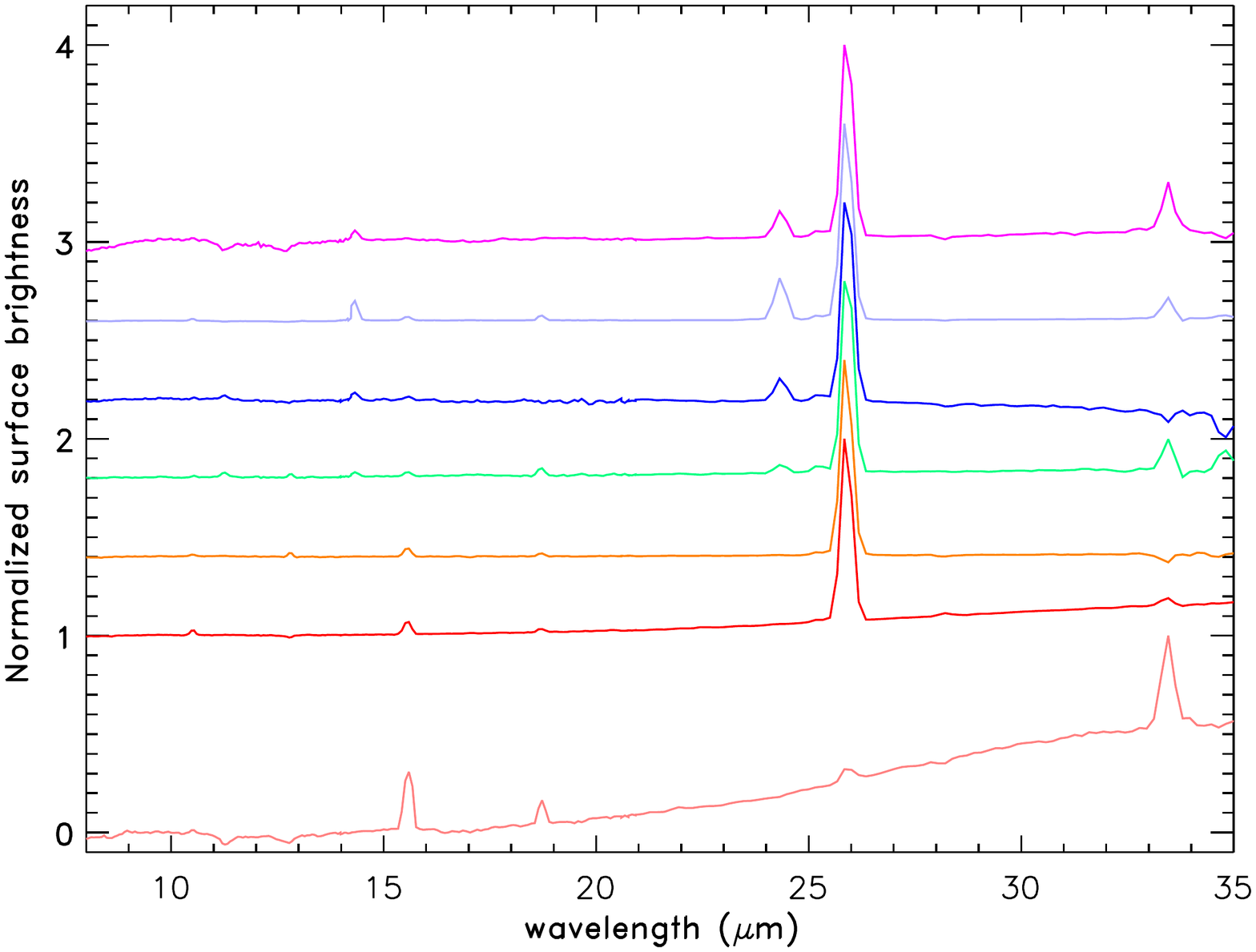}
\caption{Highly excited MBs sorted by excitation level (based mainly on the [Ne~\textsc{v}] to [Ne~\textsc{iii}] ratio). From bottom to top : MB4066, MB4021, MB4076 , MB3944, MB4376, MB4017, and MB3575. The spectra are normalized by their maximum, and an offset of 0.4 (1 for MB4021) has been introduced to 
distinguish them.}
\label{ionization_level}
\end{figure}

In Figure~\ref{ionization_level}, we combine all the spectra of the ``highly excited'' MBs, to highlight the variations in their mid-IR spectra. The dust and gas contributions to the 24\mic\ emission, as well as the neon line ratios significantly vary from the lowest to the highest excitation among the MBs. However, we do not claim there is a clear correlation between the neon line ratios and the dust contributions to the MIPS~24\mic~emission, since the number of MBs in the sample is low, and the error bars arising from the uncertainties on the background subtractions are large. We do not discuss in this paper the origin of such differences in the excitation within the MBs (e.g. age of the PNe).

\begin{table}
  \caption{Neon lines ratios}
  \label{Ne_ratio}
  \label{Ne_V_ratio}
  \begin{center}
    \begin{tabular}{c c c c c}
      \hline
      \hline
      MB & $\frac{[\textrm{Ne}~\textsc{ii}]12.8}{[\textrm{Ne}~\textsc{iii}]15.5}$ & $\frac{[\textrm{Ne}~\textsc{v}]14.3}{[\textrm{Ne}~\textsc{iii}]15.5}$ & $\frac{[\textrm{Ne}~\textsc{v}]24.3}{[\textrm{Ne}~\textsc{iii}]15.5}$ & $\frac{[\textrm{Ne}~\textsc{v}]14.3}{[\textrm{Ne}~\textsc{v}]24.3}$  \\
      \hline
      MB4066 & 0 & 0 & 0 & - \\
      MB4021 & 0 & 0 & 0 & - \\
      MB4076 & 0.41 & 0 & 0 & - \\
      MB3944 & 0.60 & 0.69 & 1.69 & 0.41\\
      MB4376 & 0 & 1.12 & 5.08 & 0.22\\
      MB4001 & 0.02 & 2.8 & 6.8 & 0.41 \\
      MB4017 & 0 & 3.58 & 11.16 & 0.32\\
      MB3575 & 0 & 3.29 & 13.71 & 0.24\\
      MB4006 & 0.06 & 4.0 & 10.5 & 0.38 \\
      \hline
    \end{tabular}
    \tablecomments{The seven ``highly-ionized'' MBs from this paper, plus two (MB4001 and MB4006) from \citet{Flagey2011} are included.}
  \end{center}
\end{table}

\subsection{\emph{Ne} ratios}

As emphasized in the previous subsection, the highly excited MBs exhibit high to very high [Ne~\textsc{v}] to [Ne~\textsc{iii}] ratios. Values of a few for the [Ne~\textsc{v}] 24.3\mic\ to [Ne~\textsc{iii}] \unit{15.5}{\micro\meter} ratio are not uncommon, but in the case of MB4017, MB3575, and MB4006, the values are above 10 and up to 13.7, which is unseen in recent literature {and difficult to obtain with photoionization models (e.g. CLOUDY\footnote{\url{http://www.nublado.org/}}, MAPPINGS\footnote{\url{http://home.strw.leidenuniv.nl/ brent/mapiii.html}})}. 

Similarly, the [Ne~\textsc{v}] 14.3\mic\ to [Ne~\textsc{v}] \unit{24.3}{\micro\meter} ratios range from 0.22 to 0.41 in the six MBs where the [Ne~\textsc{v}] lines are detected, including two MBs from \citet{Flagey2011}. {An interstellar extinction of 10 magnitudes would increase this ratio by only a few percent} (see section~\ref{results}). Such a low ratio ($\leq 1$) is not usual, although \citet{Rubin2004} have found similar values in 10 out of the 20 PNe they observed with {\it ISO}. The results of \citet{Rubin2004} and others triggered the work of \citet{Dance2013}, who obtained improved collision strengths for the mid-IR and optical transitions of Ne~\textsc{v}. This has allowed an extension of the use of this diagnostic for nebulae temperatures and densities to the low temperatures ($T_e\leq10000$~K), in which cases ratios lower than 1 can be interpreted.

However, in the case of our objects, even these improved collision strengths fail to provide a credible interpretation of the Ne ratios. According to \citet{Dance2013}, electronic temperatures significantly lower than 10,000~K and electronic densities significantly lower than $10^4~\rm{cm^{-3}}$ are required to reach [Ne~\textsc{v}]~14.3\mic\ to 24.3\mic\ line ratios below 1. {It is difficult to give tighter constraints on both parameters, which are slightly degenerate. Although, converting their Figure 4 into a contour plot of the [Ne~\textsc{v}] lines ratio, we find that for electronic densities in the range $10-10^4~\rm{cm^{-3}}$, electronic temperatures have to be as low as a few hundreds K in the six MBs that exhibit the [Ne~\textsc{v}] lines in their IRS spectra. Even though low electronic temperatures environments can be found in PNe, they are usually surrounded by larger electronic temperatures conditions \citep[e.g. the H-rich envelopes surrounding the H-deficient knots in Abell~30 ][]{Ercolano2003}. The highy excited MBs, with their very unusual Ne ratios, remain a challenge to the current understanding of collision strengths in the mid-IR, and will require additional effort in terms of photoionization modeling.}

\subsection{A common nature: planetary nebulae?}
\label{sec:pne}

The MBs in the ``highly excited'' subset seem to span a wide range of excitation conditions. However, the many characteristics that they share suggest a common nature. We look at the recent literature to find examples of similar mid-IR spectra to constrain the nature of the ``highly excited'' MBs.

%Morris2006
\subsubsection{Cepheus bubble \citep{Morris2006}}

\citet{Morris2006} used IRS low resolution observations to investigate a shell discovered in Cepheus that has a morphology similar to the ``highly excited'' MBs at 24\mic. The spectrum they obtain is almost identical to that of MB4076 or MB3944: no continuum, very strong [O~\textsc{iv}]~26\mic~line, in addition to the [Ne~\textsc{iii}], [Ne~\textsc{v}], and [S~\textsc{iii}] lines. To account for the lack of continuum, they suggest that this shell has a very high gas-to-dust ratio. They ruled out the PN hypothesis, and suggested this bubble is a young dust-free supernova remnant (SNR) because its spectrum does not show any PAH features or hydrogen lines. However, \citet{Fesen2010} ruled out this identification thanks to optical observations. They find that the spectrum of the shell shows narrow H$\alpha$ and [N~\textsc{ii}] lines with no hint of broad or high-velocity line emission. They conclude that the Cepheus bubble is not a SNR, but is likely a faint PN. They also found a potential central source with $m_V\sim22.5$ which would correspond to a white dwarf (WD) at about 2.5~kpc. Additionally, the H$\alpha$ line peaks at $v_{LSR}\simeq-70$~km~s$^{-1}$, which is consistent with the location of the Perseus Arm in the direction toward this shell, at a distance of 2-3~kpc. The Cepheus bubble has a [Ne~\textsc{iii}]~15.6\mic\ stronger that the [Ne~\textsc{v}] 24.3\mic\ line. In Table~\ref{Ne_ratio}, it would lie between MB4076 and MB3944. However, we note that no spectroscopic data have been obtained yet on the central star to further constrain the nature of this shell.

%SMP83
\subsubsection{SMP~83 \citep{Bernard-Salas2004}}
MB3944, MB4376, MB4001, and MB4017, which all exhibit a [Ne~\textsc{v}]~14.3\mic\ to [Ne~\textsc{iii}]~15.5\mic~line ratio greater than 1, are more similar to the Large Magellanic Cloud PN SMP~83 studied in \citet{Bernard-Salas2004}. The spectrum of SMP~83, acquired with the IRS high resolution and SL1 modules, is a pure emission line spectrum, with strong [O~\textsc{iv}], [Ne~\textsc{v}], and [Ne~\textsc{iii}] lines. The similarity with MB3944 in particular is supported by the [Ne~\textsc{v}]~14.3\mic\ to [Ne~\textsc{iii}]~15.5\mic\ and [Ne~\textsc{v}]~24.3\mic\ to [Ne~\textsc{iii}]~15.5\mic\ line ratio, which are 1.4 and 1.5, respectively, for SMP~83, while they are 0.7 and 1.7 respectively for MB3994.

\citet{Hamann2003} summarized the findings of almost 20 years of observations of this PN, marked in 1993-1994 by an outburst of its central source, which was then suggested as a WN4.5 \citep{Pena1995}. They suggested several interpretations regarding the nature of the central source, including single star and binary systems, but all of them contradict at least one observational fact. Nonetheless, the low or high-mass binary system -- a massive star after a common-envelope phase, and a helium-accreting white dwarf -- seem to be more favorable. Thirteen years after the first outburst, \citet{Pena2008} reported a second outburst, during which the central star was again comparable to an early WN, probably WN3, though it was much less luminous than a classical high mass WR. The favored interpretation is thus that of a variable [WN] (i.e. a central star of a PN that mimics the helium- and nitrogen-rich WN sequence of massive WR stars), probably due to a He-accreting, He-shell flash, WD that is accreting mass from a companion in a binary system.

%\subsubsection{Serpens bubble \citep{Oliveira2011}}
%\citet{Oliveira2011} found a object toward the Serpens Molecular Cloud, and identified it as a PN, thanks to a combination of observations in IR, visible and UV. This PN shows a mid-IR spectrum similar to that of the ``highly excited'' MBs: almost pure emission line spectrum, with a strong [O~\textsc{iv}]~26\mic~line, and some neon and sulfur lines. The [Ne~\textsc{v}]~14.3\mic\ to [Ne~\textsc{iii}]~15.5\mic~and [Ne~\textsc{v}]~24.3\mic\ to [Ne~\textsc{iii}]~15.5\mic~line ratios for the Serpens PN both are 2.3, which puts it close to MB3944, MB4376, MB4001, and MB4017 in Table~\ref{Ne_ratio}. The spatial extent of the PN is about 6\arcsec, as inferred from its H$\alpha$ emission. In the {\it Spitzer} images, the PN is point-like, which might be related to a large heliocentric distance. We looked at the UKIDSS \citep{Lawrence2007} data to find only a diffuse counterpart at the position of the PN, but no point source. The nature of SMP83 thus remains evasive.

\subsubsection{CSPN [WR] \citep{Hart2014}}
A {\it Spitzer}/IRS program of observations \citep[PID40115, P.I. G. Fazio, ][]{Hart2014} has been dedicated to [WC] and [WO], H-deficient, helium-, carbon- and oxygen-rich central low-mass stars that mimic the carbon-sequence of massive WR. About half of their sources show spectra similar to ours: very limited contributions of the dust continuum, high excitation gas lines. The best match to the ``highly excited'' MBs in their sample is the PN~PB6 where the [Ne~\textsc{v}]~14.3\mic\ to [Ne~\textsc{v}]~24.3\mic\ line ratio is about 0.3, while the [Ne~\textsc{v}]~14.3\mic\ to [Ne~\textsc{iii}]~15.5\mic\ line ratio is about 0.7. Those values are very similar to the ratios we measure in MB3944. The [WR] interpretation for the ``highly-excited'' MBs therefore seems plausible, and if confirmed, could provides several additional [WR] stars to progress in the study of their poorly understood evolutionary path \citep[see e.g.][and references therein]{Miszalski2012}.

\subsubsection{Classical novae \citep{Helton2012}}
Three classical novae have been studied by \citet{Helton2012} using IRS data. One of those novae (V1494~Aql) shows a spectrum (acquired in 2007) similar to the ``highly excited'' MBs, with [Ne~\textsc{v}] lines at 14.3 and 24.3\mic, though no [Ne~\textsc{iii}] line was detected at 15.5\mic. From the Spitzer Heritage Archive, we retrieved their IRS low resolution data and extracted the spectrum of V1494~Aql using a point source extraction in SPICE. We found that the ratio between the two [Ne~\textsc{v}] lines is about 0.5, in the range of the ratios measured in our data, while \citet{Helton2012} reported a ratio of 1.2 using the high-resolution observations.  The nova V1494~Aql was also reported to have an expansion speed of about 2000~km/s, which cannot be resolved with the low-resolution module of IRS. These observational characteristics make V1494~Aql a good match to the ``highly excited'' MBs. 

However, most novae, including V1494~Aql, reach very bright magnitudes at their peak ($m_V\lesssim10$), which makes them easily detectable. Indeed, about $25\%$ of the novae listed by the Central Bureau for Astronomical Telegrams\footnote{http://www.cbat.eps.harvard.edu/nova\_list.html} reached a magnitude $<6.5$ (naked-eye limit) at their peak, and about two third had a magnitude $<10$. When extrapolated to the whole catalog of \citet{Mizuno2010}, this interpretation raises a serious problem as there are more than 200 MBs with similar morphologies and that are suggested to have the same nature (see section \ref{sec:disc_extrap}). If a significant fraction of those are novae, they would all need to be distant or old enough to have been missed by astronomers during the past few centuries. However this hypothesis remains interesting to investigate further as, even though many Galactic novae are now known, the discovery of some of them through their nebulae may indicate that there is more to understand about their evolution.

\subsubsection{Conclusion}
The recent literature provides us with several objects with a mid-IR spectra similar to those of the nine ``highly excited'' MBs, although three of the MBs (MB4375, MB4017, and MB3575) are in a much higher ionization state than every other sources, as indicated by their [Ne~\textsc{v}] to [Ne~\textsc{iii}] line ratios. These objects are all identified as PNe, which strongly suggests that the ``highly excited'' MBs are also PNe. However, the exact natures of their central sources remain a mystery. Whether a few novae, some [WR], other types of WD in binary systems, or even some massive WR stars can be found among the ``highly excited'' MBs and the whole catalog of \citet{Mizuno2010} needs to be confirmed. {We emphasize here the importance of these candidates for various fields of modern astrophysics. For instance, the discovery of a handful of WD in binary systems would have ties with cosmological studies as these systems are the most favored progenitors of supernovae Ia \citep[e.g.][among many others]{Livio2000,Lepo2013}.}

The priority is thus to find the central stars, as \citet{Fesen2010} did for the Cepheus bubble. Characterizing their absolute fluxes and colors would help distinguish between the different interpretations, as the average massive WR is about 10 magnitudes more luminous than the average WD. However, unlike the Cepheus bubble, the MBs are located towards the inner Galactic plane, below 1 degree of latitude. The interstellar extinction along the lines of sight is significantly larger and may render impossible such detection. {Indeed, none of the ``highly excited'' MBs appear in the images of the Super Cosmos H$\alpha$ Survey \citep[SHS,][]{Parker2005}, probably indicating that these objects are beyond a fair amount of ISM extinction.}

\section{Dust-rich MBs}
\label{sec:disc_dust_rich}

\subsection{Massive central stars}
\label{sec:dust_rich_cs}

{The stars at the center of the four ``dust rich'' MBs are all detected in mid-IR images. The central source of MB3955 is detected in the IRAC 8\mic\ and WISE 12\mic\ images, while those of MB4121, MB4124, and MB4384 are also detected in the MIPS~24\mic\ images. They have all been previously observed and given spectral identifications. The star at the center of MB3955, identified as CD-61~3738 in SIMBAD, is a B supergiant \citep[B2 Ib or B5 Iap,][]{Reed2003}. The central sources in MB4121 and MB4124 are both LBV candidates \citep{Wachter2010, Wachter2011, Gvaramadze2010} although the resemblance with Be or B[e] stars is strong. The star at the center of MB4384 is an Oe/WN9 \citep{Wachter2010}. Therefore all are massive stars candidates.}

\subsection{Dust emission distribution}

The IRS spectra for the ``dust rich'' MBs, acquired towards their central sources and shells, are given in Figure~\ref{spectra_shell}. The spectra are dominated by a dust continuum at 24\mic: at least 98\% of their fluxes come from the continuum (see Table~\ref{dust_vs_gas}). The spectra towards the outer shells exhibit a rising continuum, {longward of} 15-20\mic, which {we associate with} ``cold'' dust emission in the nebulae. This component also dominates the spectra towards the central sources, except for MB4124, where ``warm'' dust emission contributes the most to the whole IRS spectrum. Towards the other central sources, this ``warm'' component is not as strong, and the differences between the shells' and the central sources' spectra remain subtle (see Figure~\ref{spectra_shell}). In particular, both spectra acquired towards MB3955 are extremely similar, which is in agreement with the fact that the central source of MB3955 is only detected in images at wavelengths shorter than $\sim$15\mic. A continuum at $\lambda$$<$10\mic\ is detected in the spectra towards all central sources and is associated with ``hot'' dust or stellar emission. A {distribution} of emission components is thus found, from the hottest one near the central sources to the coldest one in the outer shells. We characterize the dust components within the MBs in section \ref{sec:fit_dust}.

The spectra of the central sources in MB4124 and MB4384 are rich in emission lines{, even though the dust continuum still dominates the mid-IR emission} (see Figure~\ref{spectra_shell} and Table~\ref{table_flux_shells_2}). These spectra are very similar to those of other dust-rich {massive stars} observed with {\it Spitzer}/IRS \citep[e.g. the LBV and LBV candidates HR Car, HD 168625, G79.29+0.46, and MWC~930,][]{Umana2009, Umana2010, Agliozzo2014, Cerrigone2014}. The spectra of {the central stars in MB3955 and MB4121} are much poorer in terms of gas lines, with only a few emission lines detected towards both of them (see Figure~\ref{spectra_shell} and Table~\ref{table_flux_shells_2}).

\subsection{Characterization of the dust}
\label{sec:fit_dust}

We use the spectral energy distributions (SED) of the ``dust-rich'' MBs, towards both the central sources and the outer shells, to characterize their dust emission, {infer the interstellar extinction along the lines of sight and hence estimates of the distances to, and the dust masses in the nebulae}.

\begin{figure*}
\begin{center}
\subfloat[3955]{\includegraphics[angle=90, width=0.495\linewidth]{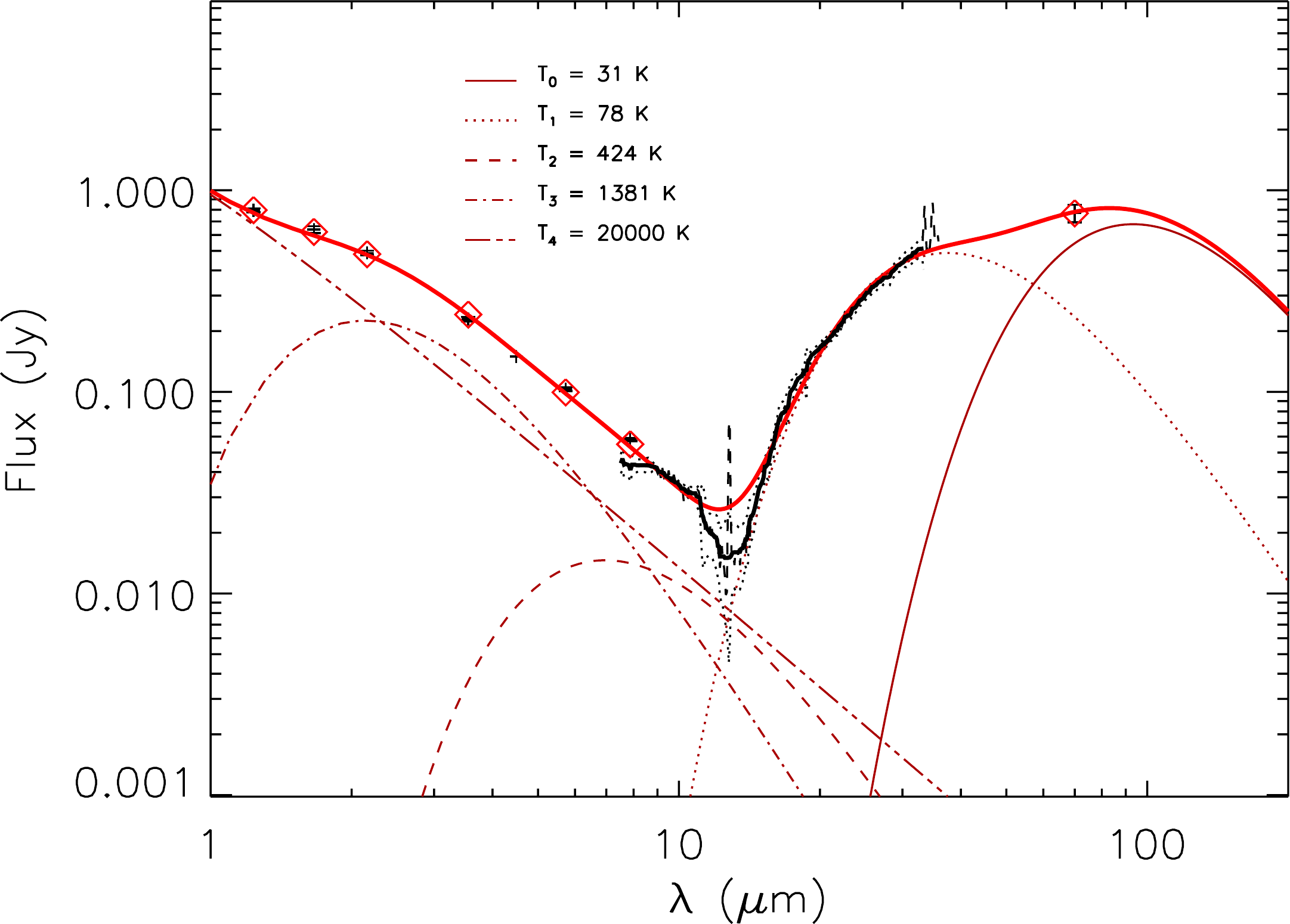}}
\hfill
\subfloat[4121]{\includegraphics[angle=90, width=0.495\linewidth]{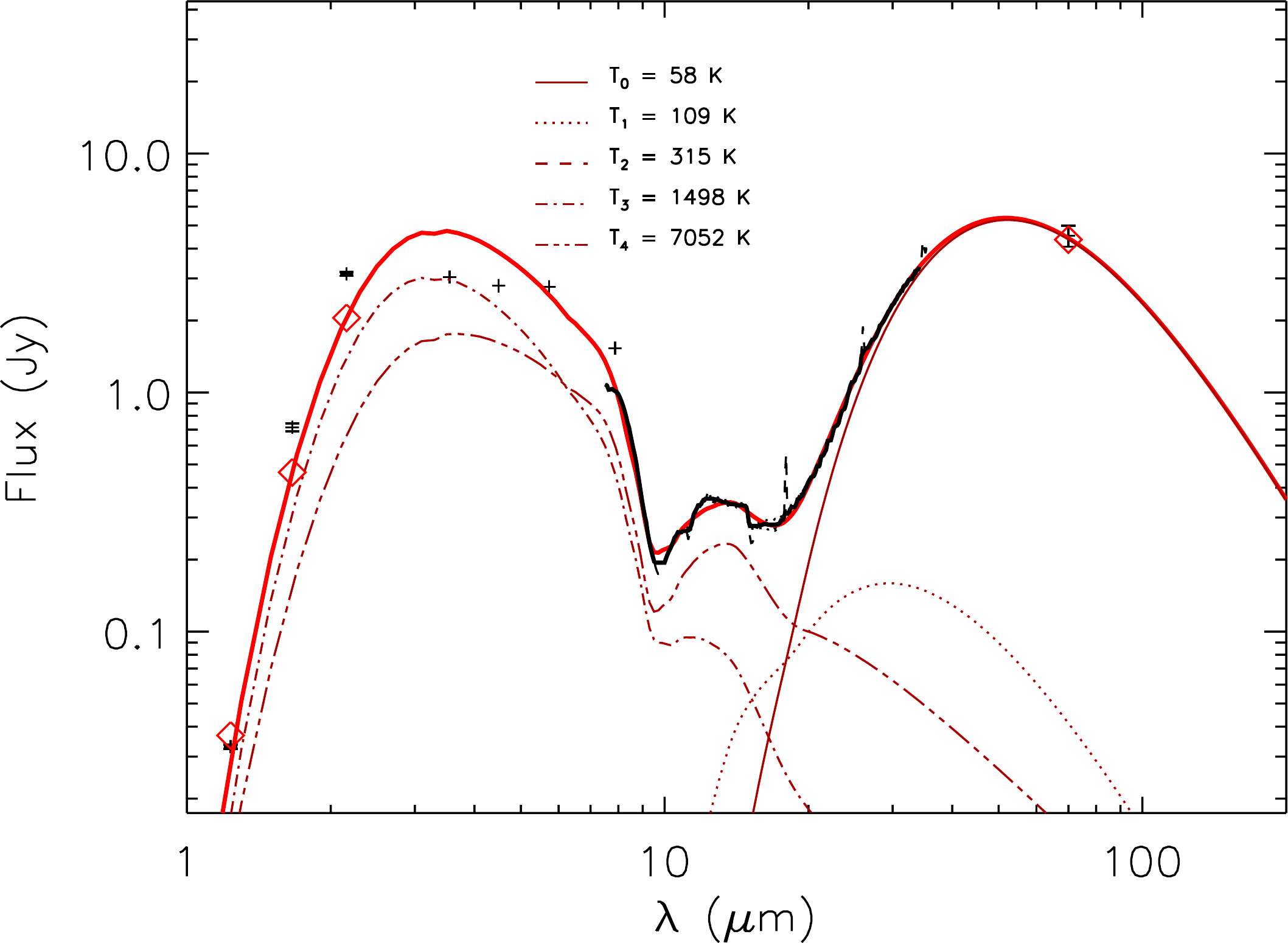}}\\
\subfloat[4124]{\includegraphics[angle=90, width=0.495\linewidth]{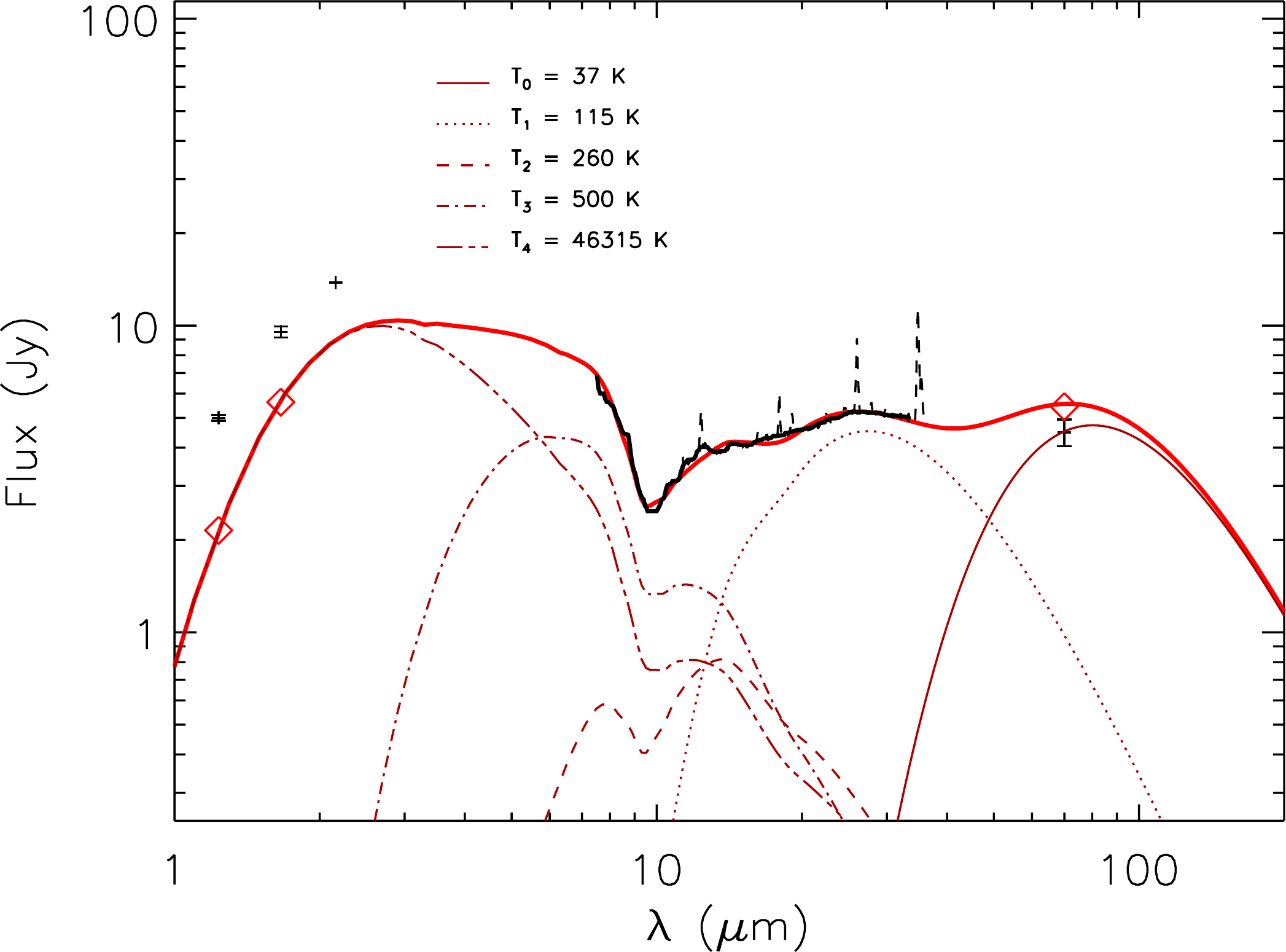}}
\hfill
\subfloat[4384]{\includegraphics[angle=90, width=0.495\linewidth]{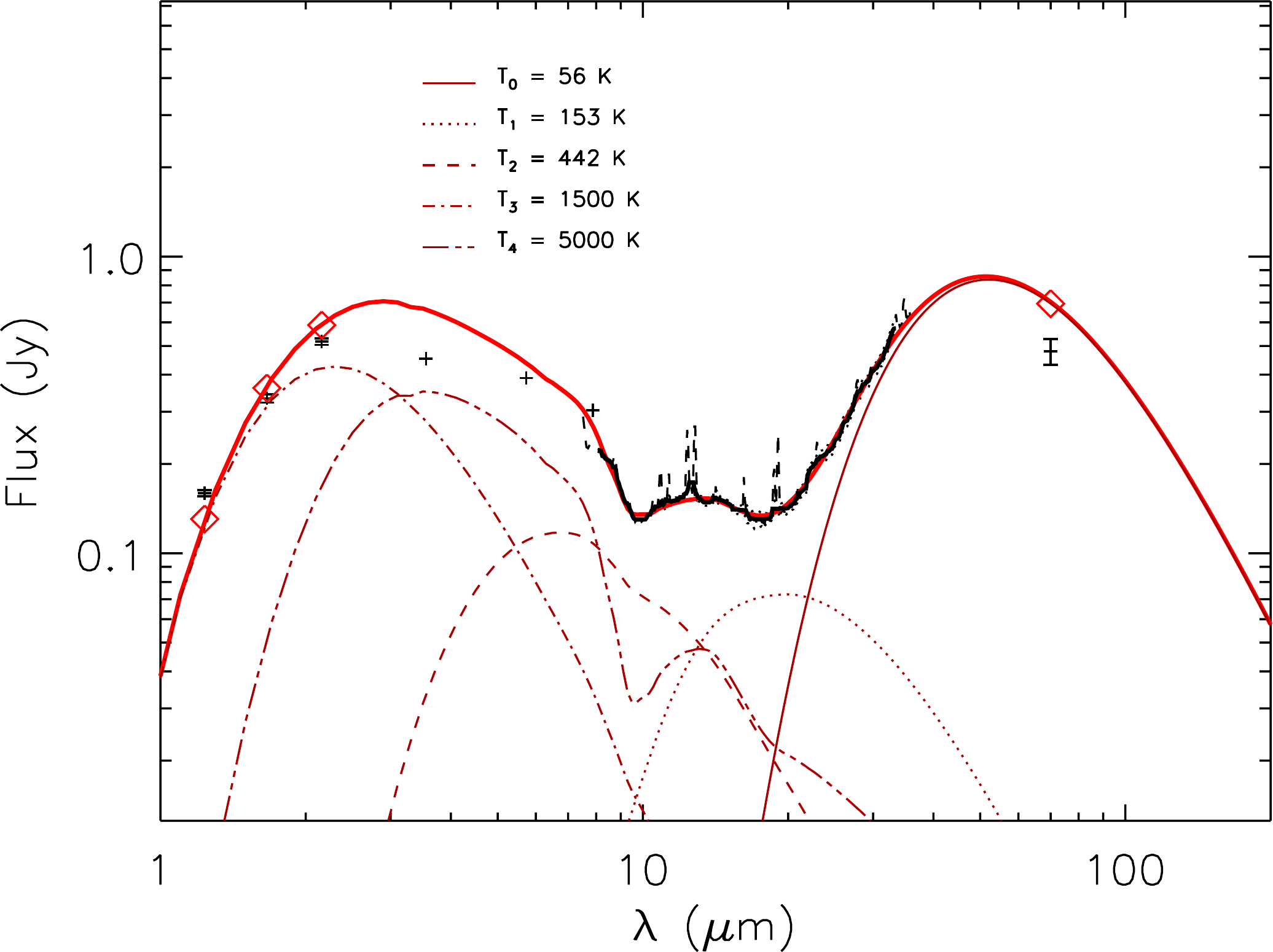}}
\caption{Best fit to the SED of the central sources in the dust-rich MBs. The black dash-line is the IRS spectrum, while the solid black line is the filtered version used for the fit, and the dotted black lines are 1$\sigma$ uncertainties. The thick solid red line is the fit, with each component using a different linestyle. Black crosses are the 2MASS and IRAC measurements, with error bars shown for the data points that we used (see text for details) while red diamonds are the results from the fits for those data points.}
\label{fig:fit_cs_b}
\end{center}
\end{figure*}

\begin{figure*}
\begin{center}
\subfloat[3955]{\includegraphics[angle=90, width=0.495\linewidth]{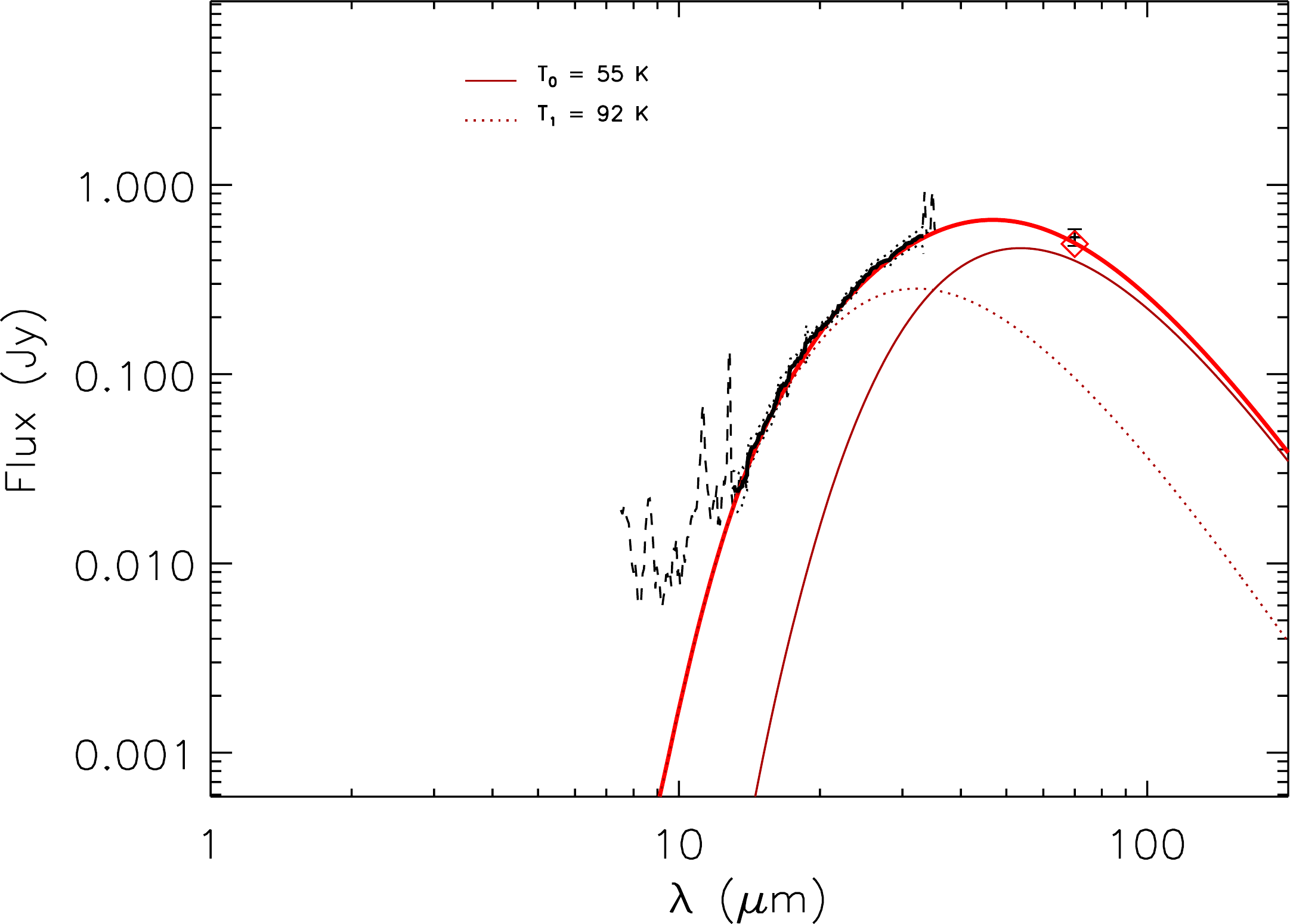}}
\hfill
\subfloat[4121]{\includegraphics[angle=90, width=0.495\linewidth]{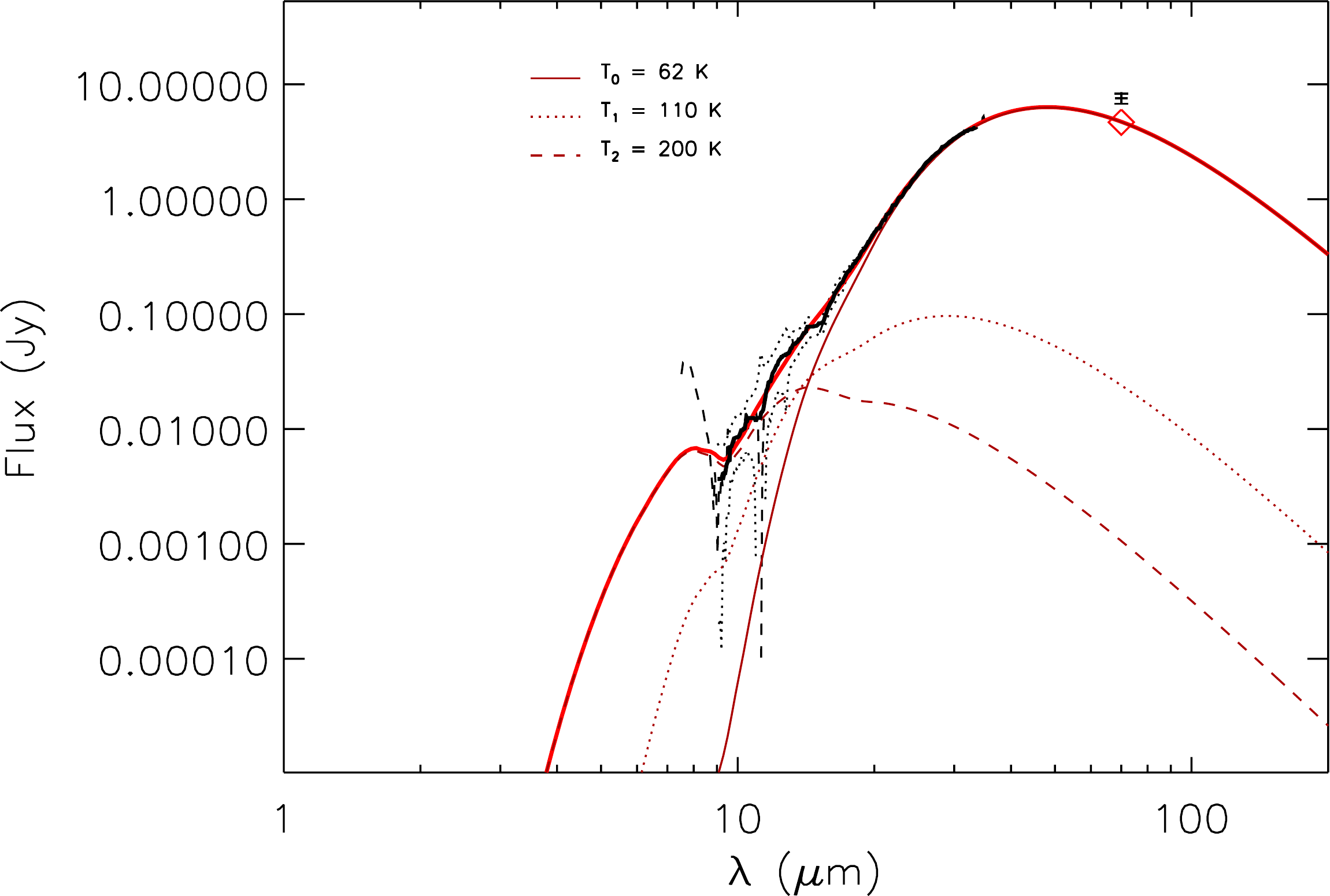}}\\
\subfloat[4124]{\includegraphics[angle=90, width=0.495\linewidth]{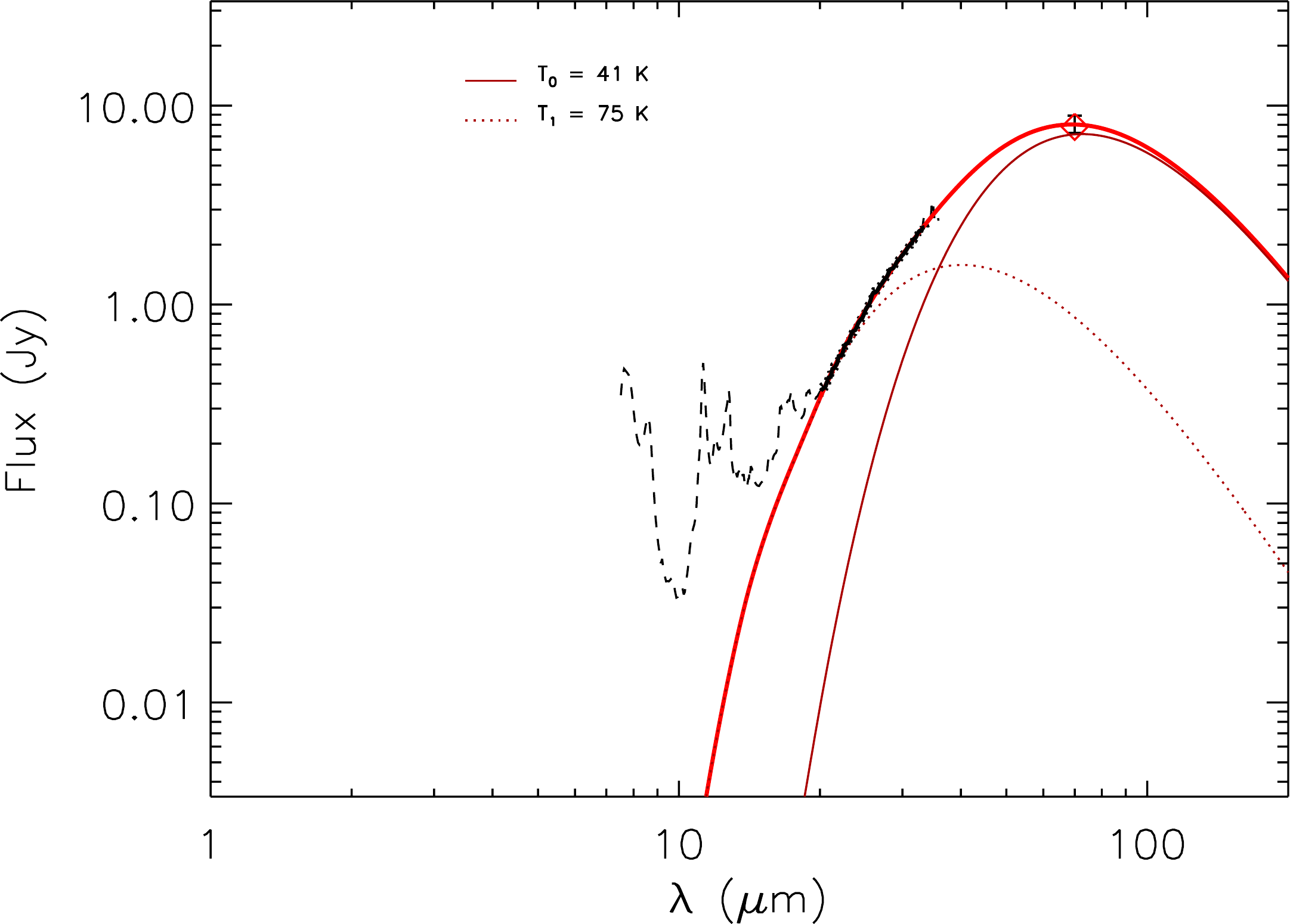}}
\hfill
\subfloat[4384]{\includegraphics[angle=90, width=0.495\linewidth]{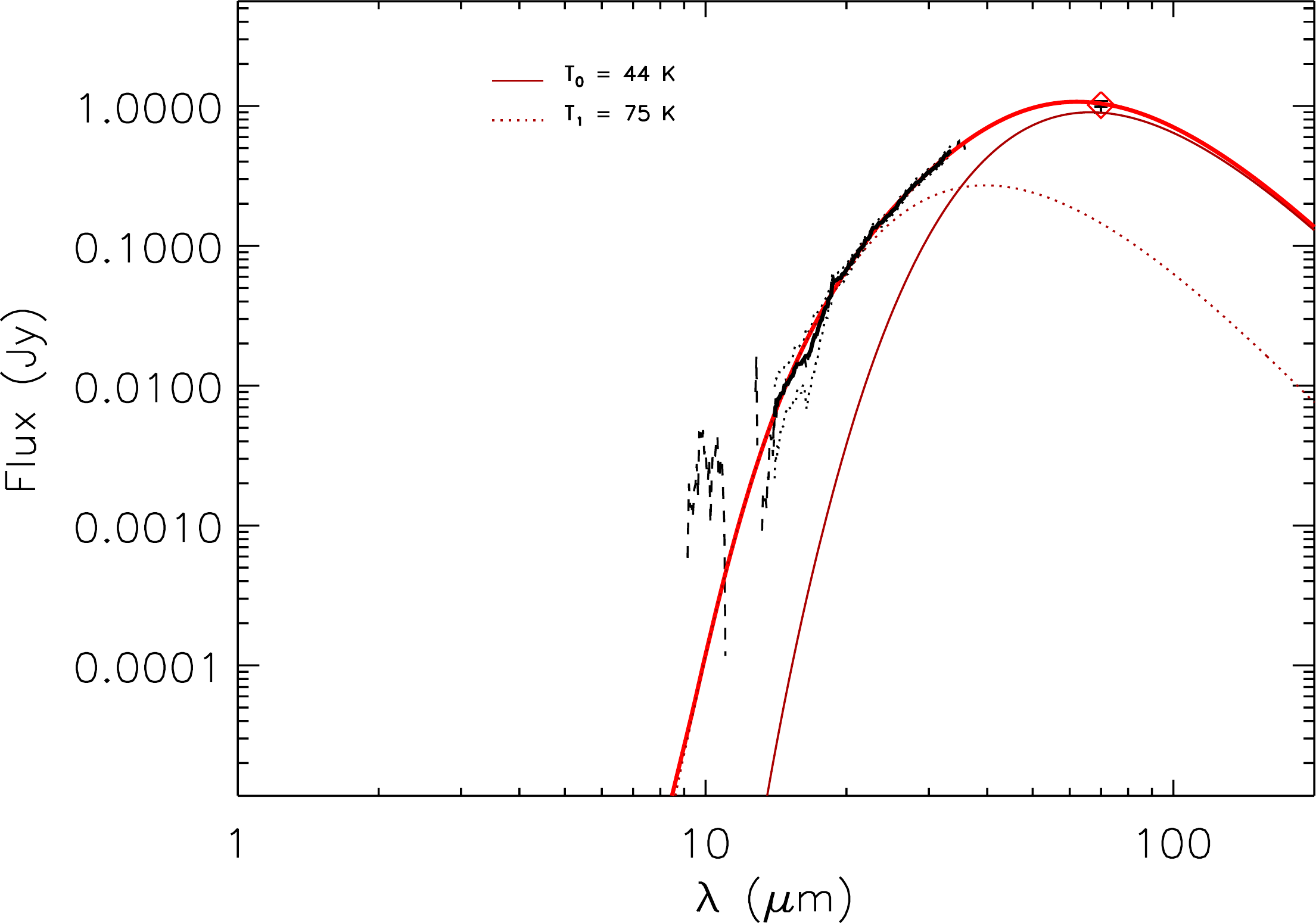}}
\caption{Same as Figure \ref{fig:fit_cs_b} for the SED of the shells of the dust-rich MBs.}
\label{fig:fit_shell}
\end{center}
\end{figure*}

\subsubsection{Method}

We fit the SEDs using the MPFIT package for IDL \citep{Markwardt2009}. For each SED we first smooth the IRS spectrum using a median filter, so that the gas lines are {almost entirely} removed from the spectrum. For the central sources, we use their IRS spectra as shown in Figure \ref{spectra_shell} (i.e. before subtraction of the outer shell spectrum {but after subtraction of the background}) and compare them with the {broadband} 2MASS and {\it Spitzer}/IRAC fluxes at shorter wavelengths. The agreement between the IRS spectra and the broadband measurements is usually good, although not perfect. For instance, the IRAC~8.0\mic\ flux of the central source in MB4121 is significantly larger than that expected from the shape of the IRS spectrum (see Figure \ref{fig:fit_cs_b}). Several interpretations are possible: {(1) some central sources have IRAC fluxes near or above the saturation limits of the instrument (from 0.2~Jy at 3.6 and 4.5\mic, to 1.4~Jy at 5.8\mic\footnote{\url{http://irsa.ipac.caltech.edu/data/SPITZER/docs/irac/irac instrumenthandbook/11/}}), (2) the measurements could be affected by the intense and inhomogeneous surrounding background due to the nebulae, and (3) the sources fluxes could actually have changed in the few years that separate the IRS observations from those done with IRAC and 2MASS, especially in the case of the two LBV candidates, at the center of MB4121 and MB4124. We also add to the SEDs the {\it Herschel}/PACS~70\mic\ measurements, derived from the publicly available {\it ``level 2.5 corrected madmap''} images from the Hi-GAL survey of the Galactic plane \citep{Molinari2010}. We derive the PACS~70\mic\ fluxes the same way we did it for the MIPS~24\mic\ fluxes used to calibrate the IRS spectra (see Section \ref{sec:match}).}

We therefore fit the SEDs of the central sources with and without IRAC and 2MASS points and take the changes in the results into account when deriving uncertainties. {However, when including the near-IR broadband photometric points, we only use those that are listed in the published catalogs with a perfect quality flag. We always use the PACS~70\mic\ data point.} For the outer shell SEDs, we use the IRS spectra {with the PACS~70\mic\ fluxes and remove spectral elements where residual features (broad PAH features, gas lines), or other glitches are still significant, even after the median filtering.}

We fit the SEDs with {a distribution of blackbody and modified blackbody} emission components. By {modified blackbody (MBB)}, we mean here a $\nu^\beta\times B_\nu(T)$ function, where $B_\nu(T)$ is the Planck blackbody (BB) function {at the temperature $T$, and $\beta$ is the emissivity spectral index. We use one BB component to represent the stellar contribution to the central sources' SEDs, and four MBB with $\beta=1.91$, as found by \citet{Bianchi2013} when using {\sc Dustem}, the dust model of \citet{Compiegne2011}, to account for the distribution of dust emission components within the nebulae.} We put constraints on the temperatures that the five components can have (see Table \ref{tab:SEDconstraints}). We include interstellar extinction along the line of sight using the {diffuse ISM} extinction curve from {\sc Dustem}. {We assume that the outer shells may contribute to an increase of the extinction towards the central sources. To represent this, the fit may use additional interstellar extinction for the stellar BB emission component. We discuss this assumption later in the paper.}

We start with the fits to the central sources' SEDs to estimate the {interstellar} extinction. Towards the central sources in MB4121, MB4124, and MB4384, the IRS spectra clearly show the silicate absorption features at 10 and 20\mic. Towards the central source in MB3955, there seems to be a feature at about 13\mic\ that is not related to silicates. It may instead be due to a gap in the temperature distribution of the dust, between the cold grains that dominate at longer wavelengths and the hot grains {or the stellar emission} that are contributing the most to the shorter wavelengths. Towards the outer shells, the emission in the wavelength range of the silicate absorption features is significantly lower than towards the central sources. {Therefore,} the strength of the silicate absorption features and hence the amount of interstellar extinction along the lines of sight, given as $\tau_{9.7}$, the optical depth at 9.7\mic, can only be inferred from the IRS spectra towards the central sources.

{We then fit the outer shells' spectra, taking advantage of MPFIT's capabilities to give upper and lower limits to the range of values that $\tau_{9.7}$ can explore. To define that range, we use the value inferred from the fit to the SEDs towards the central sources and the associated uncertainty for $\tau_{9.7}$. These ranges are indicated in Table \ref{tab:dustrichres} for each MB. Because the extinction towards the central source in MB3955 is not well constrained, we authorize $\tau_{9.7}$ to explore the range from 0 to 2. The fits to the outer shells' SEDs only use the two or three coldest MBB components.}

\begin{table}[t]
  \centering
  \caption{Constraints on the fit}
  \begin{tabular}{l c c}
    \hline
    \hline
    Parameter & Initial value & Range allowed \\
    \hline
    $\tau_{9.7}$ & 1 & 0-10 for central sources, \\
    & & see Table \ref{tab:dustrichres2} for shells \\
    T1 & 50 & 25-75 ($\beta = 1.91$) \\
    T2 & 125 & 75-200 ($\beta = 1.91$)  \\
    T3 & 350 & 200-500 ($\beta = 1.91$)  \\
    T4 & 1000 & 500-1500 ($\beta = 1.91$)  \\
    T5 & 7500 & 5000-50000 ($\beta = 0$)  \\
    & 15000 & 10000-20000 ($\beta = 0$) for MB3955 \\
    \hline
  \end{tabular}
  \label{tab:SEDconstraints}
  \tablecomments{{Since the spectral type of the star at the center of MB3955 is better known (see section \ref{sec:dust_rich_cs}), we use tighter constrains and a different initial value for T5 in that case. }}
\end{table}

\begin{table}[t]
  \centering
  \caption{Results from the fits to the central sources SEDs}
  \begin{tabular}{l c c c c c c c}
    \hline
    \hline
    Target                 & $\tau_{9.7}$    & Suggested        & Temperature \\
    &                                       & distance         &             \\
    &                                       & (kpc)            &(K)          \\
    \hline
    MB3955   & 0.0           & 7.5              & T1=32$\pm$6~K \\
    &                        &                  & T2=79$\pm$1~K \\
    &                        &                  & T3=380$\pm$50~K \\
    &                        &                  & T4=1290$\pm$60~ \\
    &                        &                  & T5=11900$\pm$300~K \\
    \hline
    MB4121   & 1.55$\pm$0.15 & 4.0              & T1=57$\pm$1~K \\
    &                        &                  & T2=98$\pm$5~K \\
    &                        &                  & no T3 required \\
    &                        &                  & T4=1500$\pm$100~K \\
    &                        &                  & T5=7000$\pm$800~K \\
    \hline
    MB4124   & 0.94$\pm$0.04 & 1.9              & T1=37$\pm$1~K \\
    &                        &                  & T2=116$\pm$2~K \\
    &                        &                  & T3=260$\pm$20~K \\
    &                        &                  & T4=500$\pm$50~K \\
    &                        &                  & T5=46000$\pm$2000~K \\
    \hline
    MB4384   & 0.45$\pm$0.25 & 14.0             & T1=57$\pm$1~K \\
    &                        &                  & T2=151$\pm$10~K \\
    &                        &                  & T3=440$\pm$60~K \\
    &                        &                  & T4=1500$\pm$150~K \\
    &                        &                  & T5=5000$\pm$500~K \\
    \hline
  \end{tabular}
  \label{tab:dustrichres}
    \tablecomments{The values found for $\tau_{9.7}$ towards the central sources define the ranges available for the fits to the shells.}
\end{table}

\begin{table*}[t]
  \centering
  \caption{Results from the fits to the outer shells SEDs}
  \begin{tabular}{l c c c c c c c}
    \hline
    \hline
    Target      & Integrated    & $\tau_{9.7}$   & Suggested        & Temperature     & Dust mass               & Dust mass \\
    &           flux            &               & distance         &                 & at suggested distance   & at 1~kpc \\
    &           (Jy)            &               & (kpc)            &(K)              & ($\times10^{-6}M_{\odot}$) & ($\times10^{-6}M_{\odot}$) \\
    \hline
    MB3955      & 0.95$\pm$0.05 & 0.2$\pm$0.9   & 7.5              & T1=55$\pm$10~K  & $760\pm310$             & $14\pm6$ \\
    &                           &               &                  & T2=93$\pm$44~K  & $40\pm110$              & $0.7\pm2$ \\
    \hline
    MB4121      & 7.3$\pm$0.1   & 1.4           & 4.0              & T1=63$\pm$1~K   & $1300\pm85$             & $83\pm5$ \\
    &                           &               &                  & T2=100$\pm$18~K & $4\pm7$                 & $0.2\pm0.4$ \\
    \hline
    MB4124      & 17$\pm$3      & 0.98          & 1.9              & T1=41$\pm$2~K   & $9800\pm3200$           & $2700\pm900$ \\
    &                           &               &                  & T2=75$\pm$6~K   & $120\pm90$              & $34\pm24$ \\
    \hline
    MB4384      & 1.60$\pm$0.05 & 0.2           & 14.0             & T1=44$\pm$2~K   & $21000\pm4900$          & $110\pm25$ \\
    &                           &               &                  & T2=75$\pm$1~K   & $500\pm90$              & $2.6\pm0.5$ \\
    \hline
  \end{tabular}
  \label{tab:dustrichres2}
    \tablecomments{The values found for $\tau_{9.7}$ towards the central sources define the ranges available for the fits to the shells. Dust masses are for the entire MBs, {both at the suggested distance (see text for details) and at a fixed distance of 1~kpc}.}
\end{table*}

\subsubsection{Results}

The fits to the SEDs are shown in Figure \ref{fig:fit_cs_b} for the central sources and in Figure \ref{fig:fit_shell} for the shells.  The best parameters of the fits are indicated in Table \ref{tab:dustrichres} and \ref{tab:dustrichres2}. The dust temperatures should be seen as a representation of the dust temperatures distributions in the MBs. We expect a range of temperatures both within the outer shells and close to the central sources, rather than a few individual temperatures. The shapes of the IRS spectra usually do not show the peak of the ``cold'' dust emission, which seems to lie somewhere around 50\mic. {The added far-IR PACS~70\mic\ measurements}, combined with the IRS data, seem sufficient to constrain the {properties} of the ``cold'' dust components.

{For the central sources, the fits are usually good from near- to far-IR. In particular, the shape of the IRS spectra is very well reproduced in most cases, which implies that the extinction is well constrained, within 10\% (see Table \ref{tab:dustrichres}). Towards MB3955, there seems to be little extinction and the fit cannot properly reproduce the spectral shape around 13\mic\ (see Figure \ref{fig:fit_cs_b}(a)). Towards MB4384, the extinction derived from the fit without the near-IR data points is almost three times larger than that from the fit with these points. In the near-IR range, the contribution of the stellar BB component to the near-IR emission is also uncertain because of the few datapoints available and the possible degeneracy with the amount of extinction applied to this emission component and the contribution of the hot dust emission component. We therefore do not discuss further the properties of the stellar BB components derived from the fits to the central sources' SED. The fits to the mid- to far-IR wavelengths range, where the outer shells dominate the emission, are more trustworthy. For instance, while the extinction towards the central source in MB4384 is poorly constrained, the dust components that contribute to the emission longward of $\sim$15\mic\ are fairly unsensitive to this uncertainty. The central source in MB4124 is the only one for which the fit to the long wavelengths is debatable, because its SED is rather flat from 15 to 70\mic, and therefore the peak of the cold dust emission cannot be determined. We use the results of the fits to the central sources to derive distances estimates in section \ref{sec:ext_n_dist}.}

{For the shells, the fits to their IRS spectra and PACS~70\mic\ fluxes are very good (reduced $\chi^2<1$). Two MBB components with temperatures between 44 and 100~K are enough to fit the SEDs. We use the results of the fits to the outer shells to derive dust masses in section \ref{sec:dust_mass}.}

\subsubsection{Extinction and distances}
\label{sec:ext_n_dist}

{The conversion $A_V = 12.1\times\tau_{9.7}$ is derived from the extinction curve of the diffuse ISM in {\sc Dustem} \citep{Compiegne2011}.} We thus infer an $A_V$ of about 17, 12, and between 3 and 8~mag towards MB4121, MB4124, and MB4384 respectively. The best-fit of the SED towards the central source of MB3955, is not satisfactory, and leads to no visual extinction. {We correct the gas lines fluxes for the extinction (see Table~\ref{table_flux_shells_3}).} Hereafter, we use the apparent magnitudes, spectral identifications, and extinction towards the central sources to derive {estimates for} their distances.

\paragraph{MB3955}
From the {\it B} and {\it V} apparent magnitudes of CD-61 3738, and the intrinsic colors of B5 supergiants, we infer a visual extinction of $\sim$2~mag and a distance of $\sim$7.5~kpc. {The fit to the SED is in agreement with a fairly low amount of extinction.} At that distance, MB3955 would be within the Perseus arm of the Galaxy and about 0.8~pc in radius.

\paragraph{MB4121}
A dark cloud, apparent in the mid-IR images, could be located in front of MB4121 and contribute significantly to the extinction along the line of sight. \citet{Flagey2011} observed MB4121 with the high resolution module of IRS from 10 to 37\mic\ range, while the low resolution observations presented in this paper reach 7.4\mic\ and thus better cover the silicate absorption features. \citet{Flagey2011} inferred $\tau_{9.7}=2$ from a fit to the IRS data, {using the extinction curves from \citet{Chiar2006}. This is significantly more than the value we derive in this paper (1.4$\pm$0.1)}. They also derived {an upper-limit} $A_V=27$~mag from the CO$_2$ absorption feature at 15\mic\ {and a lower-limit $A_V=6$~mag from the Spitzer images. The uncertainty on the interstellar extinction remains large.}

The star at the center of MB4121 is an LBV candidate detected in near-IR bands only ($J=11.7$). Assuming a {normal B supergiant} spectral type, {for which the absolute magnitude is $M_J\sim-6$ \citep[][]{Cox2000}, we infer from the extinction derived from the fit that MB4121 would be at about 4~kpc. At such distance, the derived average density of the ISM along the line of sight is 3~cm$^{-3}$, a factor of a few larger than what we expect in diffuse ISM \citep[$\sim$1~cm$^{-3}$, e.g., ][]{Whittet2003}, though it is dependent on the distance to the Galactic center, with larger densities found at smaller distances, and on the structure of the ISM along the line of sight (e.g. presence of a dark cloud). At 4~kpc from the Sun,} MB4121 would be located within the Scutum-Centaurus arm of the Galaxy {and about 0.5~pc in radius}. {We note however that} using the {\it J} magnitude rather than the {\it V} magnitude means that the estimates are contaminated by the hot dust that significantly contributes to the emission in the near-IR (see Figure \ref{fig:fit_cs_b}). {Additionally, the intrinsic colors of the central star in MB4121 could be significantly different from those of a normal B supergiant, as LBVs are known to undergo instabilities that make their temperature vary between $\sim$8000~K and that of B stars \citep[e.g.,][]{Smith2004}.}

\paragraph{MB4124} 
{Assuming that Hen~2-179, the Be/B[e]/LBV candidate at the center of MB4124, is a normal B2 supergiant, the comparison between its intrinsic and apparent magnitudes ($m_R=12.92$, $m_V=15.66$, $m_B=17.12$) leads to an extinction between 5 and 11~mag, and distances of 22 and 1.9~kpc, depending on the colors we use. The extinction derived from the $R$ and $V$ colors is in agreement with that derived from the fit to the IRS data, and the inferred distance leads to a more likely average interstellar density of 4~cm$^{-3}$. As for MB4121, the density is a factor of a few larger than what we expect in diffuse ISM, but as for MB4121, the galactocentric distance and the structure along the line  of sight may in part explain this discrepancy. Additionally, like the star at the center of MB4121, Hen~2-179 could have colors significantly different from those of a normal B supergiant. At a distance of 1.9~kpc, MB4124 would be about 0.7~pc in radius and located somewhere between the Carina-Sagittarius and the Scutum-Centaurus arms of the Galaxy.}

\paragraph{MB4384}
The star at the center of MB4384 is an Oe/WN star detected in the optical ($m_R=14.85$, $m_V=17.04$, $m_B=17.37$). {Assuming a spectral type WN9 \citep[$M_V\sim-6.7$,][]{vanderHucht2001}, we infer a distance of about 140~kpc using the extinction derived from the fit to the SED with the near-IR data points, and about 14~kpc otherwise. The former is less likely than the latter. At 14~kpc the average interstellar density along the line of sight would be about 0.3~cm$^{-3}$. This is significantly less than towards the other ``dust-rich'' MBs. However MB4384 is at 60\degree\ of longitude and close to 1\degree\ below the plane, where the amount of interstellar matter is significantly lower.} At {14~kpc}, MB4384 would have a radius of about 2~pc and would be located on the far side of the Carina-Sagittarius Arm.\\

{There are large uncertainties in the distances derived from the extinction. These are mainly due to uncertain spectral types and hence intrinsic colors of the stars (MB4121, MB4124, MB4384), although uncertain values of $\tau_{9.7}$ deduced from the fits also play a part (MB3955, MB4384). We indicate the suggested distances in Table \ref{tab:dustrichres} and \ref{tab:dustrichres2}. However, those estimates lead to average densities along the lines of sight a factor of a few higher than what is expected, in the case of MB4121 and MB4124. Therefore, in the following, when we derive dust masses and other properties of the nebulae, we assume the four MBs are at a distance of 1~kpc and indicate how the inferred values scale with distance.}

\subsubsection{Dust temperatures and masses}
\label{sec:dust_mass}

The temperatures of the dust components that are required to fit the outer shell SEDs in the four MBs roughly range from 40 to 100~K, (see Table~\ref{tab:dustrichres2}). Weighting the temperatures of the components by their mass (see details below), the {``effective''} dust temperatures are 57, 63, 42, and 44~K in MB3955, MB4121, MB4124, and MB4384, respectively. From the IRS high resolution observations of MB4121, \citet{Flagey2011} infer a dust temperature of 74~K, in fair agreement with the estimate in this paper. To derive dust masses associated with each MBB in a given shell, we use the following relation:

\begin{equation}
  f_\nu (\lambda) = M_{dust} / D^2 \times \kappa_{abs}(\lambda) \times B_\nu(T_{dust}, \lambda)
\end{equation}

\noindent{}where $f_\nu$ is the observed flux, $D$ is the distance to the MB, $\kappa_{abs}$ is the dust grain absorption cross section per unit mass, and $B_\nu(T_{dust})$ the Planck function at the temperature of the dust. A power law is usually assumed for the dust grain absorption cross section:

\begin{equation}
  \kappa_{abs}(\lambda) = \kappa_{abs}(\lambda_0) \times (\lambda/\lambda_0)^{-\beta}
\end{equation}

The mass of dust is then given by:

\begin{equation}
  M_{dust} = \frac{W \times D^2}{\kappa_{abs}(\lambda_0) \times \lambda_0^{\beta}}
\end{equation}

\noindent{}where
\begin{equation}
  W = \frac{f_\nu}{\lambda^{-\beta}\times B_\nu(T_{dust},\lambda)}
\end{equation}

\noindent{}is the amplitude of the {MBB} component, which is the actual parameter given by the fits. {For the dust grains properties, we use $\lambda_0=250\mic$, $\beta=1.91$, and $\kappa_{abs}(250\mic)=5.1~\rm{cm^2~g^{-1}}$ \citep{Compiegne2011, Bianchi2013}.}

{We scale the dust masses to the integrated fluxes over the whole nebulae in the PACS~70\mic\ images since the shells dominate the emission in that wavelengths range. For MB3955 however, we use the integrated MIPS~24\mic\ flux because its central source is not detected at 24\mic\ and, relative to the surrounding ISM, the shell is significantly brighter than at 70\mic. The fluxes are reported in Table \ref{tab:dustrichres2}. The inferred dust masses are given for both ``cold'' dust components.} We take into account the uncertainties in $W$ derived from the fits. {In the following, we assume a gas-to-dust mass ratio of 100.}

{At 1~kpc, the total dust masses range from $1.5\times10^{-5}~M_\odot$ in MB3955 to $2.7\times10^{-3}~M_\odot$ in MB4124. The masses scale with $D^2$. In the case of MB4384, a distance of 14~kpc thus leads to a dust mass of $2\times10^{-2}~M_\odot$ which seems overestimated, as this would lead to total mass of $2~M_\odot$ in the outer shell.} The ``cold'' dust components are at least an order of magnitude more massive than the ``warm'' dust components. \citet{Flagey2011} estimated the mass of dust in the shell of MB4121. Taking into account the integrated flux of the shell, and assuming a dust size distribution dominated by large grains (a few 10 to a few 100~nm), the dust mass they found corresponds to $3.2\times10^{-5}~M_\odot$ at 1~kpc{, which is a less than a factor of 3 smaller than that derived in the present paper using different observations and methods.}

Given the masses of dust in the shells and their physical sizes, we estimate the amounts of local extinction they produce and compare them to the interstellar extinction along the whole lines of sight. We assume that the dust is isotropically distributed. The dust column densities thus are {1.5, 14, 52, and $12\times 10^{-8}~\rm{g~cm^{-2}}$} in the shells of MB3955, MB4121, MB4124, and MB4384, respectively. {Therefore, hydrogen column densities $N_H$ of 1.0, 8.9, 32, and $7.3\times10^{18}\rm{cm^{-2}}$ are derived. The column densities are independent of the distances. From the diffuse ISM extinction curve of {\sc Dustem}, we infer that $\tau_{9.7}=0.004$ (or $A_v=0.05$~mag) for $N_H=10^{20}\rm{cm^{-2}}$. This means that the contributions from the dust in the outer shells to the total extinction are at most 0.02\%.}

{At 1~kpc, the total masses of the shells are between $1.4\times10^{-3}~M_\odot$ in MB3955 to $2.7\times10^{-1}~M_\odot$ in MB4124. Assuming average expansion velocities of 100~km/s, ages of the shells, which scale with the distance, range from 1200 years for MB4121 to 3600 years for  MB4124. The dust production rates, which scale with the distance, are between $9.5\times10^{-9}~\rm{M_\odot/yr}$ for MB3955 and $7.7\times10^{-7}~\rm{M_\odot/yr}$ for MB4124, while total mass loss rates are 100 times larger. In the case of MB4384, at a distance of 14~kpc, the total mass loss rate would be 1.4$\times10^{-3}~\rm{M_\odot/yr}$, which again highlights the overestimated distance to MB4384.}

The properties we derive for the dust-rich MBs, in terms of dust masses, total masses, and mass loss rates, are highly uncertain because they depend on the distance to the MBs, and their expansion velocities. Uncertainties of an order of magnitude are therefore expected on these parameters. However, the estimates we derive from the fits to the IRS spectra are in the range of values usually found for WR and LBV stars \citep[$10^{-6}-10^{-4}~\rm{M_\odot/yr}$,][]{Crowther2007, Humphreys1994}.

\section{Morphological identification and extrapolation to the whole catalog}
\label{sec:disc_extrap}

{\citet{Mizuno2010} suggested a possible correlation between the morphologies and the natures of the MBs. Their conclusion, based on the natures of 64 MBs identified in SIMBAD as more than IRAS sources, was that most of the objects without central sources at \unit{24}{\micro\meter} MBs were probably PNe, whereas MBs with central sources were mostly emission line stars. Our {\it Spitzer}/IRS observations of 14 MBs \citep[including the four MBs from][]{Flagey2011} supports this suggestion}. The five ``ring-like'' MBs with central sources in the mid-IR have all been identified or suggested as massive stars, and ``disk-like'' MBs without central sources in the mid-IR have all been suggested as highly-excited PNe. {This paper also suggests that this correlation encompasses the origin of the mid-IR emission: the ``disk-like'' PNe are surrounded by dust-poor highly-excited nebulae while the ``ring-like'' massive stars have dust-rich shells.}

\begin{figure}
\begin{center}
\includegraphics[width=\linewidth]{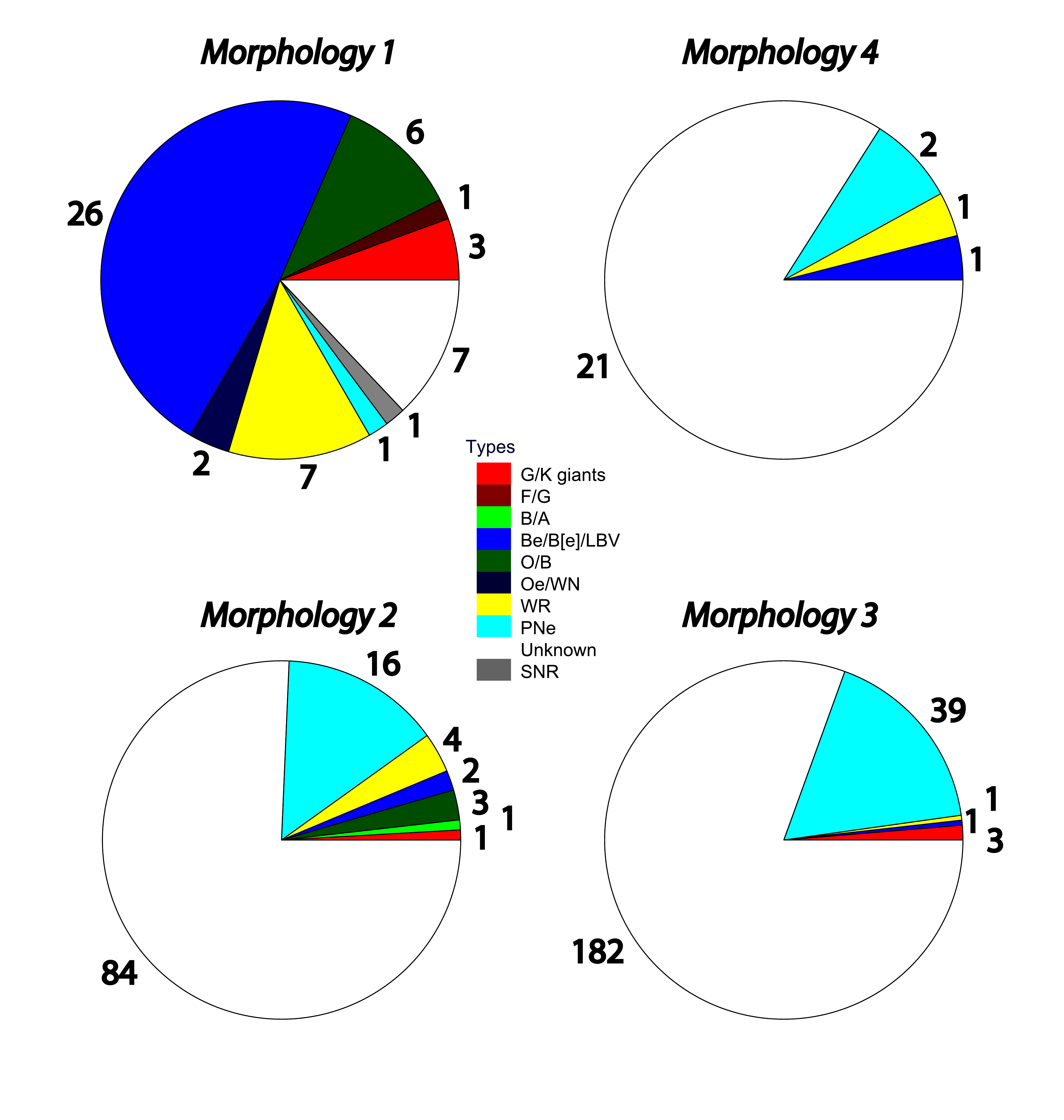}
\caption{Type distribution for the MBs with (morphology 1) and without central sources detected at 24\mic\ (morphologies 2, 3, and 4). Each color in the pie-charts is associated with a type, as indicated by the legend. The numbers next to the pie-charts indicate the number of MBs for each morphology and nature.}
\label{fig:piemorph}
\end{center}
\end{figure}

\begin{figure}
\begin{center}
\includegraphics[width=\linewidth]{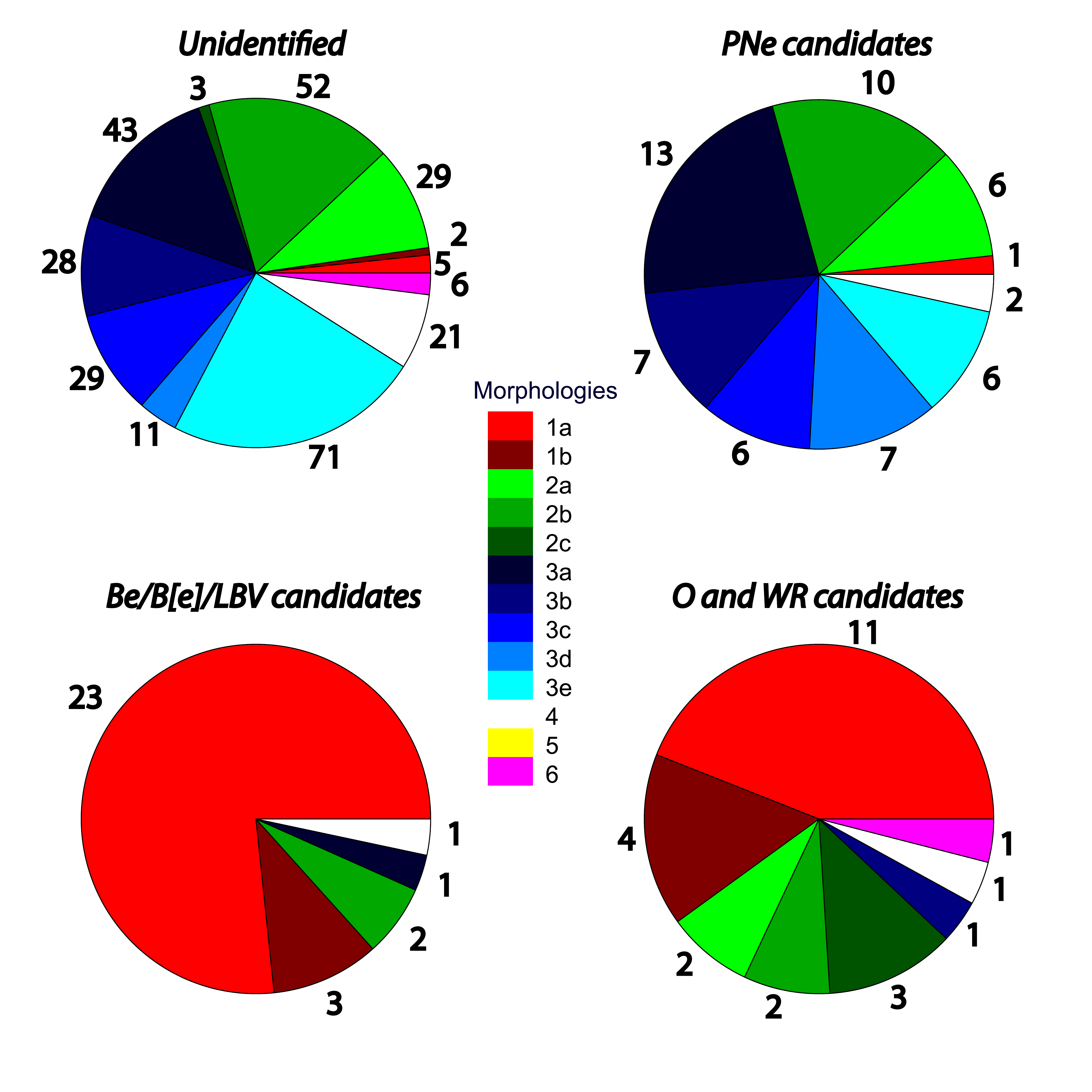}
\caption{Morphology distribution for the unidentified MBs, the PNe candidates, the Be/B[e]/LBV candidates, and the WR candidates. Each color in the pie-charts is associated with a morphology, as indicated by the legend. The numbers next to the pie-charts indicate the number of MBs for each morphology and nature.}
\label{fig:piestype}
\end{center}
\end{figure}

{To understand to what extent the correlation between nature and morphologies of the MBs holds true}, we compile here the recent findings on the nature of the MBs, as published in \citet{Gvaramadze2010}, \citet{Mauerhan2011}, \citet{Wachter2010}, \citet{Wachter2011}, and \citet{Flagey2014a}. {These authors have doubled the number of identified MBs since the publication of the \citet{Mizuno2010} catalog.} The details about the morphologies classification can be found in \citet{Mizuno2010} and are summarized in Table \ref{tab:morph_def}. Table \ref{tab:morph} summarizes the identifications of the MBs sorted in 10 distinct types (e.g., G/K supergiants and giants, O/B stars, WR stars). Figures~\ref{fig:piemorph} and \ref{fig:piestype} show several pie-charts for the distributions of morphologies and natures of the MBs.

Most of the known MBs in morphology 1 are Be/B[e]/LBV candidates (55\%), while 76\% of the known MBs in morphologies 2 to 4 are PNe candidates. Reciprocally, most of the PNe candidates (67\%) are in morphology 3, while most of the Be/B[e]/LBV candidates (87\%) are among the morphology 1. On the other hand, only 3\% of the Be/B[e]/LBV candidates are in morphology 3, and only 2\% of the PNe candidates are in morphology 1. Morphologies 1 and 3 thus seem rather exclusive of each other in terms of the natures of the MBs they include. The correlation suggested in \citet{Mizuno2010} thus remains true on the larger sample of 128 identified MBs. 

\begin{table}[t]
  \centering
  \caption{Morphologies defined in \citet{Mizuno2010}.}
  \begin{tabular}{l l l}
    \hline
    \hline
    \multicolumn{3}{c}{Morphology} \\
    \hline
    1 & \multicolumn{2}{l}{rings with central sources detected at 24\mic} \\
    & 1a & regular \\
    & 1b & irregular \\
    \hline
    2 & \multicolumn{2}{l}{rings without central sources detected at 24\mic} \\
    & 2a & complete \\
    & 2b & irregular  \\
    & 2b & bilaterally symmetric \\
    \hline
    3 & \multicolumn{2}{l}{disks} \\
    & 3a & flat \\
    & 3b & peaked \\
    & 3c & bilaterally symmetric \\
    & 3d & oblong \\
    & 3e & irregular \\
    \hline
    4 & \multicolumn{2}{l}{two-lobed} \\
    \hline
    5 & \multicolumn{2}{l}{filamentary} \\
    \hline
    6 & \multicolumn{2}{l}{miscellaneous} \\
    \hline
  \end{tabular}
  \label{tab:morph_def}
\end{table}

The MBs in morphology 2 and the OB, Oe/WN, or WR candidates are not as clearly defined. In morphology~2, 60\% of the known MBs are PNe candidates and 26\% are OB, Oe/WN, or WR candidates. Reciprocally, there are 28\% of all the known PNe, and 28\% of the OB, Oe/WN, or WR candidates in morphology 2. The predominance of PNe could be interpreted by their overall larger number in the whole sample of 428 MBs, while OB, Oe/WN, and WR candidates are significantly less common. Additionally, the PNe comprises many MBs whose central sources have not yet been detected and/or identified, some of which might be massive WR stars. {This seems consistant with the similarity between the pie-charts for the PNe candidates and for the unidentified MBs (see Figure~\ref{fig:piestype}).}

The OB, Oe/WN, and WR candidates are more scattered than the Be/B[e]/LBV candidates across different morphologies (see Figure~\ref{fig:piestype}). In particular, looking at the balance between morphologies 1a and 1b, it seems more likely that the central sources in Be/B[e]/LBV candidates are detected in the mid-IR, and that their shells have a more regular shape than those of O or WR stars.

Extrapolating from this analysis, we suggest that the 7 unidentified MBs in morphology 1 surely are Be/B[e]/LBV, or, less likely, OB, Oe/WN, or WR candidates, while most of the 182 unidentified MBs in morphology 3 are likely PNe. The 84 unknown MBs in morphology 2 might be mostly PNe, although a fraction of LBV, WR and O stars might also be found.

\begin{table*}
  \begin{center}
    \caption{Distribution of the MBs among the different morphologies and types.\label{tab:morph}}
    \begin{tabular}{ c c | c c c | c c c c | c c c c c c | c c c | c }
      \hline
      \hline
      & & \multicolumn{16}{c|}{{Morphologies$^{a}$}} & \\
      &       & \multicolumn{3}{c|}{{1}} & \multicolumn{4}{c|}{{2}} & \multicolumn{6}{c|}{{3}} & {4} & {5} & {6} & total\\
      &       & a & b & total & a & b & c & total & a & b & c & d & e & total &  &  &  & \\
      \hline
      \multirow{10}{*}{\rotatebox{90}{{Types}}}
      & G/K (super)giants &  3 &    &    3    &    &    &  1 &    1    &  1 &    &  1 &    &  1 &    3    &    &    &    &   7\\
      & F/G               &  1 &    &    1    &    &    &    &         &    &    &    &    &    &         &    &    &    &   1\\
      & B/A               &    &    &         &    &    &  1 &    1    &    &    &    &    &    &         &    &    &  1 &   2\\
      & Be, B[e], LBV     & 23 &  3 &   26    &    &  2 &    &    2    &  1 &    &    &    &    &    1    &  1 &    &    &  30\\
      & O/B               &  5 &  1 &    6    &  1 &  1 &  1 &    3    &    &    &    &    &    &         &    &    &  1 &  10\\
      & Oe/WN             &  2 &    &    2    &    &    &    &         &    &    &    &    &    &         &    &    &    &   2\\
      & WR                &  4 &  3 &    7    &  1 &  1 &  2 &    4    &    &  1 &    &    &    &    1    &  1 &    &    &  13\\
      & Galaxies          &    &    &         &    &    &    &         &    &    &    &    &    &         &    &    &  2 &   2\\
      & PNe               &  1 &    &    1    &  6 & 10 &    &   16    & 13 &  7 &  6 &  7 &  6 &   39    &  2 &    &    &  58\\
      & SNR               &    &  1 &    1    &    &    &    &         &    &    &    &    &    &         &    &  2 &    &   3\\
      \hline
      & Known total       & 39 &  8 &   47    &  8 & 14 &  5 &   27    & 15 &  8 &  7 &  7 &  7 &   44    &  4 &  2 &  4 & 128\\
      & Unknown           &  5 &  2 &    7    & 29 & 52 &  3 &   84    & 43 & 28 & 29 & 11 & 71 &  182    & 21 &    &  6 & 300\\
      \hline
      & Total             & 44 & 10 &   54    & 37 & 66 &  8 &  111    & 58 & 36 & 36 & 18 & 78 &  226    & 25 &  2 & 10 & 428\\
      \tableline
    \end{tabular}
    \tablecomments{$^{(a)}$ Morphologies are from \citet{Mizuno2010}.}
  \end{center}
\end{table*}

\section{Conclusions}
\label{sec:ccl}

In this paper, we present the results of the {\it Spitzer}/IRS observations of 11 MIPSGAL compact ``bubbles'' obtained with the low resolution modules. They complement the 4 MBs observed with the high resolution modules by \citet{Flagey2011} in the quest to identify the nature of the MBs and characterize their mid-IR emission. Before the launch of the {\it James Webb Space Telescope}, these data are the only mid-IR spectroscopic observations available on the extended emission of the MBs. Here are the conclusions of our paper:

\begin{itemize}
\item The seven MBs without central sources in the mid-IR have a highly excited emission lines dominated spectrum, with mainly neon, sulfur, and oxygen lines. In five of these MBs, the [O~\textsc{iv}] line at 25.9\mic\ and, to a lesser extent, the [Ne~\textsc{v}] line at 24.3\mic, account for a significant part of the MIPSGAL~24\mic\ flux, while the dust continuum contributes to less than 60\% of the emission. Based on comparison with published spectra, we suggest that these seven MBs are PNe.

\item The Neon lines ratios are unusually high in those seven MBs: in four of them, the [Ne~\textsc{v}]~24.3\mic\ to [Ne~\textsc{iii}]~15.6\mic\ ratio is greater than 1 and as high as 15. This suggests high excitation states in the nebulae. We found several objects with similar characteristics, all of them some type of PNe with central sources that include [WR] stars, novae and other WD in binary systems.

\item The other four MBs have central sources detected in the mid-IR (8, 12, and/or 24\mic). These stars have been previously suggested as massive stars (B5 supergiant, Oe/WN star, LBV candidates). The mid-IR emission in these MBs is almost entirely (>98\%) due to a warm dust continuum. Some emission lines (e.g., H~\textsc{i}, [Fe~\textsc{ii}]) are detected towards some central sources, but remain very weak in the shell.

\item We characterize the dust temperatures and masses in the shells surrounding the central sources. We find that the dust temperatures in the outer shells range from $\sim40$ to $\sim100$~K, while they are significantly higher close to the inner sources detected at 24\mic. The {\it Spitzer}/IRS spectra  enable us to constrain the extinction along the line of sight and to derive rough distance estimates, given the apparent magnitudes and spectral types of the central stars. We infer dust masses {at an arbitrary distance of 1 kpc} that range between a few $10^{-4}$ and a few $10^{-3}~M_\odot$. Assuming average expansion velocities of 100~km/s, we estimate ages of a few 1000~years for the nebulae and dust production rates of a few $10^{-6}~M_\odot/$yr to a few $10^{-4}~M_\odot/$yr.

\item We summarize the identifications of 128 MBs and discuss the correspondence with their morphologies. We confirm that Be/B[e]/LBV candidates, and to a lesser extent WR candidates, appear predominantly like ``rings'' with central sources in the mid-IR images, while PNe candidates and the remaining unidentified MBs are mostly ``disks''. Extrapolating to the 428 MBs of the \citet{Mizuno2010} catalog, we expect that among the 300 MBs that are still unidentified, only a few might be massive star candidates, while the very large majority will be classified as PNe.
\end{itemize}

\acknowledgments The authors would like to express their gratitude to the anonymous referee for comments that helped improved the paper, to M.~Shara, R.~Fesen, and M.~Pe\~na for valuable discussion, to J.~Hora and A.~Hart for sharing their {\it Spitzer}/IRS observations of [WR], and to E.~Palay and A.~Pradhan for providing a digitized version of their results.

This work is based on observations made with the Spitzer Space Telescope, which is operated by the Jet Propulsion Laboratory, California Institute of Technology under a contract with NASA. Support for this work was provided by NASA through an award issued by JPL/Caltech.

\bibliographystyle{aa}
\bibliography{bubbles}

\end{document}

%% file: tables_flux.tex
\begin{center}
\begin{sidewaystable}
  \begin{scriptsize}
    \caption{Table of the detected line fluxes in the objects that show no central source at \unit{24}{\micro\meter} (in $10^{-16}$\unit{}{\watt\per\meter\squaren{}})}
    \label{table_flux_shells_1}
    \begin{center}
      \begin{tabular}{c c c c c c c c c c c c}
        \hline
        \hline
        Object & [Ar~\textsc{iii}] & H~\textsc{ii} & [S~\textsc{iv}] & [Ne~\textsc{ii}] & [Ne~\textsc{v}] & [Ne~\textsc{iii}] & [S~\textsc{iii}] & [Ne~\textsc{v}] & [O~\textsc{iv}] & [S~\textsc{iii}] \\
        & \unit{9.0}{\micro\meter} & \unit{9.7}{\micro\meter} & \unit{10.5}{\micro\meter} & \unit{12.8}{\micro\meter} & \unit{14.3}{\micro\meter} & \unit{15.6}{\micro\meter} & \unit{18.7}{\micro\meter} & \unit{24.3}{\micro\meter} & \unit{25.9}{\micro\meter} & \unit{33.5}{\micro\meter}\\
        \hline
        MB4376 &$ - $&$ 0.07\pm{}0.01 $&$ 0.10\pm{}0.04 $&$ - $&$ 0.28\pm{}0.03 $&$ 0.25\pm{}0.08 $&$ - $&$ 1.27\pm{}0.06 $&$ 11.6\pm{}0.2 $&$ - $\\
        MB3944 &$ - $&$ 0.12\pm{}0.01 $&$ 0.15\pm{}0.01 $&$ 0.21\pm{}0.08 $&$ 0.24\pm{}0.01 $&$ 0.35\pm{}0.05 $&$ 0.6\pm{}0.3 $&$ 0.59\pm{}0.06 $&$ 15.7\pm{}0.4 $&$ 1.6\pm{}0.4 $\\
        MB4021 &$ - $&$ - $&$ 0.61\pm{}0.06 $&$ - $&$ - $&$ 1.39\pm{}0.02 $&$ 0.3\pm{}0.1 $&$ - $&$ 21.0\pm{}0.2 $&$ 0.6\pm{}0.1 $\\
        MB4017 &$ - $&$ 0.2\pm{}0.1 $&$ 2.8\pm{}0.8 $&$ - $&$ 11.6\pm{}1.3 $&$ 3.2\pm{}0.3 $&$ 4.5\pm{}0.6 $&$ 36\pm{}1 $&$ 162\pm{}3 $&$ 12.8\pm{}0.7 $\\
        MB4066 &$ 0.14\pm{}0.07 $&$ 0.09\pm{}0.03 $&$ 0.2\pm{}0.1 $&$ - $&$ - $&$ 2.9\pm{}0.2 $&$ 0.9\pm{}0.4 $&$ - $&$ 0.8\pm{}0.5 $&$ 3\pm{}1 $\\
        MB4076 &$ - $&$ 0.10\pm{}0.03 $&$ 0.35\pm{}0.06 $&$ 0.6\pm{}0.1 $&$ - $&$ 1.39\pm{}0.07 $&$ 0.5\pm{}0.2 $&$ - $&$ 32.6\pm{}0.2 $&$ - $\\
        MB3575 &$ - $&$ 0.06\pm{}0.02 $&$ - $&$ - $&$ 0.23\pm{}0.02 $&$ 0.07\pm{}0.02 $&$ - $&$ 1.0\pm{}0.1 $&$ 7.54\pm{}0.04 $&$ 1.5\pm{}0.3 $\\
      \end{tabular}
    \end{center}
  \end{scriptsize}
\end{sidewaystable}
\end{center}

\begin{sidewaystable}
  \begin{scriptsize}
    \caption{Table of the detected line fluxes in the objects with detected central sources (in $10^{-16}$\unit{}{\watt\per\meter\squaren{}})}
    \label{table_flux_shells_2}
    \begin{center}
      \begin{tabular}{l c c c c c c c c c c}
        \hline
        \hline
        Object & [Ar~\textsc{iii}] & H$_{\rm 2}$ & [S~\textsc{iv}] & H~\textsc{i} & H~\textsc{i} & H$_{\rm 2}$ & [Ne~\textsc{ii}] & [Ar~\textsc{v}] & He~\textsc{ii} \\
        & \unit{9.0}{\micro\meter} & \unit{9.7}{\micro\meter} & \unit{10.5}{\micro\meter} & \unit{10.8}{\micro\meter} & \unit{11.5}{\micro\meter} & \unit{12.3}{\micro\meter} & \unit{12.8}{\micro\meter} & \unit{13.1}{\micro\meter} & \unit{16.2}{\micro\meter}\\
        \hline
        MB4384 cent. src &$ - $&$ - $&$ 0.29\pm{}0.02 $&$ 1.1\pm{}0.1 $&$ 0.9\pm{}0.2 $&$ 1.70\pm{}0.06 $&$ 1.7\pm{}0.3 $&$ - $&$ 0.49\pm{}0.02 $\\
        MB4384 shell &$ - $&$ 0.03\pm{}0.01 $&$ 0.03\pm{}0.01 $&$ - $&$ 0.04\pm{}0.02 $&$ - $&$ 0.28\pm{}0.05 $&$ - $&$ - $\\
        MB3955 cent. src &$ - $&$ 0.04\pm{}0.02 $&$ - $&$ - $&$ - $&$ - $&$ 0.84\pm{}0.07 $&$ - $&$ - $\\
        MB3955 shell &$ - $&$ 0.05\pm{}0.01 $&$ - $&$ - $&$ - $&$ - $&$ 1.4\pm{}0.1 $&$ - $&$ - $\\
        MB4121 cent. src &$ 1.3\pm{}0.3 $&$ - $&$ - $&$ 0.6\pm{}0.1 $&$ 0.42\pm{}0.02 $&$ 0.9\pm{}0.2 $&$ - $&$ 0.30\pm{}0.03 $&$ - $\\
        MB4121 shell &$ - $&$ - $&$ - $&$ - $&$ 0.06\pm{}0.02 $&$ - $&$ - $&$ 0.11\pm{}0.03 $&$ - $\\
        MB4124 cent. src &$ 2.2\pm{}0.3 $&$ - $&$ - $&$ 2.4\pm{}0.1 $&$ 8.66\pm{}0.07 $&$ 21\pm{}2 $&$ - $&$ 1.6\pm{}0.3 $&$ 4.4\pm{}0.4 $\\
        MB4124 shell &$ - $&$ 0.2\pm{}0.1 $&$ - $&$ - $&$ 0.6\pm{}0.2 $&$ 0.21\pm{}0.05 $&$ 1.3\pm{}0.6 $&$ - $&$ - $\\
        \\
        \hline
        \hline
        Object & H~\textsc{i}/H$_{\rm 2}$ & [Fe~\textsc{ii}] & [S~\textsc{iii}] & H~\textsc{i} & [Fe~\textsc{iii}] & [Fe~\textsc{ii}] & [Fe~\textsc{ii}] & He~\textsc{ii} & H$_{\rm 2}$ & [S~\textsc{iii}] \\
        & \unit{16.9}{\micro\meter} & \unit{17.9}{\micro\meter} & \unit{18.7}{\micro\meter} & \unit{19.0}{\micro\meter} & \unit{22.9}{\micro\meter} & \unit{24.5}{\micro\meter} & \unit{25.9}{\micro\meter} & \unit{27.9}{\micro\meter} & \unit{28.2}{\micro\meter} & \unit{33.5}{\micro\meter}\\
        \hline
        MB4384 cent.src &$ 0.16\pm{}0.04 $&$ - $&$ 0.85\pm{}0.09 $&$ 1.48\pm{}0.08 $&$ 0.90\pm{}0.07 $&$ - $&$ - $&$ 1.10\pm{}0.04 $&$ - $&$ - $\\
        MB4384 shell &$ - $&$ - $&$ 0.18\pm{}0.05 $&$ - $&$ 0.22\pm{}0.04 $&$ - $&$ - $&$ 0.07\pm{}0.02 $&$ 0.19\pm{}0.06 $&$ - $\\
        MB3955 cent.src &$ - $&$ - $&$ - $&$ 0.05\pm{}0.02 $&$ - $&$- $&$ - $&$ - $&$ - $&$ 4\pm{}2 $\\
        MB3955 shell &$ - $&$ - $&$ - $&$ - $&$ - $&$ - $&$ - $&$ 0.3\pm{}0.1 $&$ - $&$ 4\pm{}2 $\\
        MB4121 cent.src &$ 0.2\pm{}0.1 $&$ 4.3\pm{}0.2 $&$ 1.18\pm{}0.09 $&$ - $&$ - $&$ 1.2\pm{}0.2 $&$ 12.8\pm{}0.4 $&$ - $&$ 2.9\pm{}0.3 $&$ - $\\
        MB4121 shell &$ - $&$ - $&$ 0.27\pm{}0.07 $&$ - $&$ - $&$ - $&$ 2.2\pm{}0.2 $&$ 0.4\pm{}0.1 $&$ 2.2\pm{}0.3 $&$ - $\\
        MB4124 cent.src &$ 1.1\pm{}0.3 $&$ 21\pm{}3 $&$ 1.5\pm{}0.3 $&$ 9.8\pm{}0.4 $&$ - $&$ 6.3\pm{}0.4 $&$ 57.1\pm{}0.7 $&$ 6.9\pm{}0.1 $&$ - $&$ - $\\
        MB4124 shell &$ - $&$ - $&$ - $&$ 0.12\pm{}0.04 $&$ - $&$ - $&$ 1.4\pm{}0.6 $&$ - $&$ - $&$ 5\pm{}1 $\\
      \end{tabular}
    \end{center}
  \end{scriptsize}
\end{sidewaystable}

%% file: table_flux_corrected.tex
\begin{sidewaystable}
  \begin{scriptsize}
    \caption{Table of the line fluxes corrected for reddening in the objects with detected central sources (in $10^{-16}$\unit{}{\watt\per\meter\squaren{}})}
    \label{table_flux_shells_3}
    \begin{center}
      \begin{tabular}{l c c c c c c c c c c}
        \hline
        \hline
        Object & [Ar~\textsc{iii}] & H$_{\rm 2}$ & [S~\textsc{iv}] & H~\textsc{i} & H~\textsc{i} & H$_{\rm 2}$ & [Ne~\textsc{ii}] & [Ar~\textsc{v}] & He~\textsc{ii} \\
        & \unit{9.0}{\micro\meter} & \unit{9.7}{\micro\meter} & \unit{10.5}{\micro\meter} & \unit{10.8}{\micro\meter} & \unit{11.5}{\micro\meter} & \unit{12.3}{\micro\meter} & \unit{12.8}{\micro\meter} & \unit{13.1}{\micro\meter} & \unit{16.2}{\micro\meter}\\
        \hline
        MB4384 cent. src &$ - $&$ - $&$ 0.42\pm{}0.03 $&$ 1.5\pm{}0.1 $&$ 1.2\pm{}0.3 $&$ 2.04\pm{}0.07 $&$ 2.0\pm{}0.3 $&$ - $&$ 0.56\pm{}0.02 $\\
        MB4384 shell &$ - $&$ 0.04\pm{}0.01 $&$ 0.04\pm{}0.01 $&$ - $&$ 0.04\pm{}0.02 $&$ - $&$ 0.30\pm{}0.05 $&$ - $&$ - $&\\
        MB3955 cent. src &$ - $&$ 0.04\pm{}0.02 $&$ - $&$ - $&$ - $&$ - $&$ 0.84\pm{}0.07 $&$ - $&$ - $\\
        MB3955 shell &$ - $&$ 0.06\pm{}0.01 $&$ - $&$ - $&$ - $&$ - $&$ 1.5\pm{}0.1 $&$ - $&$ - $\\
        MB4121 cent. src &$ 5\pm{}1 $&$ - $&$ - $&$ 1.8\pm{}0.3 $&$ 0.98\pm{}0.04 $&$ 1.7\pm{}0.4 $&$ - $&$ 0.48\pm{}0.05 $&$ - $\\
        MB4121 shell &$ - $&$ - $&$ - $&$ - $&$ 0.13\pm{}0.04 $&$ - $&$ - $&$ 0.17\pm{}0.05 $&$ - $\\
        MB4124 cent. src &$ 4.7\pm{}0.7 $&$ - $&$ - $&$ 4.7\pm{}0.2 $&$ 14.5\pm{}0.1 $&$ 31\pm{}3 $&$ - $&$ 2.1\pm{}0.4 $&$ 5.9\pm{}0.5 $\\
        MB4124 shell &$ - $&$ 0.5\pm{}0.3 $&$ - $&$ - $&$ 1.0\pm{}0.3 $&$ 0.31\pm{}0.07 $&$ 1.8\pm{}0.8 $&$ - $&$ - $\\
        \\
        \hline
        \hline
        Object & H~\textsc{i}/H$_{\rm 2}$ & [Fe~\textsc{ii}] & [S~\textsc{iii}] & H~\textsc{i} & [Fe~\textsc{iii}] & [Fe~\textsc{ii}] & [Fe~\textsc{ii}] & He~\textsc{ii} & H$_{\rm 2}$ & [S~\textsc{iii}] \\
        & \unit{16.9}{\micro\meter} & \unit{17.9}{\micro\meter} & \unit{18.7}{\micro\meter} & \unit{19.0}{\micro\meter} & \unit{22.9}{\micro\meter} & \unit{24.5}{\micro\meter} & \unit{25.9}{\micro\meter} & \unit{27.9}{\micro\meter} & \unit{28.2}{\micro\meter} & \unit{33.5}{\micro\meter}\\
        \hline
        MB4384 cent.src &$ 0.19\pm{}0.05 $&$ - $&$ 1.0\pm{}0.1 $&$ 1.7\pm{}0.1 $&$ 1.00\pm{}0.08 $&$ - $&$ - $&$ 1.18\pm{}0.04 $&$ - $&$ - $\\
        MB4384 shell &$ - $&$ - $&$ 0.19\pm{}0.05 $&$ - $&$ 0.23\pm{}0.04 $&$ - $&$ - $&$ 0.07\pm{}0.02 $&$ 0.20\pm{}0.06 $&$ - $\\
        MB3955 cent.src &$ - $&$ - $&$ - $&$ 0.05\pm{}0.02 $&$ - $&$- $&$ - $&$ - $&$ - $&$ 4\pm{}2 $\\
        MB3955 shell &$ - $&$ - $&$ - $&$ - $&$ - $&$ - $&$ - $&$ 0.3\pm{}0.1 $&$ - $&$ 4\pm{}2 $\\
        MB4121 cent.src &$ 0.3\pm{}0.2 $&$ 7.5\pm{}0.4 $&$ 2.1\pm{}0.2 $&$ - $&$ - $&$ 1.7\pm{}0.3 $&$ 17.0\pm{}0.5 $&$ - $&$ 3.7\pm{}0.4 $&$ - $\\
        MB4121 shell &$ - $&$ - $&$ 0.4\pm{}0.1 $&$ - $&$ - $&$ - $&$ 2.9\pm{}032 $&$ 0.5\pm{}0.1 $&$ 2.7\pm{}0.4 $&$ - $\\
        MB4124 cent.src &$ 1.5\pm{}0.4 $&$ 29\pm{}4 $&$ 2.1\pm{}0.4 $&$ 13.6\pm{}0.5 $&$ - $&$ 7.7\pm{}0.5 $&$ 68.0\pm{}0.8 $&$ 8.0\pm{}0.1 $&$ - $&$ - $\\
        MB4124 shell &$ - $&$ - $&$ - $&$ 0.17\pm{}0.06 $&$ - $&$ - $&$ 1.7\pm{}0.7 $&$ - $&$ - $&$ 6\pm{}1 $\\
      \end{tabular}
    \end{center}
  \end{scriptsize}
\end{sidewaystable}